\documentclass[aps,prd,10pt,notitlepage,nofootinbib]{revtex4-1}

\usepackage[utf8]{inputenc}
\usepackage{amsmath,amssymb,amsfonts}
\usepackage{newtxtext,newtxmath}

\usepackage{bbold}
\usepackage{bm}
\usepackage{graphicx}
\usepackage[usenames,dvipsnames]{xcolor}
\usepackage{color}
\usepackage[colorlinks=true,linkcolor=blue,urlcolor=blue,citecolor=blue]{hyperref}

\usepackage{slashed}
\usepackage[english]{babel}
\usepackage{dcolumn}
\usepackage{pifont}
\usepackage{dsfont,mathrsfs}
\usepackage{cancel}
\usepackage{bigints}
\usepackage{accents}
\usepackage{soul}
\usepackage{multirow}

\usepackage{amsmath, amssymb}
\usepackage{bbold}
\usepackage{makecell}
\usepackage{simpler-wick}

\usepackage{orcidlink} 

\newcommand{\nn}{\nonumber}
\newcommand{\ket}[1]{\left|#1\right\rangle}

\newcommand{\ensembleaverage}[1]{\left\langle#1\right\rangle}

\newcommand{\MB}[1]{\left|#1\right|}

\newcommand{\FB}[1]{\left(#1\right)}
\newcommand{\fb}[1]{(#1)}
\newcommand{\SB}[1]{\left\{#1\right\}}
\newcommand{\TB}[1]{\left[#1\right]}
\newcommand{\AB}[1]{\left<#1\right>}

\newcommand{\scrM}{\mathscr{M}}
\newcommand{\scrMtil}{\tilde{\mathscr{M}}}

\newcommand{\munu}{{\mu\nu}}

\newcommand{\alphabeta}{{\alpha\beta}}
\newcommand{\IM}{\text{Im}}
\newcommand{\RE}{\text{Re}}
\newcommand{\Tr}{\text{Tr}}

\newcommand{\psibar}{\overline{\psi}}
\newcommand{\qbar}{\bar{q}}
\newcommand{\ubar}{\bar{u}}
\newcommand{\dbar}{\bar{d}}

\newcommand{\del}{\partial}
\newcommand{\identity}{\mathds{1}}

\newcommand{\wpr}{\omega^r_{\bm{p}}}
\newcommand{\wps}{\omega^s_{\bm{p}}}
\newcommand{\wkr}{\omega^r_{\bm{k}}}
\newcommand{\wks}{\omega^s_{\bm{k}}}
\newcommand{\fpkr}{f_+^{\bm{k}r}}
\newcommand{\fmkr}{f_-^{\bm{k}r}}
\newcommand{\fpps}{f_+^{\bm{p}s}}
\newcommand{\fmps}{f_-^{\bm{p}s}}
\newcommand{\fpks}{f_+^{\bm{k}s}}
\newcommand{\fmks}{f_-^{\bm{k}s}}

\newcommand{\wk}{\omega_{\bm{k}}}
\newcommand{\fpk}{f_+^{\bm{k}}}
\newcommand{\fmk}{f_-^{\bm{k}}}

\newcommand{\mans}{\mathfrak{s}}
\newcommand{\mant}{\mathfrak{t}}
\newcommand{\manu}{\mathfrak{u}}

\allowdisplaybreaks

\begin{document}
	\title{Effect of chiral imbalance on the electrical conductivity of hot and dense quark matter using Green-Kubo Method within the 2-flavour gauged NJL model}

\author{Snigdha Ghosh\orcidlink{0000-0002-2496-2007}$^{a}$}
\email{snigdha.ghosh@bangla.gov.in}
\email{snigdha.physics@gmail.com}
\thanks{Corresponding Author}

\author{Nilanjan Chaudhuri\orcidlink{0000-0002-7776-3503}$^{b,d}$}
\email{sovon.nilanjan@gmail.com}
\email{n.chaudhri@vecc.gov.in}

\author{Sourav Sarkar\orcidlink{0000-0002-2952-3767}$^{b,d}$}
\email{sourav@vecc.gov.in}

\author{Pradip Roy$^{c,d}$}
\email{pradipk.roy@saha.ac.in}

\affiliation{$^a$Government General Degree College Kharagpur-II, Paschim Medinipur - 721149, West Bengal, India}	
\affiliation{$^b$Variable Energy Cyclotron Centre, 1/AF Bidhannagar, Kolkata - 700064, India}
\affiliation{$^c$Saha Institute of Nuclear Physics, 1/AF Bidhannagar, Kolkata - 700064, India}
\affiliation{$^d$Homi Bhabha National Institute, Training School Complex, Anushaktinagar, Mumbai - 400085, India}

\begin{abstract}
The electrical conductivity of hot and dense quark matter is calculated using the 2-flavour gauged Nambu-Jona--Lasinio (NJL) model in the presence of a chiral imbalance quantified in terms of a chiral chemical potential (CCP). To this end, the in-medium spectral function corresponding to the vector current correlator is evaluated employing the real time formulation of finite temperature field theory. Taking the long wavelength limit of the spectral function we extract the electrical conductivity using the Green-Kubo relation. The thermal widths of the quarks/antiquarks that appear in the expression of electrical conductivity are calculated by considering the $2\to2$ scattering in the NJL model. The scattering amplitudes containing the polarization functions of the mesonic modes in the scalar and pseudoscalar channels are also evaluated by considering finite value of CCP. We find that the ratio of electrical conductivity to temperature has significant dependence on CCP especially in the low temperature region.
\end{abstract}

\maketitle

\section{Introduction}
The experimental heavy-ion programs conducted at facilities such as the Relativistic Heavy Ion Collider (RHIC), the Large Hadron Collider (LHC) and the upcoming Nuclotron-based Ion Collider Facility (NICA) and the Facility for Antiproton and Ion Research (FAIR) provide captivating opportunities to explore the behavior of strongly interacting matter under extreme conditions characterized by high temperature and/or density. Under these circumstances, hadronic matter is expected to undergo a transition into a quark-gluon plasma (QGP) state where quarks are liberated from their confinement within the hadrons. These experiments establish a crucial link with Quantum Chromodynamics(QCD) which is the underlying theory of quarks and gluons. This connection is facilitated through phenomenology which serves as the bridge connecting quantities that can be computed from first principles with the observables acquired from the experiments. Experimental observations of collective behavior~\cite{STAR:2005gfr} as well as theoretical and phenomenological investigations involving viscous hydrodynamics and parton transport, have provided strong evidence that the QGP behaves like a strongly interacting fluid~\cite{Csernai:2006zz,Romatschke:2007mq,Luzum:2008cw}. This fluid exhibits a very small shear viscosity to entropy density ratio, denoted as $\eta/s$, which is close to the theoretical lower bound of $1/4\pi$ predicted by AdS/CFT \cite{Kovtun:2004de}. Given that the QGP created in Heavy Ion Collisions (HICs) is far-from-equilibrium, the study of its transport coefficients holds paramount importance in characterizing the hydrodynamic evolution of the transient  matter  formed in such collisions.

It is well known that in QCD there exists a multitude of degenerate vacuum states each characterized by an integer-valued winding number and separated by potential energy barriers~\cite{Shifman:1988zk}. At lower temperatures these vacuum states can be explored through non-trivial topological gauge field configurations known as instantons~\cite{Belavin:1975fg,tHooft:1976rip,tHooft:1976snw}. However, at high temperatures, such as in the QGP phase, there is an expectation of abundant production of another type of gluon configuration called sphalerons~\cite{Manton:1983nd,Klinkhamer:1984di}. These topologically non-trivial gauge field configurations have the ability to change the helicity of quarks during interactions. This leads to the localized breaking of parity (P) and charge-parity (CP) symmetries through the axial anomaly of QCD~\cite{Adler:1969gk,Bell:1969ts}, creating an asymmetry between left and right-handed quarks. This locally induced chirality imbalance is quantified using a chiral chemical potential (CCP) which essentially represents the difference between the numbers of right and left-handed quarks/antiquarks.

There is a growing interest in understanding the impact of strong electromagnetic (EM) fields generated during the initial stages of non-central HICs on the strongly interacting matter~\cite{Tuchin:2013ie,Gursoy:2014aka}. To estimate how electrically charged QGP responds to these EM fields, one needs to examine the electrical conductivity ($\sigma_\text{el}$). Numerous recent efforts have been made to understand $\sigma_\text{el}$ within the context of EM fields in RHIC experiments~\cite{Tuchin:2013ie, McLerran:2013hla, Gursoy:2014aka, Satow:2014lia}. Additionally, estimations of electrical conductivity by measuring charge fluctuations in HICs can be found in Ref.~\cite{Ling:2013ksb}. There are also proposals to extract electrical conductivity from flow parameters observed in HICs~\cite{Hirono:2012rt}. Moreover, conductivity plays a pivotal role in charge transport, particle production, and the time evolution of electromagnetic fields in HICs, underscoring its phenomenological importance~\cite{Deng:2012pc,Hirono:2012rt,Tuchin:2013ie,Yin:2013kya,Kharzeev:2015znc,Pratt:2019pnd}. The usual definition of conductivity, employing the Kubo formula can be related to the Green’s function in quantum field theory in a specific limit. On the theoretical front, conductivity can be computed through various methods, ranging from weak-coupling approaches using Feynman diagrams~\cite{Aarts:2002tn,Gagnon:2006hi} and kinetic theory in QCD~\cite{Arnold:2000dr,Arnold:2003zc} or effective models~\cite{Cassing:2013iz,Marty:2013ita,Mitra:2016zdw} to strong-coupling methods employing holography~\cite{Policastro:2002se,Teaney:2006nc}. Significant efforts have also been made to estimate $\sigma_\text{el}$ in Lattice QCD calculations~\cite{Gupta:2003zh,Aarts:2007wj,Ding:2010ga,Francis:2011bt,Amato:2013naa,Aarts:2014nba}. The electromagnetic spectral function vis-a-vis the electrical conductivity can be extracted from the photon and dilepton spectra in a HIC experiment~\cite{Gupta:2003zh,Floerchinger:2021xhb,Atchison:2024lmf} as well.

In situations where chiral imbalance coexists with a magnetic field, such as in non-central HICs, an interesting phenomenon known as the chiral magnetic effect (CME) can occur~\cite{Kharzeev:2007jp,Skokov:2009qp}. This leads to the separation of positive and negative charges with respect to the reaction plane resulting in the generation of an electric current along the direction of the magnetic field. Extensive efforts have been made for the detection of the CME in HIC experiments  at RHIC at Brookhaven. However, a recent analysis by the STAR Collaboration has not provided any evidence of the CME occurring in these collisions~\cite{STAR:2021mii}. Consequently, new experimental techniques for detecting the CME have been proposed~\cite{An:2021wof,STAR:2022zpv}. Given the expected presence of locally generated chirality imbalance in QGP, an extensive research efforts have been dedicated to comprehending various aspects of this phenomenon. These investigations include microscopic transport phenomena ~\cite{Vilenkin:1979ui,Vilenkin:1980fu,Fukushima:2008xe,Son:2009tf}, collective oscillations ~\cite{Akamatsu:2013pjd,Carignano:2018thu,Carignano:2021mrn}, fermion damping rate~\cite{Carignano:2019ivp} and collisional energy loss of fermions~\cite{Carignano:2021mrn} in chirally imbalanced media. The phase structure of strongly interacting medium and its thermodynamic properties have been investigated in presence of CCP in  Refs.~\cite{Ruggieri:2016lrn,Ruggieri:2016asg,Ruggieri:2020qtq,Chaudhuri:2021lui,Fukushima:2010fe,Andrianov:2013dta,Braguta:2016aov,Azeredo:2024sqc,Espriu:2020dge} using various effective models, and in Refs.~\cite{Braguta:2015zta,Braguta:2015owi} using Lattice QCD. Moreover, the properties of the electromagnetic spectral function~\cite{Ghosh:2022xbf}, dilepton production rate~\cite{Chaudhuri:2022rwo}, mesonic modes~\cite{Andrianov:2012dj,GomezNicola:2023ghi,Ghosh:2023rft} are also studied in the presence of chiral imbalance. Additionally, chirally asymmetric plasmas are expected to form in the gap regions of the magnetospheres of pulsars and black holes~\cite{Gorbar:2021tnw} as well as in other stellar astrophysical scenarios~\cite{Charbonneau:2009ax,Akamatsu:2013pjd,Yamamoto:2015gzz,Shovkovy:2021yyw}. It is important to highlight that CME has been observed in condensed matter systems, specifically in 3D Dirac and Weyl semimetals~\cite{Li:2014bha,Kharzeev:2015znc}. Hence, the exploration of the characteristics of chirally imbalanced matter continues to be a subject of great interest. It is thus expected that the presence of a CCP will influence the transport properties of the medium and should be taken into consideration in cases where such a scenario exists.

The non-perturbative nature of QCD makes it challenging to study properties of QGP matter as well as its response functions from first principle calculations. Much of our current understanding is derived from Lattice QCD (LQCD) simulations~\cite{deForcrand:2006pv,Aoki:2006br,Aoki:2009sc,Bazavov:2009zn,Cheng:2007jq,Muroya:2003qs}. However, direct simulations of QCD at finite density encounter the so-called sign problem. This arises because the fermion determinant, necessary for calculating observables, becomes complex and hinders the application of conventional Monte Carlo methods. To circumvent this challenge, effective models become valuable tools for probing low-energy QCD. In this study, we employ the Nambu-Jona--Lasinio (NJL) model~\cite{Nambu:1961fr,Nambu:1961tp} which preserves the global symmetries of QCD, particularly the crucial chiral symmetry. The NJL model, as an effective model, provides a useful framework to investigate key non-perturbative properties of the QCD vacuum within its applicable range~\cite{Klevansky:1992qe,Vogl:1991qt,Buballa:2003qv,Volkov:2005kw}. In this effective description, the gluonic degrees of freedom are integrated out which results in point-like interactions among quarks~\cite{Klevansky:1992qe}. This makes the NJL model non-renormalizable. Therefore, it is necessary to choose a proper regularization scheme to deal with the divergent integrals. Subsequently, model parameters can be determined by matching to a set of phenomenological quantities, such as the pion-decay constant and quark condensate, among others. To this end, considerable number of works can be found in the literature aiming to calculate the transport coefficients (like viscosities and conductivities) of hot and/or dense QCD medium using NJL-like models such as in Refs.~\cite{Lang:2013lla,Lang:2015nca,Iwasaki:2007iv,Alberico:2007fu,Ghosh:2015mda}. However, to the best of our knowledge, we have not come across any calculation of the electrical conductivity of hot quark matter at finite CCP using NJL-like models.

In this work, we aim to calculate the electrical conductivity of hot and dense quark matter using the 2-flavour gauged NJL model at finite CCP employing the one-loop Green-Kubo formalism. The essential ingredient to calculate the $\sigma_\text{el}$ is the in-medium spectral function corresponding to the vector current correlator which will be evaluated employing the real time formulation (RTF) of thermal field theory (TFT). $\sigma_\text{el}$ will then be extracted from the long wavelength limit of the spectral function using Green-Kubo relation. The main dynamical input that enters in the expression of the electrical conductivity is the momentum dependent thermal widths of the quarks/antiquarks (which may be interpreted as the inverse of the relaxation time); which will be calculated from the the $2\to2$ scattering among the quarks/antiquarks. In the NJL model, the matrix elements of the scattering amplitude contain the polarization functions of the mesonic modes in the scalar and pseudoscalar channels. These polarization functions correspond to the one-loop graphs containing a quark/antiquark in the loop which will also be evaluated employing the RTF by considering finite value of CCP.

The paper is organized as follows. In Sec.~\ref{sec.gap}, the gap equation is discussed in which the constituent quark mass is obtained at finite temperature ($ T $), density ($ \mu $) and CCP ($ \mu_5 $). In Sec.~\ref{sec.kubo}, general formalism of the Green-Kubo method is used to calculate the conductivity tensor vis-a-vis the scalar electrical conductivity ($\sigma_\text{el}$) in presence of CCP employing the RTF. Sec.~\ref{sec.width} is devoted to the calculation of the thermal widths of the quark/antiquark which goes as the dynamical input in the expression of $\sigma_\text{el}$. The polarization functions in the scalar and pseudoscalar channels are evaluated at finite CCP in Sec.~\ref{sec.pola} which are essential ingredients in the calculation of the thermal widths of quarks and antiquarks. In Sec.~\ref{sec.results}, we show the numerical results and finally we summarize and conclude in Sec.~\ref{sec.summary}. Detailed calculations of the invariant amplitudes for $2\to2$ scattering processes among quarks/antiquarks are provided in Appendix~\ref{App.Amplitudes}.

\section{Gap Equation at Finite CCP} \label{sec.gap}
Let us start with the standard expression of the Lagrangian (density) for a 2-flavour gauged NJL model at finite CCP and quark chemical potential as~\cite{Ghosh:2022xbf,Chaudhuri:2022rwo,Ghosh:2023rft}:
\begin{eqnarray}
	\mathscr{L}_\text{NJL} = \psibar(i\cancel{\del}-m+\mu\gamma^0+\mu_5\gamma^0\gamma^5)\psi + G_s\TB{(\psibar\psi)^2+(\psibar i\gamma^5\bm{\tau}\psi)^2} -\frac{1}{4}F^\munu F_\munu - J^\mu_\text{em}A_\mu \label{lagrangian}
\end{eqnarray} 
where $\psi = \FB{u~d}^T$ is the quark iso-doublet, $m$ is the current quark mass, $A^\mu$ is the photon field, $F^\munu = (\del_\mu A_\nu - \del_\nu A_\mu)$ is the electromagnetic field strength tensor, $\bm{\tau}$ is the Pauli isospin matrices and $G_s$ represents the scalar coupling. The last term in the above Lagrangian represents the minimal coupling of the Fermion fields with the electromagnetic field in which the photon field couples with the conserved vector current $J^\mu_\text{em} = \psibar \hat{Q} e \gamma^\mu \psi $ where, $\hat{Q} = \text{diag}(Q_u, Q_d)$ is the charge fraction matrix in flavour space ($Q_u=2/3$ and $Q_d=-1/3$ respectively) and $e>0$ is the charge of a proton. Throught the article, we will be using metric tensor $g^\munu=\text{diag}(1,-1,-1,-1)$.

In the mean field approximation (MFA), the constituent quark mass $M$ is obtained by solving the gap equation,
\begin{eqnarray}
	M = m - 2G_s\ensembleaverage{\psibar \psi} \label{gap}~,
\end{eqnarray}
where the chiral condensate in the MFA is given by
\begin{eqnarray}
	\ensembleaverage{\psibar \psi} = \RE~i\int\frac{d^4p}{(2\pi)^4} \Tr_\text{d,c,f}\TB{S_{11}(p)} \label{condensate}~,
\end{eqnarray}
in which the $\Tr_\text{d,c,f}[...]$ implies taking trace in Dirac, colour and flavour spaces and $S_{11}(p)$ is the $11$ component of the real time thermal quark propagator. The explicit expression of $S_{11}(p)$ is given by~\cite{Mallik:2016anp}
\begin{eqnarray}
	S_{11}(p) = \mathcal{D}(p^0+\mu,\bm{p}) \sum_{r\in\{\pm\}}^{}\frac{1}{4|\bm{p}|r\mu_5}\TB{\frac{-1}{(p^0+\mu)^2-(\wpr)^2+i\epsilon}-2\pi i\eta(p^0+\mu)\delta\SB{(p^0+\mu)^2-(\wpr)^2}}\otimes\identity_\text{colour}\otimes\identity_\text{flavour}
	\label{S11}~,
\end{eqnarray}
where $r$ represents the helicity of the propagating quark, $\wpr = \sqrt{(|\bm{p}|+r\mu_5)^2+M^2}$ is the single particle energy in presence of CCP, $\eta(x)=\theta(x)f_+(x)+\theta(-x)f_-(-x)$, $\theta(x)$ is the unit step function and $f_\pm(x) = \dfrac{1}{e^{(x\mp\mu)/T}+1}$ is the Fermi-Dirac thermal distribution function and $\mathcal{D}$ contains complicated Dirac structure as follows:
\begin{eqnarray}
	\mathcal{D}(p^0,\bm{p}) = \sum_{j \in \{\pm\}} \mathscr{P}_j \TB{ p_{-j}^2\cancel{p}_j - M^2 \cancel{p}_{-j} + M (p_j\cdot p_{-j}-M^2) - i M \sigma_\munu p_j^\mu p_{-j}^\nu  } \label{D}~,
\end{eqnarray}
in which $\mathscr{P}_j = \frac{1}{2}(\identity+j\gamma^5)$ and $p^\mu_j \equiv (p^0 +j\mu_5,\bm{p})$.  The sign of the $\sigma_\munu$-term in Eq~\eqref{D} has been corrected from a typographic error present in our earlier work in Ref.~\cite{Ghosh:2022xbf}. Substituting the propagator from Eq.~\eqref{S11} into Eq.~\eqref{condensate} and evaluating the $dp^0$ integration we get,
\begin{eqnarray}
	\ensembleaverage{\psibar\psi} = -N_c N_f M \sum_{r \in \{\pm\}} \int \frac{d^3p}{(2\pi)^3} \frac{1}{\wpr} \FB{1-f_+^{\bm{p}r} -f_-^{\bm{p}r}}
	\label{condensate.2}~,
\end{eqnarray}
where $N_c=3$ is the number of colors, $N_f=2$ is the number of flavours and $f_\pm^{\bm{p}r} = f_\pm(\wpr)$. 

The MFA mentioned here is the usual Hartree or in the Hartree-Fock approximation. The effective four-fermion coupling $ G_s $ is large as it is supposed to include non-perturbative (gluonic) dynamics. Hence choice of $ G_s $ as an expansion parameter is not meaningful~\cite{Quack:1993ie,Lang:2015nca}. However, calculations beyond mean-field theory can be carried out by expanding in powers of the inverse number of colors, $1/N_c$, which serves as the organizing scheme for the contributing diagrams. It is important to note that, in QCD, the color gauge symmetry is local, while in the NJL model, it is reduced to a global symmetry. In this large-$N_c$ expansion, a hierarchy of Dyson-Schwinger equations emerges, where the leading order, $\mathcal{O}(N_c^0)$, corresponds to the NJL gap equation in the Hartree approximation. The next-to-leading order in $1/N_c$ accounts for corrections to the quark self-energy, originating from the `meson cloud' that surrounds the quark and can either screen or anti-screen the bare four-point interaction. However, as discussed in Ref.~\cite{Quack:1993ie}, the meson cloud introduces minor quantitative changes but no significant qualitative changes.

The first term within the parenthesis in the above integral is the temperature independent vacuum contribution to the chiral condensate which is UV divergent. We will be using the smooth three-momentum cutoff regulator and replace the vacuum terms as
\begin{eqnarray}
\int\dfrac{d^3p}{(2\pi)^3}1 \to \int\dfrac{d^3p}{(2\pi)^3} f_\Lambda^{\bm{p}} \label{regularization}~,
\end{eqnarray}
where $f_\Lambda^{\bm{p}} = \sqrt{\dfrac{\Lambda^{20}}{\Lambda^{20} + |\bm{p}|^{20}}}$. Substituting Eq.~\eqref{condensate.2} in Eq.~\eqref{gap} and making use of Eq.~\eqref{regularization}, we finally obtain the gap equation as
 \begin{eqnarray}
 	M = m + 2G_s N_c N_f M \sum_{r \in \{\pm\}} \int \frac{d^3p}{(2\pi)^3} \frac{1}{\wpr} \FB{f_\Lambda^{\bm{p}} -f_+^{\bm{p}r} -f_-^{\bm{p}r}}~. \label{gap.final}
 \end{eqnarray}

Few comments on the use of a $\mu_5$-independent smooth three-momentum cutoff regularization prescription are in order here. Expanding the first term within the bracket of Eq.~\eqref{gap.final}, we get 
	\begin{eqnarray}
		\int d^3p f_\Lambda^{\bm{p}} \frac{1}{\wpr} = \int d^3p f_\Lambda^{\bm{p}} \TB{ \frac{1}{|\bm{p}|} -\frac{M^2}{2|\bm{p}|^3} - \frac{r\mu_5}{|\bm{p}|^2} + \frac{r^2\mu_5^2}{|\bm{p}|^3} +\mathcal{O}\FB{\frac{1}{|\bm{p}|}}^4 }~.
	\end{eqnarray}
	It can be seen that at large three-momentum $|\bm{p}|\to\infty$, the third and fourth terms within the square brackets, both of which depend on $\mu_5$, exhibit different types of divergent behavior. Notably, the third term demonstrates a linear UV divergence, whereas the fourth term manifests a logarithmic UV divergence. The NJL model being non-renormalizable, we have restricted ourselves to regularization of the above expression by using a smooth three-momentum cutoff via the introduction of an overall form- factor $f_\Lambda^{\bm{p}}$ containing only one parameter $\Lambda$ on the LHS of the above equation. The regularization procedure is not unique and one can equally consider a $\mu_5$-dependent expression of the form-factor. However, qualitatively the results are expected to remain unchanged by such $\mu_5$-dependent regulator and is similar to using different regulators (such as the Pauli-Villars, sharp-cutoff, proper-time, etc.)~\cite{Klevansky:1992qe}.
\section{Electrical Conductivity at finite CCP using one-loop Green-Kubo Formalism} \label{sec.kubo}
In the Green-Kubo formalism, the electrical conductivity can be calculated from the in-medium spectral function $\rho^\munu$ which is related to the Fourier transform of the $11$-component of time-ordered two-point vector current correlation function $ \ensembleaverage{\mathcal{T}_C J^\mu_\text{em}(x)J^\nu_\text{em}(0)}_{11}$ given by
\begin{eqnarray}
	\rho^\munu(q) = \tanh\FB{\frac{q^0}{2T}} \IM~ i \int d^4x e^{iq\cdot x} \ensembleaverage{\mathcal{T}_C J^\mu_\text{em}(x)J^\nu_\text{em}(0)}_{11} \label{sp.fn}~,
\end{eqnarray} 
where $\mathcal{T}_C$ is the time-ordering with respect to the symmetric Schwinger-Keyldish type complex time contour as used in the RTF of TFT. Substituting the expression of current $J^\mu_\text{em}(x)$ defined earlier into Eq.~\eqref{sp.fn} and applying the Wick's theorem, we obtain
\begin{eqnarray}
	\rho^\munu(q) = \tanh\FB{\frac{q^0}{2T}} \IM~ i \int d^4x e^{iq\cdot x} (-)\Tr_\text{d,c,f} \TB{\gamma^\mu \hat{Q} S_{11}(x)\gamma^\nu \hat{Q} S_{11}(-x)} \label{sp.fn.2}~,
\end{eqnarray} 
where $S_{11}(x) = \ensembleaverage{\mathcal{T}_C\psi(x)\psibar(0)}_{11}$ is the $11$-component of the real-time thermal quark propagator in coordinate space which can be written as 
\begin{eqnarray}
	S_{11}(x) = \int\frac{d^4p}{(2\pi)^4} e^{-ip\cdot x} \{-i S_{11}(p)\} \label{S11x}
\end{eqnarray}
in which $S_{11}(p)$ is the corresponding momentum space propagator given in Eq.~\eqref{S11}. Substituting Eq.~\eqref{S11x} into Eq.~\eqref{sp.fn.2}, we get the following expression of the spectral function
\begin{eqnarray}
	\rho^\munu(q) = \tanh\FB{\frac{q^0}{2T}} \IM~ i \int\frac{d^4k}{(2\pi)^4} \Tr_\text{d,c,f} \TB{\gamma^\mu \hat{Q} S_{11}(p=q+k)\gamma^\nu \hat{Q} S_{11}(k)} \label{sp.fn.3}~.
\end{eqnarray} 
Now substituting the quark propagator from Eq.~\eqref{S11} into Eq.~\eqref{sp.fn.3} and performing the $dk^0$ integral, we obtain
\begin{eqnarray}
		\rho^\munu (q_0,\bm{q}) &=& -\tanh\FB{\frac{q^0}{2T}}N_c e^2\sum_{f \in \{u,d\} }Q_f^2 \pi \int \frac{d^3k}{(2\pi)^3} \sum_{r \in \{\pm\}} \sum_{s \in \{\pm\}} 
	\frac{1}{16rs\mu_5^2 |\bm{p}||\bm{k}|} \frac{1}{4\wkr\wps} \nn \\ && \hspace{-2cm}
	 \times \Big[ \mathcal{N}^\munu(k^0=-\wkr) \fb{1-\fmkr -\fpps +2\fmkr\fpps}\delta(q_0-\wkr-\wps) 
 + \mathcal{N}^\munu(k^0=\wkr)  \fb{1-\fpkr -\fmps +2\fpkr\fmps} \delta(q_0+\wkr+\wps) \nn \\ && \hspace{-2cm}
	 + \mathcal{N}^\munu(k^0=\wkr) \fb{-\fpkr -\fpps +2\fpkr\fpps} \delta(q_0+\wkr-\wps) 
	+ \mathcal{N}^\munu(k^0=-\wkr) \fb{-\fmkr -\fmps +2\fmkr\fmps} \delta(q_0-\wkr+\wps) \Big],
	\label{sp.fn.4}
\end{eqnarray}
where 
\begin{eqnarray}
	\mathcal{N}^\munu (k;q) = \Tr_\text{d}\TB{\gamma^\mu\mathcal{D}(p=q+k)\gamma^\nu\mathcal{D}(k)} ~,\label{N}
\end{eqnarray}
 in which $\mathcal{D}(p^0,\bm{p})$ is defined in Eq.~\eqref{D}.

The vector current spectral function $\rho^\munu(q)$ in Eq.~\eqref{sp.fn.4} contains sixteen Dirac delta functions specifying the branch cuts of the thermal self energy function in the complex energy ($q_0$) plane. The terms with $\delta(q_0-\wkr-\wps)$ and $\delta(q_0+\wkr+\wps)$ are respectively called Unitary-I and Unitary-II cuts; whereas the terms with $\delta(q_0+\wkr-\wps)$ and $\delta(q_0-\wkr+\wps)$ are called Landau-I and Landau-II cuts respectively. Each of these cuts again contains four sub-cuts corresponding to the different values of helicities $ (r,s) $. These Unitary and Landau cuts correspond to different physical processes like decay and scattering (absorption/emission). A detailed analysis on the analytic structure of the spectral function can be found in~\cite{Ghosh:2022xbf,Ghosh:2023rft}. 

Since we are only interested in calculating the transport coefficient (the electrical conductivity), we need to consider the static or long-wavelength limit ($\bm{q}=\bm{0}, q^0\to0$) of the spectral function $\rho^\munu (q_0,\bm{q})$ in which case the Unitary cuts as well as the Landau cuts with $r\ne s$ do not contribute and we are left with
\begin{eqnarray}
	\rho^\munu (q_0,\bm{0}) &=&  \lim\limits_{\Gamma^r_{q, \bar{q}}\to 0} \tanh\FB{\frac{q^0}{2T}}N_c e^2\sum_{f \in \{u,d\} }Q_f^2 \int \frac{d^3k}{(2\pi)^3} \sum_{r \in \{\pm\}}  \frac{1}{32\mu_5^2 |\bm{k}|^2(\wkr)^2}   \nn \\ && \times \Big[  \mathcal{N}^\munu(k^0=\wkr) \fpkr\fb{1-\fpkr} \frac{\Gamma^r_q}{(\Gamma^r_q)^2+(q^0)^2}
	+ \mathcal{N}^\munu(k^0=-\wkr)\fmkr \fb{1-\fmkr } \frac{\Gamma^r_{\bar q}}{(\Gamma^r_{\bar q})^2+(q^0)^2} \Big]
	\label{sp.fn.5}~,
\end{eqnarray}
where a Breit-Wigner representation of the Dirac delta function $\delta(x) = \lim\limits_{\Gamma \to 0} \frac{1}{\pi} \FB{\frac{\Gamma}{\Gamma^2+x^2}}$ has been used. 

Now the electrical conductivity tensor $\sigma^\munu$ is obtained from the spectral function using the formula: 
\begin{eqnarray}
	\sigma^\munu = \frac{\del\rho^\munu(q^0,\bm{0})}{\del q^0}\Big|_{q^0\to 0} &=&
	\lim\limits_{\Gamma^r_{q, \bar{q}}\to 0} \frac{1}{2T}N_c e^2\sum_{f \in \{u,d\} }Q_f^2 \int \frac{d^3k}{(2\pi)^3} \sum_{r \in \{\pm\}} \frac{1}{32\mu_5^2 |\bm{k}|^2(\wkr)^2}   \nn \\ && \times \Big[  \mathcal{N}^\munu(k^0=\wkr;q=0) \fpkr\fb{1-\fpkr}\frac{1}{\Gamma^r_q} + \mathcal{N}^\munu(k^0=-\wkr;q=0)\fmkr \fb{1-\fmkr }\frac{1}{\Gamma^r_{\bar q}} \Big] 	\label{sp.fn.6}~,
\end{eqnarray}
where the explicit expression of $\mathcal{N}^\munu(k;q=0)$ is given by 
\begin{eqnarray}
	\mathcal{N}^\munu(k;q=0) =  \Tr_\text{d}\TB{\gamma^\mu\mathcal{D}(k)\gamma^\nu\mathcal{D}(k)}  &=&  
	-2g^\munu(k_+^2+k_-^2-2M^2)(k_+^2k_-^2-2M^2k_+\cdot k_- +M^4) + 4k_+^\mu k_+^\nu (k_-^2-M^2)^2 \nn \\ &&
	+ 4k_-^\mu k_-^\nu (k_+^2-M^2)^2 - 4M^2(k_+^\mu k_-^\nu+k_+^\nu k_-^\mu)(k_+-k_-)^2~, \label{N.2}
\end{eqnarray}
which is obtained by substituting Eq.~\eqref{D} in Eq.~\eqref{N}.
The electrical conductivity $\sigma_\text{el}$ is now extracted from the conductivity tensor using proper projection tensor~\cite{Harutyunyan:2017ttz} as
\begin{eqnarray}
	\sigma_\text{el} = - \frac{1}{3}g^\alphabeta \Delta_{\alpha\mu}\Delta_{\beta\nu}\sigma^\munu \label{cond}~,
\end{eqnarray}
where $\Delta^\munu = (g^\munu-u^\mu u^\nu)$ and $u^\mu$ is the medium four-velocity which in the local rest frame (LRF) becomes $u^\mu_\text{LRF} \equiv (1,\bm{0})$. Substituting Eq.~\eqref{sp.fn.6} in Eq.~\eqref{cond} yields
\begin{eqnarray}
	\sigma_\text{el} = \lim\limits_{\Gamma^r_{q, \bar{q}}\to 0} \frac{1}{2T}N_c e^2\sum_{f \in \{u,d\} }Q_f^2 \int \frac{d^3k}{(2\pi)^3} \sum_{r \in \{\pm\}} \frac{1}{32\mu_5^2 |\bm{k}|^2(\wkr)^2}  \TB{  \mathcal{N}_+ \fpkr\fb{1-\fpkr}\frac{1}{\Gamma^r_q} + \mathcal{N}_-\fmkr \fb{1-\fmkr }\frac{1}{\Gamma^r_{\bar q}} } ~,	\label{cond.2}
\end{eqnarray}
where
\begin{eqnarray}
	\mathcal{N}_\pm = - \frac{1}{3}g^\alphabeta \Delta_{\alpha\mu}\Delta_{\beta\nu}\mathcal{N}^\munu(k^0=\pm \wkr; q=0)~. 
	\label{N.3}
\end{eqnarray}
Making use of Eq.~\eqref{N.2} $\mathcal{N}_\pm$ of Eq.~\eqref{N.3} simplifies to  $\mathcal{N}_\pm = \frac{64}{3} |\bm{k}|^2 \mu_5^2 \FB{|\bm{k}|+r\mu_5}^2 $ and using  this in Eq.~\eqref{cond.2} we get the final expression of electrical conductivity in the presence of CCP as
\begin{eqnarray}
	\sigma_\text{el} (T,\mu,\mu_5) = \lim\limits_{\Gamma^r_{q, \bar{q}}\to 0} \frac{1}{3T}N_c e^2\sum_{f \in \{u,d\} }Q_f^2 \int \frac{d^3k}{(2\pi)^3} \sum_{r \in \{\pm\}} \frac{\FB{|\bm{k}|+r\mu_5}^2}{(\wkr)^2}  \TB{ \fpkr\fb{1-\fpkr}\frac{1}{\Gamma^r_q} + \fmkr \fb{1-\fmkr }\frac{1}{\Gamma^r_{\bar q}} }~. \label{cond.3}
\end{eqnarray}

It is to be noted that in Eq.~\eqref{cond.3}, $\Gamma^r_{q, \bar{q}}$ are infinitesimal parameters of the Breit-Wigner representation of the Dirac delta function which is put by hand and as $\Gamma^r_{q, \bar{q}}\to0$, the electrical conductivity diverges. This actually corresponds to the well known `pinch singularity' problem~\cite{Hosoya:1983id,Mallik:2016anp} of the one-loop Kubo formalism for a non-interacting system. Thus, in order to get a finite electrical conductivity, we must take finite values of $\Gamma^r_{q, \bar{q}}$ which is equivalent to considering interaction in the medium. The $\Gamma^r_{q, \bar{q}} = \Gamma^r_{q, \bar{q}}(\bm{k};T,\mu,\mu_5)$ can then be identified as the momentum-dependent thermal width of the quarks/aniquarks with helicity $r$ in the medium which we will calculate in the next section.

It is important to note that taking the limit $\mu_5\to0$ of the above equation one gets back the well known expression of 	$\sigma_\text{el}(T,\mu)$ in absence of $\mu_5$: 
\begin{eqnarray}
	\sigma_\text{el} (T,\mu,\mu_5=0) = \lim\limits_{\Gamma_{q, \bar{q}}\to 0} \frac{2}{3T}N_c e^2\sum_{f \in \{u,d\} }Q_f^2 \int \frac{d^3k}{(2\pi)^3}  \frac{|\bm{k}|^2}{\wk^2}   \TB{ \fpk\fb{1-\fpk}\frac{1}{\Gamma_q} + \fmk \fb{1-\fmk }\frac{1}{\Gamma_{\bar q}} }~, \label{cond.4}
\end{eqnarray}
with $\wk = \sqrt{|\bm{k}|^2+M^2}$ and $f_\pm^{\bm{k}} = f_\pm(\wk)$. Eq.~\eqref{cond.4} can either be obtained from kinetic theory employing the relaxation time approximation (RTA) (on identifying relaxation time of quark/antiquark $\tau_{q, \bar{q}} = \Gamma_{q, \bar{q}}^{-1}$)~\cite{Harutyunyan:2016rxm} or from Green-Kubo method at $\mu_5=0$~\cite{Fernandez-Fraile:2005bew}.

Few comments on the applicability of one-loop Kubo formalism for the estimation of the transport coefficients in NJL-type effective models are in order here. The two-point correlation function $\ensembleaverage{\mathcal{T}_C J^\mu_\text{em}(x)J^\nu_\text{em}(0)}_{11}$ appearing in Eq.~\eqref{sp.fn} in general contains the Heisenberg current $J^{\mu,H}_\text{em}(x)$ for an interacting system. This would imply that Eq.~\eqref{sp.fn.2} is only the leading order (one-loop) contribution to the spectral function coming from the first term of a perturbative expansion of the correlator $\ensembleaverage{\mathcal{T}_C J^{\mu,H}_\text{em}(x)J^{\nu,H}_\text{em}(0)}_{11}$. As shown in Refs.~\cite{Jeon:1994if,Jeon:1995zm,ValleBasagoiti:2002ir}, such perturbative treatment of the spectral function for the calculation of transport coefficients will break-down because of the specific coupling-dependence of the thermal widths $\Gamma_{q, \bar{q}}$ that was introduced to avoid the pinching singularities. As a cure, an infinite sum of ladder diagrams should be considered~\cite{Jeon:1994if,Jeon:1995zm,Carrington:2001ms}. However, the resummation of ladder diagrams can be implemented in NJL-type models by considering large $ N_c $-expansion. But these are found to be of subleading order in $ 1/N_c  $  because each rank in the ladder gives rise to a suppression factor $ G_s N_c \sim 1/N_c $. Thus in the NJL-type model as used in this work, the use of constituent quark mass $M=M(T,\mu_5)$ in the quark propagator along with  the consideration of finite widths $\Gamma_{q, \bar{q}}$ of the quark spectral function (the bare spectral function is Dirac delta functions) actually corresponds to a large $N_C$ expansion~\cite{Lang:2013lla}.
%
%
\section{Thermal Width of the quark: Dynamical Inputs} \label{sec.width}
In this section we will calculate the thermal widths $\Gamma^s_{q,\bar{q}}(\bm{p};T,\mu,\mu_5)$ of the quarks/antiquarks using the NJL model. Considering only the $2\to2$ scattering processes among the quarks and antiquarks, $\Gamma^s_{q,\bar{q}}$ of quark and antiquark with helicity $s$ are respectively given by~\cite{Chakraborty:2010fr}
\begin{eqnarray}
	\Gamma^s_q(\bm{p}) &=& \tau^s_q(\bm{p})^{-1} = g \int \frac{d^3k}{(2\pi)^3} \sum_{r \in \{\pm\}} \fpkr\fb{1-\fpkr} v^{rs}_\text{rel} \sigma_{qq\to qq}(\mans) + g \int \frac{d^3k}{(2\pi)^3} \sum_{r \in \{\pm\}}  \fmkr\fb{1-\fmkr} v^{rs}_\text{rel} \sigma_{q\qbar\to q\qbar}(\mans), \label{gam.p}\\
	\Gamma^s_{\bar q}(\bm{p}) &=& \tau^s_{\bar q}(\bm{p})^{-1} = g \int \frac{d^3k}{(2\pi)^3} \sum_{r \in \{\pm\}} \fpkr\fb{1-\fpkr} v^{rs}_\text{rel} \sigma_{\qbar q\to \qbar q}(\mans) + g \int \frac{d^3k}{(2\pi)^3} \sum_{r \in \{\pm\}} \fmkr\fb{1-\fmkr} v^{rs}_\text{rel} \sigma_{\qbar\qbar\to \qbar\qbar}(\mans) \label{gam.m}
\end{eqnarray}
where $g=N_cN_f$ is the degeneracy factor, $v^{rs}_\text{rel} = \dfrac{1}{2\wkr\wps} \lambda^\frac{1}{2}\FB{(\wkr+\wps)^2,M^2,M^2}$ is the relative velocity of the initial state particles, and $\lambda(x,y,z)=x^2+y^2+z^2-2xy-2yz-2zx$ is the K\"all\'en function. In Eqs.~\eqref{gam.p} and \eqref{gam.m} $\sigma_{AB\to AB}$ (with $A\in\{q,\qbar\}$ and $B\in\{q,\qbar\}$) denotes the isospin averaged cross sections for the process $A(k)B(p)\to A(k')B(p')$ which can be calculated using
\begin{eqnarray}
	\sigma_{AB\to AB} (\mans) &=& \frac{1}{16\pi\lambda(\mans,M^2,M^2)}\frac{1}{(1+\delta_{A,B})}\int_{\mant_\text{min}}^{0}d\mant \overline{\MB{\scrM_{AB}}^2} \label{xsection}~,
\end{eqnarray}
where $\mant_\text{min} = -\lambda(\mans,M^2,M^2)/\mans$, 
$\mans=(k+p)^2=(\wkr+\wps)^2-(\bm{k}+\bm{p})^2$ and $\mant=(k-k')^2$ are the Mandlestam variables, $\overline{\MB{\scrM_{AB}}^2} $ is the spin-isospin averaged squared invariant amplitude for the process $A(k)B(p)\to A(k')B(p')$ and the factor $\frac{1}{(1+\delta_{A,B})}$ on the right hand side takes care of the identical initial state particles. The calculation of the matrix elements $\overline{\MB{\scrM_{AB}}^2} $ is provided in Appendix~\ref{App.Amplitudes} and we can read off from Eqs.~\eqref{M.qq} and \eqref{M.qqb} the following expressions for the matrix elements:
\begin{eqnarray}
	\overline{\MB{\scrM_{qq}}^2} = \overline{\MB{\scrM_{\qbar\qbar}}^2} &=& 3\mant^2|D_\mant^\pi|^2 + 3\manu^2|D_\manu^\pi|^2 + (4M^2-\mant)^2|D_\mant^\sigma|^2 + (4M^2-\manu)^2|D_\manu^\sigma|^2 + \frac{1}{2N_C} \RE \big[ 3\mant\manu D_\mant^\pi D_\manu^{\pi*} \nn \\
	&& - 3\mant(4M^2-\manu) D_\mant^\pi D_\manu^{\sigma*} - 3\manu(4M^2-\mant) D_\mant^\sigma D_\manu^{\pi*} + \{(4M^2-\mant)(4M^2-\manu)+2\mant\manu\} D_\mant^\sigma D_\manu^{\sigma*}  \big], \label{M2qq}\\
	\overline{\MB{\scrM_{q\qbar}}^2} = \overline{\MB{\scrM_{\qbar q}}^2} &=& 3\mans^2|D_\mans^\pi|^2 + 3\mant^2|D_\mant^\pi|^2 + (4M^2-\mant)^2|D_\mant^\sigma|^2 + (4M^2-\mans)^2|D_\mans^\sigma|^2 + \frac{1}{2N_C} \RE \big[ 3\mans\mant D_\mans^\pi D_\mant^{\pi*} \nn \\
	&& - 3\mans(4M^2-\mant) D_\mans^\pi D_\mant^{\sigma*}  - 3\mant(4M^2-\mans) D_\mans^\sigma D_\mant^{\pi*} + \{(4M^2-\mans)(4M^2-\mant)-2\mans\mant\} D_\mans^\sigma D_\mant^{\sigma*}  \big] \label{M2qqb}~,
\end{eqnarray}
where $D_{\mans,\mant,\manu}^h = D^h(q_{\mans,\mant,\manu})$, $q_\mans=(k+p)$, $q_\mant=(k-k')$, $q_\manu=(k-p')$ and $D^h(q;T,\mu,\mu_5)$ is defined in Eq.~\eqref{Dh}:
\begin{eqnarray}
	D^h(q;T,\mu,\mu_5) = \frac{2G}{1-2G\Pi_h(q;T,\mu,\mu_5)} \label{Dh.f}~,
\end{eqnarray}
in which $\Pi_h(q;T,\mu_5)$ is the polarization function in the mesonic channel $h$. The symmetry of the system enables us to write $\Pi_h(q) = \Pi_h(q^0,|\bm{q}|)$ at finite temperature and CCP (i.e. $\Pi_h(q)$ separately depends on $q^0$ and $|\bm{q}|$ but does not depend on the direction of $\bm{q}$). Moreover, in the calculation of the scattering cross section we do not actually require $\Pi_h(q^0,|\bm{q}|)$ for arbitrary values of $q^0$ and $|\bm{q}|$, rather the values of $  \Pi_h $ for the following special cases are sufficient for our purpose~\cite{Zhuang:1995uf}: (i) $\Pi^h(q^0,|\bm{q}|=0)$, and (ii) $\Pi_h(q^0=0,|\bm{q}|)$. This is done by exploiting the Lorentz invariance. In particular if the momentum of the exchanged mesonic mode $q$ is time-like ($q^2>0$) we can choose $(q^0=\sqrt{q^2},|\bm{q}|=0)$ and if $q$ is space-like ($q^2<0$) we choose $(q^0=0,|\bm{q}|=\sqrt{-q^2})$. The detailed calculation of $\Pi_h(q)$ at finite temperature and CCP will be described in the next section.

%
\section{Polarization Functions at finite CCP and temperature} \label{sec.pola}
In this section we will briefly discuss the calculation of one-loop  polarization function $\Pi_h(q;T,\mu,\mu_5)$ at finite CCP and temperature using the RTF of finite temperature field theory following~\cite{Ghosh:2023rft}. Being a two-point correlation function, the real time polarization function becomes a $2\times2$ matrix structure (in thermal space) whose $11$-component is given by
\begin{eqnarray}
	\Pi_h^{11}(q;T,\mu,\mu_5) = i\int\!\!\frac{d^4k}{(2\pi)^4}\Tr_\text{d,c,f}\TB{\Gamma_hS_{11}(p=q+k)\Gamma_hS_{11}(k)} \label{Pi.11}~,
\end{eqnarray}
where $h\in \{\pi,\sigma\}$, $\Gamma_\pi = i\gamma^5$, $\Gamma_\sigma = \identity$ and the quark propagator $S_{11}(p)$ is defined in Eq.~\eqref{S11}. Now substituting Eq.~\eqref{S11} in Eq.~\eqref{Pi.11} and performing the $dk^0$ integration, we obtain the real and imaginary parts of the 11-component of mesonic polarization function as
\begin{eqnarray}
		\RE\Pi_h^{11}(q_0,\bm{q}) &=& N_cN_f \int \frac{d^3k}{(2\pi)^3} \sum_{r \in \{\pm\}} \sum_{s \in \{\pm\}} 
	\frac{1}{64rs\mu_5^2 |\bm{p}||\bm{k}|} 
	 \mathcal{P} \Big[ \frac{\mathcal{N}_h(k^0=\wkr)(f_\Lambda^{\bm{k}}-2\fpkr)}{\wkr \SB{(q^0+\wkr)^2-(\wps)^2}}	
	+ \frac{\mathcal{N}_h(k^0=-\wkr)(f_\Lambda^{\bm{k}}-2\fmkr)}{\wkr \SB{(q^0-\wkr)^2-(\wps)^2}} \nn \\	
	&& + \frac{\mathcal{N}_h(k^0=-q^0+\wps)(f_\Lambda^{\bm{k}}-2\fpps)}{\wps \SB{(q^0-\wps)^2-(\wkr)^2}}
	+ \frac{\mathcal{N}_h(k^0=-q^0-\wps)(f_\Lambda^{\bm{k}}-2\fmps)}{\wps \SB{(q^0+\wps)^2-(\wkr)^2}} \Big] \label{RePi.1} ~,\\
	\IM\Pi_h^{11}(q_0,\bm{q}) &=& -N_cN_f \pi \int \frac{d^3k}{(2\pi)^3} \sum_{r \in \{\pm\}} \sum_{s \in \{\pm\}} 
	\frac{1}{16rs\mu_5^2 |\bm{p}||\bm{k}|} \frac{1}{4\wkr\wps} \nn \\ && \hspace{-2.5cm}
	\times \Big[ \mathcal{N}_h(k^0=-\wkr) \fb{1-\fmkr -\fpps +2\fmkr\fpps}\delta(q_0-\wkr-\wps) 
	+ \mathcal{N}_h(k^0=\wkr)  \fb{1-\fpkr -\fmps +2\fpkr\fmps} \delta(q_0+\wkr+\wps) \nn \\ && \hspace{-2.5cm}
	+ \mathcal{N}_h(k^0=\wkr) \fb{-\fpkr -\fpps +2\fpkr\fpps} \delta(q_0+\wkr-\wps) 
	+ \mathcal{N}_h(k^0=-\wkr) \fb{-\fmkr -\fmps +2\fmkr\fmps} \delta(q_0-\wkr+\wps) \Big]
	\label{ImPi.1}~,
\end{eqnarray}
where the smooth-cutoff $f_\Lambda^{\bm{k}}$ has been used to regulate the UV-divergent part of $\RE\Pi_h^{11}(q_0,\bm{q})$. $\mathcal{P}$ denotes Cauchy principal value integration and
\begin{eqnarray}
	\mathcal{N}_h(k) &=& 4a_hM^6 + 2M^4 \big\{-2a_h (k_+\cdot k_-)+(k_+\cdot p_-)+(k_-\cdot p_+)-2a_h(p_+\cdot p_-)\big\}-2M^2 \big\{ -2a_h(k_+\cdot p_-)(k_+\cdot p_+) \nn \\ 
	 && -2a_h(k_+\cdot k_-)(p_+\cdot p_-) + 2a_h(k_+\cdot p_+)(k_-\cdot p_-) + (k_-\cdot p_-)(k_+^2+p_+^2) + (k_+\cdot p_+)(k_-^2+p_-^2) \big\} \nn \\
	 && + 2 \big\{ (k_+\cdot p_-)k_-^2p_+^2 + (k_-\cdot p_+)k_+^2p_-^2 \big\} \label{Nh}~,
\end{eqnarray} 
in which $a_\pi=-1$ and $a_\sigma=1$. Having calculated the 11-components, it is now easy to diagonalize the real-time thermal polarization matrix~\cite{Mallik:2016anp} and extract the diagonal components which are analytic functions that appear in Eq.~\eqref{Dh.f}. The real and imaginary parts of the analytic function $\Pi_h(q)$ are related to the 11-components in the following way:
\begin{eqnarray}
	\RE\Pi_h(q) = \RE\Pi_h^{11}(q) ~~\text{and}~~~ \IM\Pi_h(q) = \text{sign}(q^0)\tanh\FB{\frac{q^0}{2T}}\IM\Pi_h^{11}(q) \label{bar.to.11}~.
\end{eqnarray}
As already mentioned in the previous section, while calculating the scattering cross section, we only need to calculate $ \Pi_h $ for two special cases. For $q^0\ne0,\bm{q}=\bm{0}$, substitution of Eqs.~\eqref{RePi.1} and \eqref{ImPi.1} in Eq.~\eqref{bar.to.11} yields: 
\begin{eqnarray}
	\RE\Pi_h(q_0,\bm{q}=\bm{0};T,\mu,\mu_5) &=& \frac{N_cN_f}{2\pi^2} \int_{0}^{\infty}d|\bm{k}| \sum_{r \in \{\pm\}} \sum_{s \in \{\pm\}} 
	\frac{1}{64rs\mu_5^2} 
	 \mathcal{P}\Big[ \frac{\mathcal{N}_h(k^0=\wkr)(f_\Lambda^{\bm{k}}-2\fpkr)}{\wkr \SB{(q^0+\wkr)^2-(\wks)^2}}	
	+ \frac{\mathcal{N}_h(k^0=-\wkr)(f_\Lambda^{\bm{k}}-2\fmkr)}{\wkr \SB{(q^0-\wkr)^2-(\wks)^2}} \nn \\	
	&& + \frac{\mathcal{N}_h(k^0=-q^0+\wks)(f_\Lambda^{\bm{k}}-2\fpks)}{\wks \SB{(q^0-\wks)^2-(\wkr)^2}}
	+ \frac{\mathcal{N}_h(k^0=-q^0-\wks)(f_\Lambda^{\bm{k}}-2\fmks)}{\wks \SB{(q^0+\wks)^2-(\wkr)^2}} \Big]_{\bm{q}=\bm{0}}~, \\
	\label{RePi.2}
	\IM\Pi_h(q_0,\bm{q}=\bm{0};T,\mu,\mu_5) &=& - \text{sign}(q^0)\tanh\FB{\frac{q^0}{2T}}\frac{N_cN_f}{2\pi} \int_{0}^{\infty}d|\bm{k}| \sum_{r \in \{\pm\}} \sum_{s \in \{\pm\}} 
	\frac{1}{16rs\mu_5^2} \frac{1}{4\wkr\wks} \nn \\ && \hspace{-3.7cm}
	\times \Big[ \mathcal{N}_h(k^0=-\wkr) \fb{1-\fmkr -\fpks +2\fmkr\fpks}\delta(q_0-\wkr-\wks) 
	+ \mathcal{N}_h(k^0=\wkr)  \fb{1-\fpkr -\fmks +2\fpkr\fmks} \delta(q_0+\wkr+\wks) \nn \\ && \hspace{-3.7cm}
	+ \mathcal{N}_h(k^0=\wkr) \fb{-\fpkr -\fpks +2\fpkr\fpks} \delta(q_0+\wkr-\wks) 
	+ \mathcal{N}_h(k^0=-\wkr) \fb{-\fmkr -\fmks +2\fmkr\fmks} \delta(q_0-\wkr+\wks) \Big]_{\bm{q}=\bm{0}}~.
	\label{ImPi.2} \nn \\
\end{eqnarray}
On the other hand, for $q^0=0,\bm{q}\ne\bm{0}$, Eq.~\eqref{bar.to.11} gives $\IM\Pi_h(q_0=0,\bm{q}) = 0$ and $\RE\Pi_h(q_0=0,\bm{q})$ becomes
\begin{eqnarray}
	\RE\Pi_h(q_0=0,\bm{q};T,\mu,\mu_5) &=& \frac{N_cN_f}{2\pi^2} \int_{0}^{\infty} \!\! d|\bm{k}||\bm{k}| \sum_{r \in \{\pm\}} \sum_{s \in \{\pm\}} 
	\frac{\fb{ f_\Lambda^{\bm{k}}-\fpkr -\fmkr}}{32rs\mu_5^2\wkr}  \int_{-1}^{1}d(\cos\theta) \frac{1}{|\bm{p}|}
	\mathcal{P}\Big[ \frac{\mathcal{N}_h(k^0=\wkr,q^0=0)}{(\wkr)^2-(\wps)^2} \Big]
	\label{RePi.3} \nn \\
\end{eqnarray}
where $\theta$ is the angle between $\bm{q}$ and $\bm{k}$. We note that the remaining $d|\bm{k}|$ integral in Eq.~\eqref{ImPi.2} can be performed using the Dirac delta functions present in the integrand. Moreover, the $d(\cos\theta)$ integral in Eq.~\eqref{RePi.3} can also be performed by noting the fact that $\mathcal{N}_h(k^0=\wkr,q^0=0)$ is a polynomial in $\cos\theta$ of degree 2 (see Eq.~\eqref{Nh}).
$\IM\Pi_h(q_0,\bm{q}=\bm{0})$ in Eq.~\eqref{ImPi.2} contains sixteen Dirac delta functions and they give rise to branch cuts of the polarization function in the complex energy plane. The terms with $\delta(q_0-\wkr-\wks)$ and $\delta(q_0+\wkr+\wks)$ are the Unitary-I and Unitary-II cuts respectively, whereas the terms with $\delta(q_0+\wkr-\wks)$ and $\delta(q_0-\wkr+\wks)$ are respectively termed as Landau-I and Landau-II cuts. Each of these Unitary and Landau cuts, in turn consists of four sub-cuts corresponding to different helicities $ (r,s) $. Each of the sixteen Dirac delta functions in Eq.~\eqref{ImPi.2} are non-vanishing at different kinematic domains which are given in Table~\ref{tab.kin}
\begin{center}
	\begin{table} [h]
	\begin{tabular}{|c|c|}%
		\hline 
		Dirac Delta Function & \makecell[c]{ ~ \\ Kinematic Regions \\ ~ } \\
		\hline \hline
		$\delta(q_0\mp\wkr\mp\wks)$ & \makecell[c]{ $ 2M \le \pm q_0 < \infty$ ~~if~~ $(r,s)=(-,-)$,\\ 
			$ 2\sqrt{M^2+\mu_5^2} \le \pm q_0 < \infty$ ~~otherwise~~ } \\
		\hline
			$\delta(q_0\pm\wkr\mp\wks)$ & \makecell[c]{ $ 0 \le \pm q_0 \le 2\mu_5$ ~~if~~ $(r,s)=(-,+)$,
				\\ $ -2\mu_5 \le \pm q_0 \le 0$ ~~if~~ $(r,s)=(+,-)$ } \\
		\hline 
	\end{tabular}
\caption{Kinematic domains where the sixteen Dirac delta functions in Eq.~\eqref{ImPi.2} are non-vanishing.}
\label{tab.kin}
	\end{table}
\end{center}
\begin{figure}[h]
	\includegraphics[angle=0,scale=0.5]{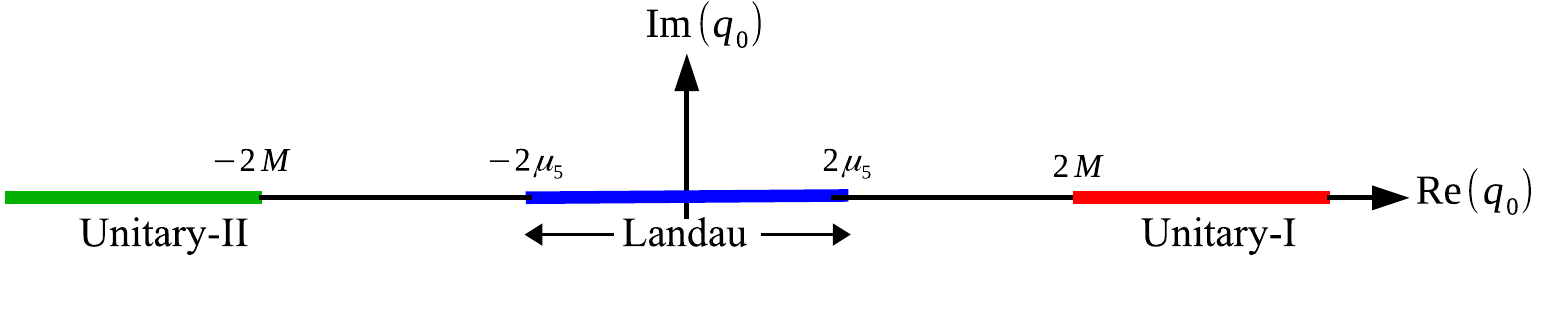} 
	\caption{(Color Online) The analytic structure of the polarization function $\Pi_h(q^0,\bm{q}=\bm{0})$ showing its branch cuts in the complex $q_0$ plane.}
	\label{fig.analytic}
\end{figure}

In Eq.~\eqref{ImPi.2}, when helicity indices $(r,s)$ are summed over, we find from Table~\ref{tab.kin} that the kinematic domain for Unitary-I cut is $2M\leq q^0<\infty$ whereas the same for Unitary-II cut is $-\infty<q^0\le2M$. On the other hand, both the Landau cuts lie in the region $|q^0|\le2\mu_5$. The complete analytic structure of the polarization function $\Pi_h(q^0,\bm{q}=\bm{0})$ has been shown in Fig.~\ref{fig.analytic}. It is evident that for non-zero $\mu_5$ the Landau cuts contribute to the physical time-like region defined in terms of $q^0>0$ and $q^2>0$ which is a purely finite CCP effect. Moreover, for small values of $\mu_5<2M$, there exists a forbidden gap between the Unitary and Landau cuts. However, for sufficiently high values of $\mu_5>2M$ this forbidden gap vanishes which is again a purely finite CCP effect.
%

\section{Numerical Results \& Discussions} \label{sec.results}
To discuss the numerical results we first specify the values of the parameters of the NJL model that are used in this work. We have taken the current quark mass $m=5.6$ MeV, coupling constant $G=5.742$ GeV$^{-2}$ and smooth three-momentum cutoff $\Lambda=568.69$ MeV. The choice of these parameters reproduces the experimental/phenomenological vacuum ($T=\mu=\mu_5=0$) values of the quark-condensate per flavour $\frac{\ensembleaverage{\psibar\psi}}{N_f} = -(245.1~\rm MeV)^3$, pion decay constant $f_\pi = 91.9$ MeV, constituent quark mass in vacuum $M = 343.6$ MeV and the mass of pion $m_\pi=140$ MeV. Moreover, with these choice of parameters, the vacuum  $s$-wave $\pi-\pi$ scattering lenghths $a^I$ in the isospin channel $I$ come out to be $a^0=0.170~m_\pi^{-1}$ and $a^2=-0.045~m_\pi^{-1}$ which are consistent with the corresponding Weinberg values~\cite{Schulze:1995rb,Bernard:1990ye} $a^0_W = \frac{7m_\pi^2}{32\pi f_pi^2} = 0.162 ~m_\pi^{-1}$ and $a^2_W = -\frac{2m_\pi^2}{32\pi f_pi^2} = -0.046 ~m_\pi^{-1}$. 
\begin{figure}[h]
	\includegraphics[angle=-90,scale=0.35]{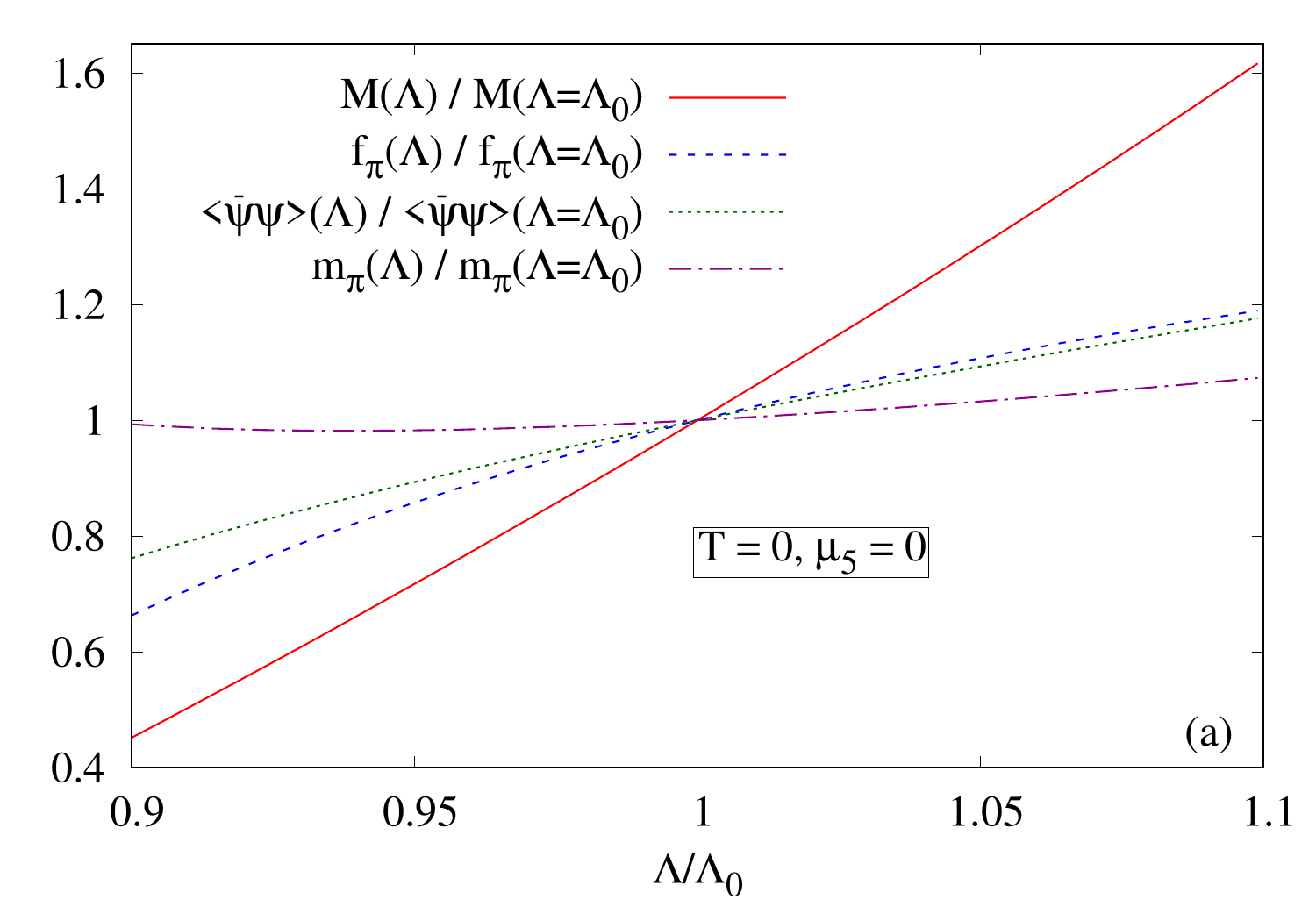} \includegraphics[angle=-90,scale=0.35]{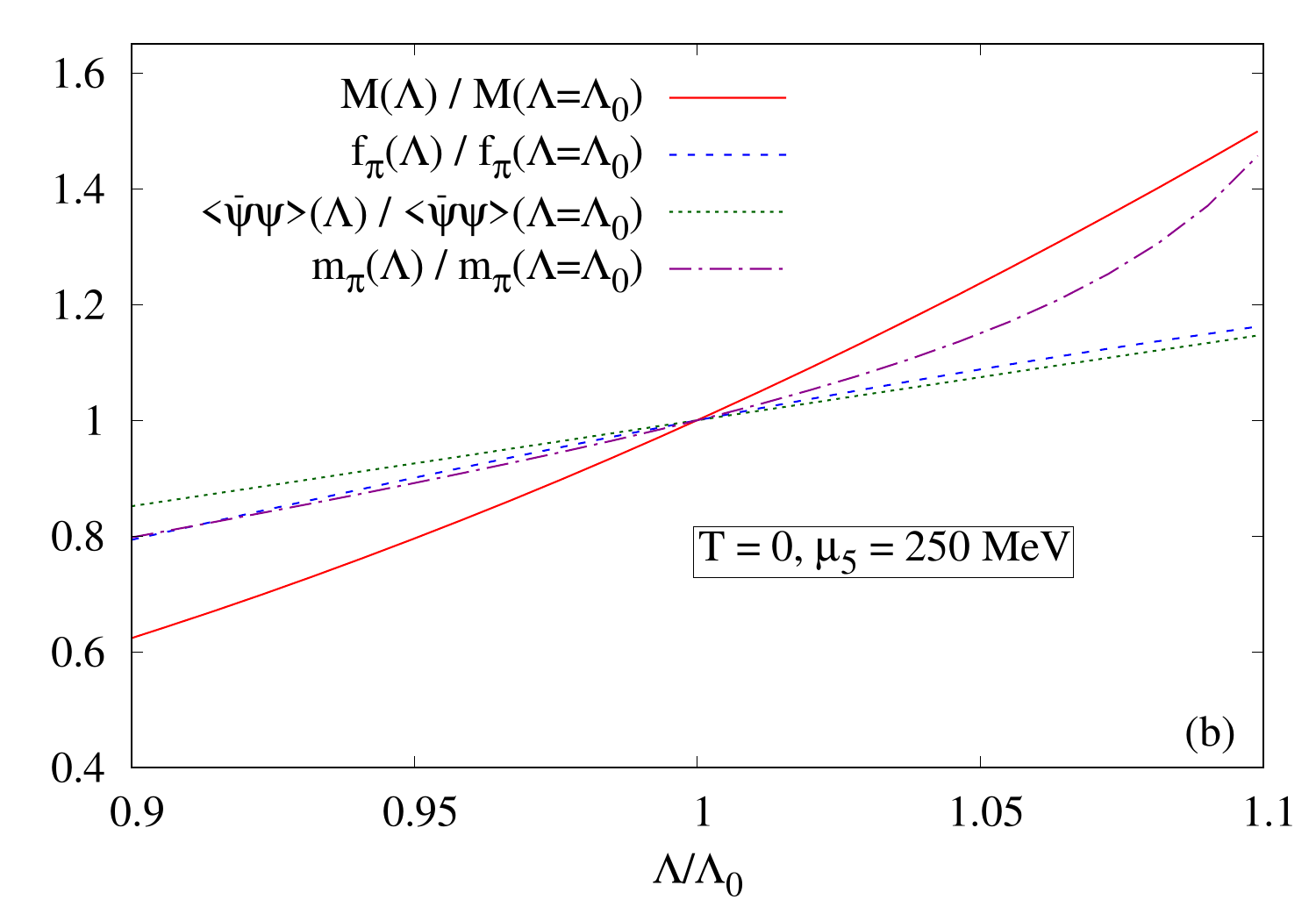}
	\caption{(Color Online) Relative cutoff parameter $\Lambda$-dependence of constituent quark mass $M$, pion decay constant $f_\pi$, quark-condensate $\ensembleaverage{\psibar\psi}$ and pion mass $m_\pi$ in (a) vacuum ($T=\mu=\mu_5=0$), and (b) at $T=\mu=0, \mu_5=250$ MeV, scaled to their values at $\Lambda=\Lambda_0 = 568.69$ MeV.}
	\label{fig.Lam}
\end{figure}

Except for the pion mass, these experimental/phenomenological vacuum values ($M$, $f_\pi$ and $\ensembleaverage{\psibar\psi}$) are quite sensitive to the smooth cutoff parameter $\Lambda$ which can be observed in Figs.~2(a) where we have shown the variation of the vacuum values of the constituent quark mass $M$, pion decay constant $f_\pi$, quark-condensate $\ensembleaverage{\psibar\psi}$ and the pion mass $m_\pi$ scaled to their values at $\Lambda=\Lambda_0 = 568.69$ MeV as a function of $\frac{\Lambda}{\Lambda_0}$ at $T=\mu=\mu_5=0$. It can be noticed that, as the cutoff parameter $\Lambda$ is changed by 10\%, the maximum change in $M$, $f_\pi$, $\ensembleaverage{\psibar\psi}$ and $m_\pi$ are respectively 60\%, 30\%, 20\% and 10\%. Thus $M$ is the most sensitive to the change in $\Lambda$ whereas  $m_\pi$ has the least sensitivity to the change in $\Lambda$. 
In order to see the $\Lambda$-dependencies of the physical parameters ($M$, $f_\pi$, $\ensembleaverage{\psibar\psi}$ and $m_\pi$) at non-zero CCP, we have shown their relative   $\Lambda$-dependencies in Fig.~\ref{fig.Lam}(b) at $T=\mu=0$ and $\mu_5=250$ MeV. As can be observed from this figure, at non-zero CCP the $\Lambda$-dependence is slightly suppressed as compared to the $\mu_5=0$ case for all the parameters except for the pion mass which is now significantly affected. In particular, while changing $\Lambda$ by 10\% at $\mu_5=250$ MeV, the maximum changes in $M$, $f_\pi$, $\ensembleaverage{\psibar\psi}$ and $m_\pi$ come out to be respectively 50\%, 20\%, 20\% and 40\%. From both the Figs.~\ref{fig.Lam}(a) and (b), we thus conclude that, more than one physical parameters have deviated beyond their acceptable ranges when $\Lambda$ is changed by more than 5\% owing to their considerable sensitivity on $\Lambda$. Thus, the results with such values of $\Lambda$ can not be trusted. Hence, the parameter $\Lambda$ can not be tuned alone in NJL type model; in particular the NJL model as used in this work has two more parameters $m$ and $G$ in addition to the cutoff parameter $\Lambda$ which are kept constant while obtaining Figs.~\ref{fig.Lam}(a) and (b). In Ref.~\cite{Kohyama:2015hix}, the dependence of various results on the full three-dimensional parameter space (m, G, $ \Lambda $) without considering chiral imbalance and using different regularization schemes has been explored. Similar exercise  at finite CCP  will be of significant interest but is too exhaustive to be pursued here. In this work we restrict ourselves to $ \Lambda= \Lambda_0 $ and focus on the studies of the CCP dependence of various quantities.
\begin{figure}[h]
	\includegraphics[angle=-90,scale=0.35]{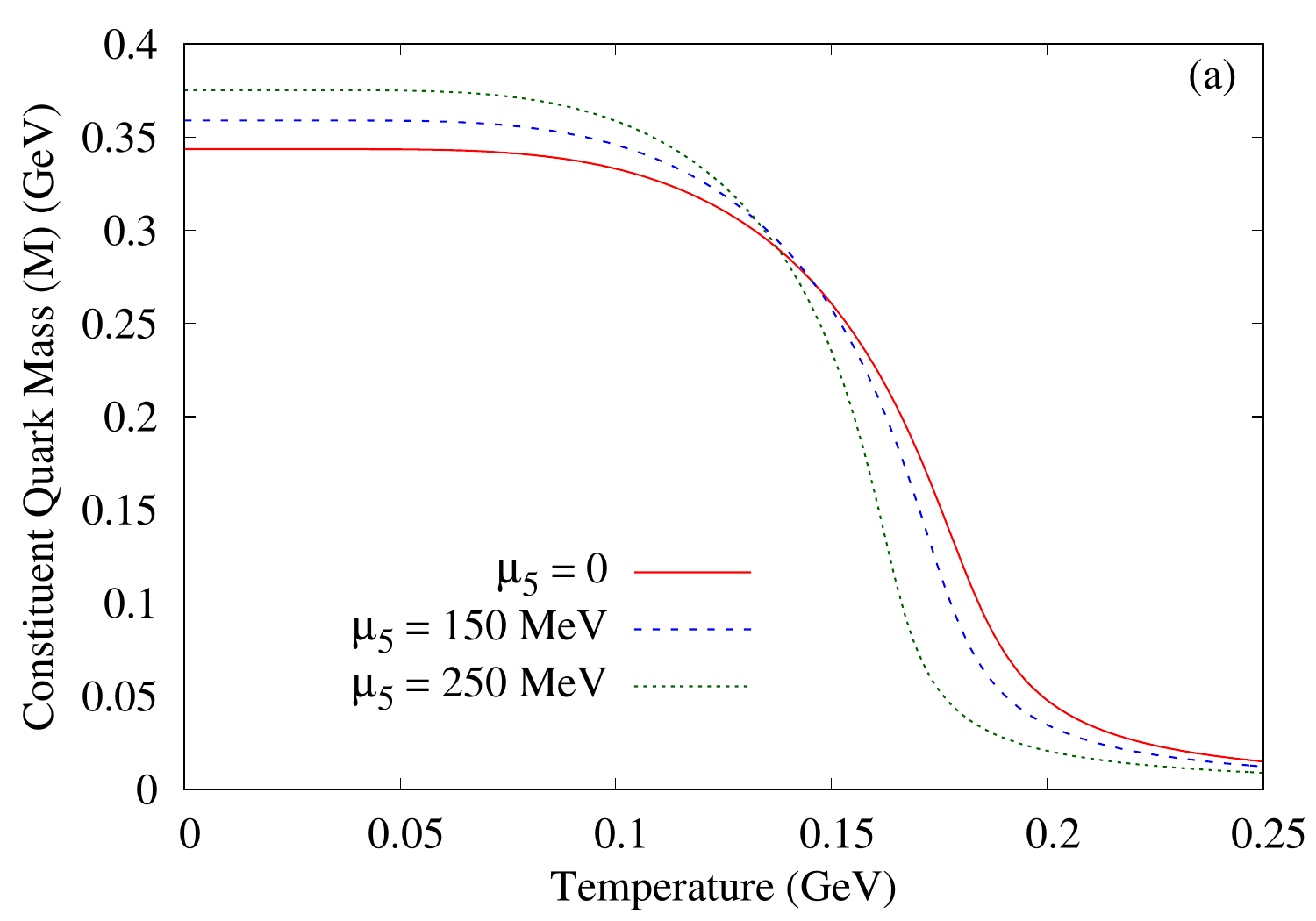} \includegraphics[angle=-90,scale=0.35]{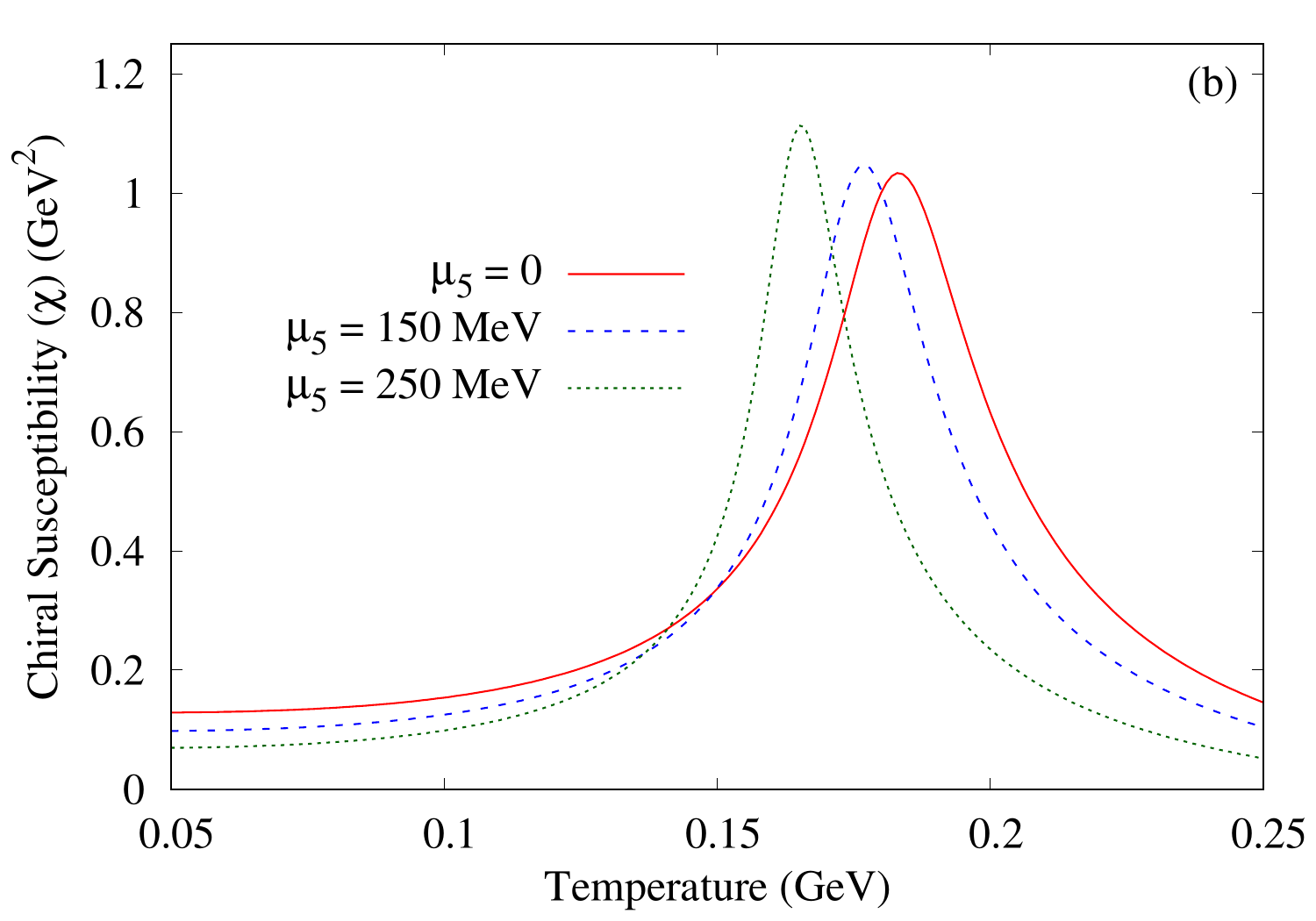} 
	\includegraphics[angle=-90,scale=0.35]{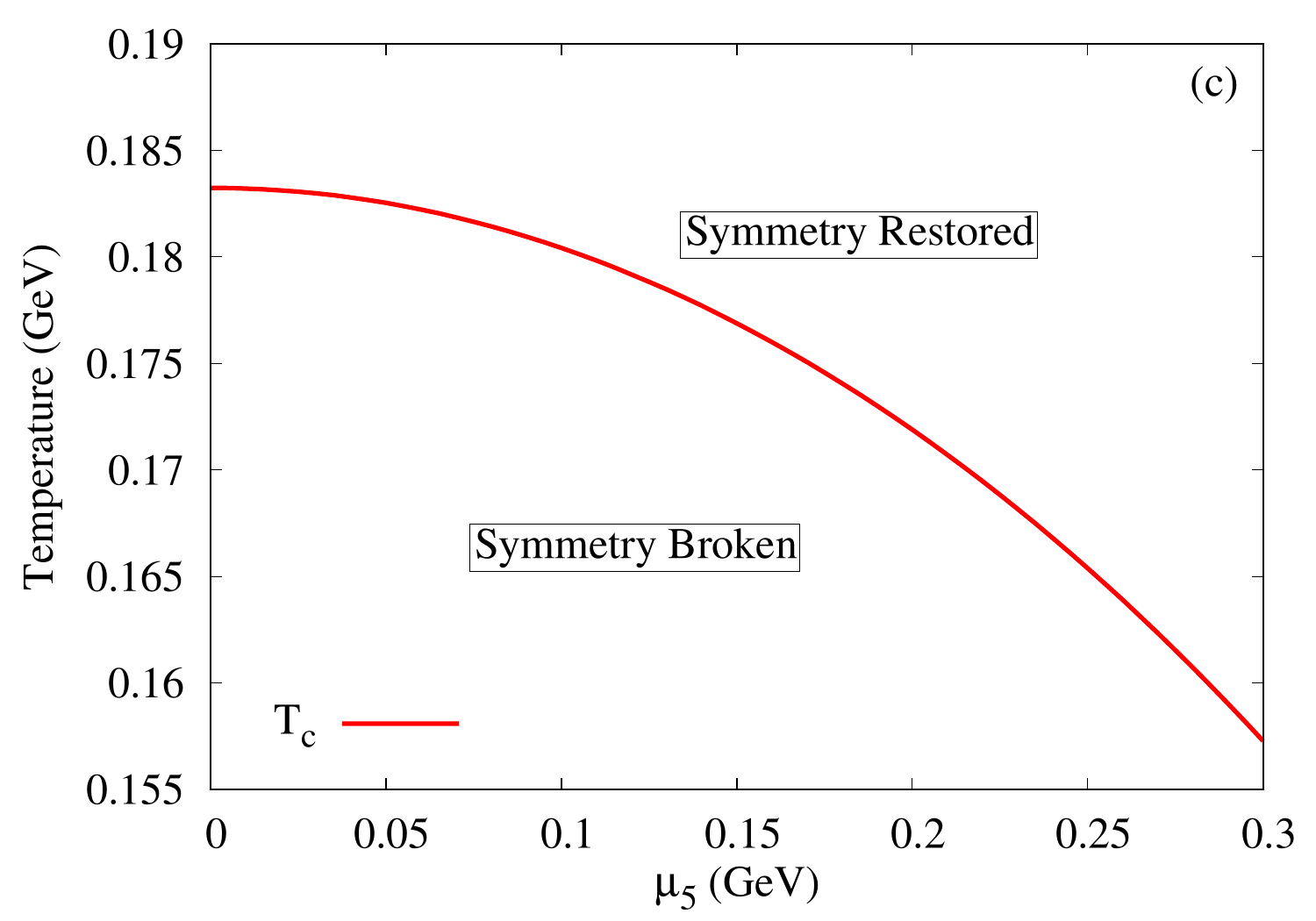} \includegraphics[angle=-90,scale=0.35]{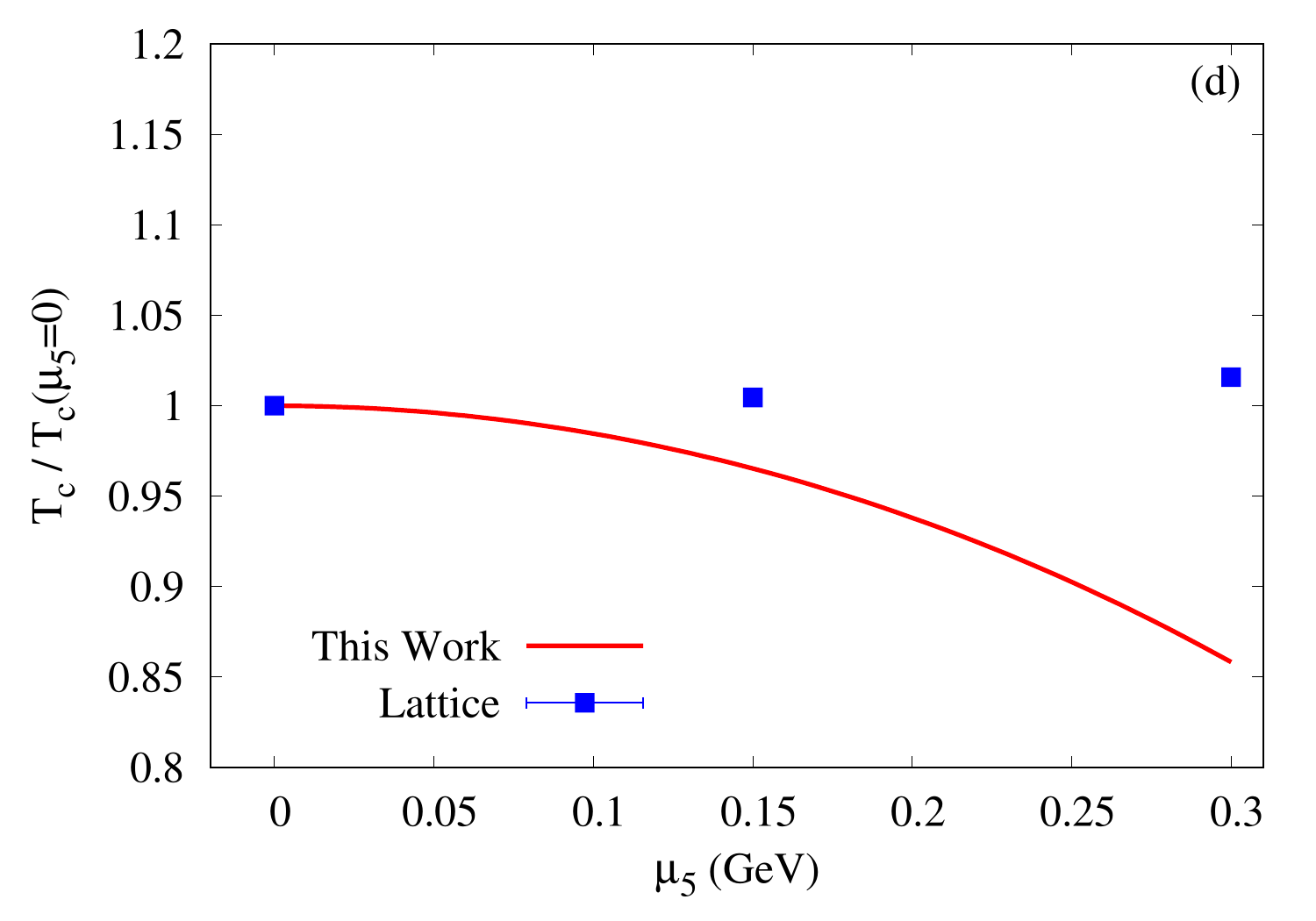}
	\caption{(Color Online) The variation of the (a) constituent quark mass $M$, and (b) chiral susceptibility $\chi$ as a function of temperature $T$ at different values of CCP. The variation of the (c) transition temperature $T_c$ in the chiral phase diagram on $T-\mu_5$ plane, and (d) scaled transition temperature $\frac{T_c}{T_c(\mu_5=0)}$ in comparison with LQCD results of Ref.~\cite{Braguta:2015zta} as a function of CCP.}
	\label{fig.M}
\end{figure}

It is important to note that all the calculations presented in Secs.~\ref{sec.gap}-\ref{sec.pola} have been performed considering a finite value of the quark chemical potential $\mu$ which is one-third of the baryon chemical potential $\mu_B = 3\mu$. However, when displaying numerical results in this section, we have chosen to set $\mu$ to zero. This choice has been made to reduce the parameter space from ($T-\mu-\mu_5$) to ($T-\mu_5$) which allows us to focus on studying the interplay between thermal and CCP effects.

We first show the temperature dependence of the constituent quark mass $M$ at different values of CCP in Fig.~\ref{fig.M}(a). $M$ has been obtained by solving the gap equation~\eqref{gap.final}. As can be seen in Fig.~\ref{fig.M}(a), irrespective of the value of CCP, $M$ is large ($\sim 300$ MeV) in the low temperature region due to the spontaneous breaking of the chiral symmetry corresponding to large values of the quark condensate. As the temperature increases $M$ first remains constant up to a certain value of temperature after which $M$ suddenly drops as a consequence of the pseudo-chiral phase transition. In the high temperature region $M$ asymptotically approaches to the current quark mass $m$ irrespective of the values of $\mu_5$. 

By comparing the different CCP curves of Fig.~\ref{fig.M}(a), we find that $M$ increases in the low temperature region with the increase in CCP. Moreover, with the increase in CCP the transition temperature ($T_c$) of the pseudo-chiral phase transition decreases and the sudden drop of $M$ occurs at relatively smaller temperature. Hence, CCP has the effect of making the chiral condensate stronger at low temperature region indicating a ``chiral catalysis'' (CC) in which the chiral imbalance (characterized by $\mu_5$) catalyzes the dynamical symmetry breaking. On the contrary, at high temperature region an opposite effect is seen in which the CCP makes the chiral condensate weaker so that chiral symmetry is restored at a relatively lower temperature as compared to the $\mu_5=0$ case. This implies an ``inverse chiral catalysis'' (ICC) where the dynamical symmetry breaking is opposed by the presence of a chiral imbalance~\cite{Ghosh:2022xbf,Chaudhuri:2022rwo,Ghosh:2023rft}. 

In order to understand the quantitative variation of the pseudo-chiral phase transition temperature ($T_c$) with the CCP, we have calculated the chiral susceptibility $\chi = \dfrac{1}{2G}\FB{\dfrac{\del M}{\del m}-1}$	and shown its variation as a function of $\mu_5$ in Fig.~\ref{fig.M}(b). The position of the peak of the chiral susceptibility correspond to the pseudo-chiral phase transition temperature $T_c$. As can be clearly noticed in the figure, with the increase in CCP, the position of the peak shifts towards lower values of temperature indicating ICC as described earlier. Next in Fig.~\ref{fig.M}(c), we have depicted the chiral phase diagram on the $T-\mu_5$ plane in which we notice a monotonically decreasing trend of $T_c$ with the increase in CCP. Finally, in Fig.~\ref{fig.M}(d), the variation of the scaled transition temperature $\frac{T_c}{T_c(\mu_5=0)}$ as a function of $\mu_5$ has been shown and it has been compared with the LQCD results of Ref.~\cite{Braguta:2015zta}. 
The LQCD findings~\cite{Braguta:2015zta,Braguta:2015owi} indicate a chiral catalysis (CC) effect in which the transition temperature increases with the increase in CCP which is opposite to our results predicting an ICC effect consistent with the results reported earlier in Refs.~\cite{Gatto:2011wc,Chernodub:2011fr,Ruggieri:2011xc,Yu:2015hym,Fukushima:2010fe} using NJL-type effective models. On the other hand, in Ref.~\cite{Yu:2015hym}, it has been shown that the conventional hard-cutoff regularization scheme can give a CC effect as a regularization-artifact if a complete contribution from the thermal fluctuations is not included. Similarly a CC effect is shown in Ref.~\cite{Andrianov:2013dta} by using two-flavor NJL model with broken $U(1)_A$ symmetry and in Ref.~\cite{Ruggieri:2016ejz} using non-local NJL model with different interaction kernels. We thus believe that, the reproduction of the LQCD data at finite CCP may not be possible alone by adjusting the constant parameters ($m,G,\Lambda$) in the basic NJL-type model, and would require finer treatments as mentioned before. The contradiction of the results of the NJL-type models with the LQCD calculations are also found in presence of an external magnetic field; in particular, near the transition temperature the basic NJL-type models predicts a magnetic catalysis (MC) effect whereas the LQCD results show the inverse magnetic catalysis (IMC) effect~\cite{Farias:2016gmy}. Such mismatch with the LQCD findings in presence of magnetic field $B$ has been cured in Ref.~\cite{Ferreira:2014kpa,Farias:2016gmy} by using a parametrized temperature and magnetic field dependent NJL-coupling $G=G(T,B)$. In a similar way, in our case perhaps a CC effect may also be reproduced by considering a temperature and CCP dependent NJL-coupling $G=G(T,\mu_5)$ and tuning the other parameters $m$ and $\Lambda$ to reproduce the LQCD data. It is to be noted that the lattice results are obtained by assuming higher values of pion mass without extrapolation to the physical limit. So	 until those extrapolations  are provided, it is illustrative to compare tendencies obtained within different approaches	 based on effective theories. In particular, regarding chiral catalysis, apart from those based on the NJL formalism, the ChPT analysis in Refs.~\cite{Espriu:2020dge,GomezNicola:2023ghi} also favour a CC behaviour. In the present scenario varying the cut-off parameter by 10\% (keeping $G$ and $m$ fixed) could render a $T_c(\mu_5)$ curve closer to lattice data. This may be due to the increase in pion mass as we change $\Lambda$ alone. However, in present work we restrict ourselves to use the basic NJL model with constant parameters $(m,G,\Lambda)$ which are well accepted at $\mu_5=0$ and focus on the study of the CCP dependence of the electrical conductivity.
\begin{figure}[h]
	\includegraphics[angle=-90,scale=0.69]{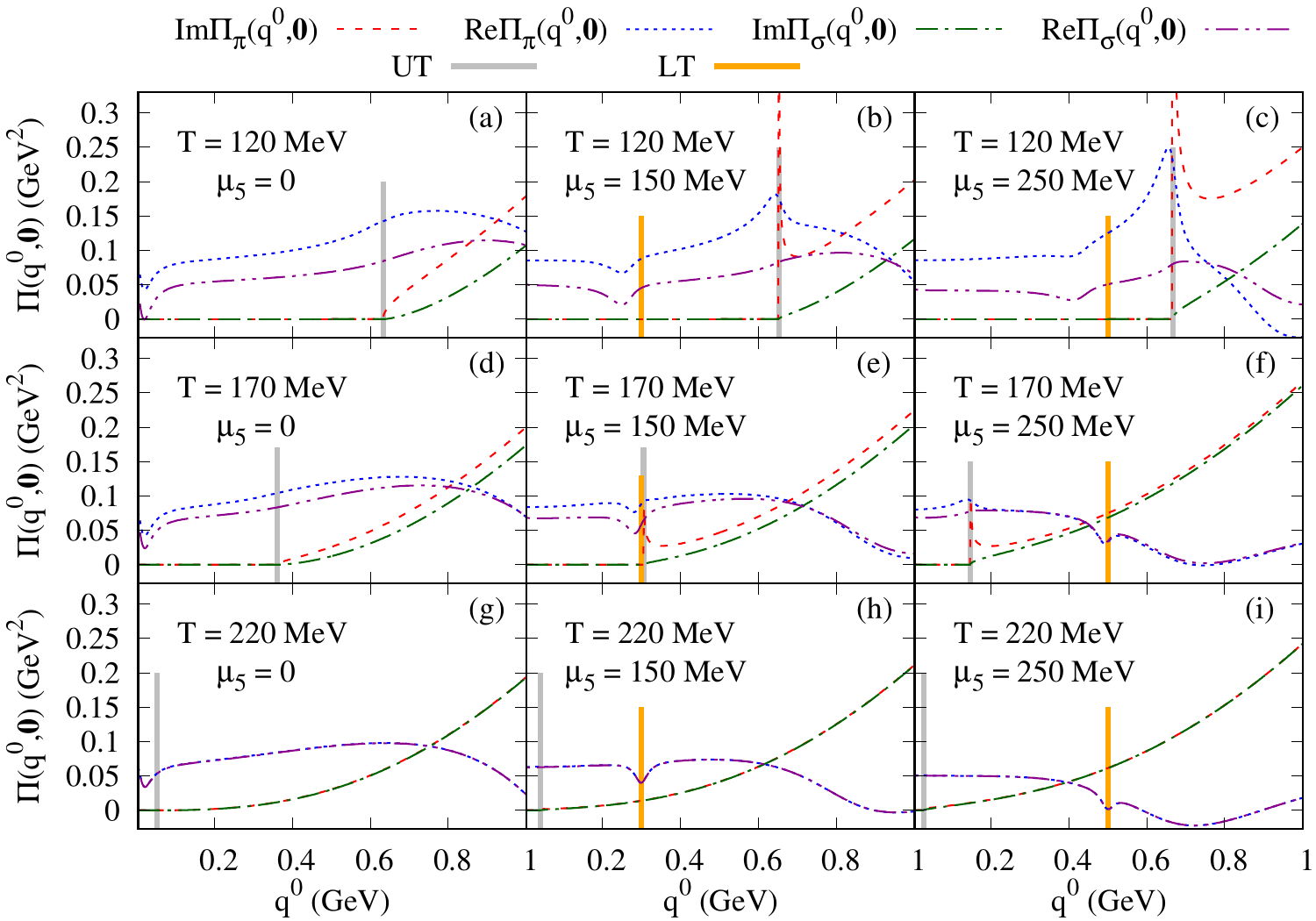} 
	\caption{(Color Online) The variation of polarization functions of the scalar ($\sigma$) and pseudo-scalar ($\pi$) mesons as a function of energy $q^0$ at vanishing mesonic spatial momenta and at different values of temperature and CCP. `UT' and `LT' refers to the Unitary and Landau cut thresholds respectively; and they are shown by grey and orange colored vertical lines.}
\label{fig.q0}
\end{figure}
\begin{figure}[h]
	\includegraphics[angle=-90,scale=0.69]{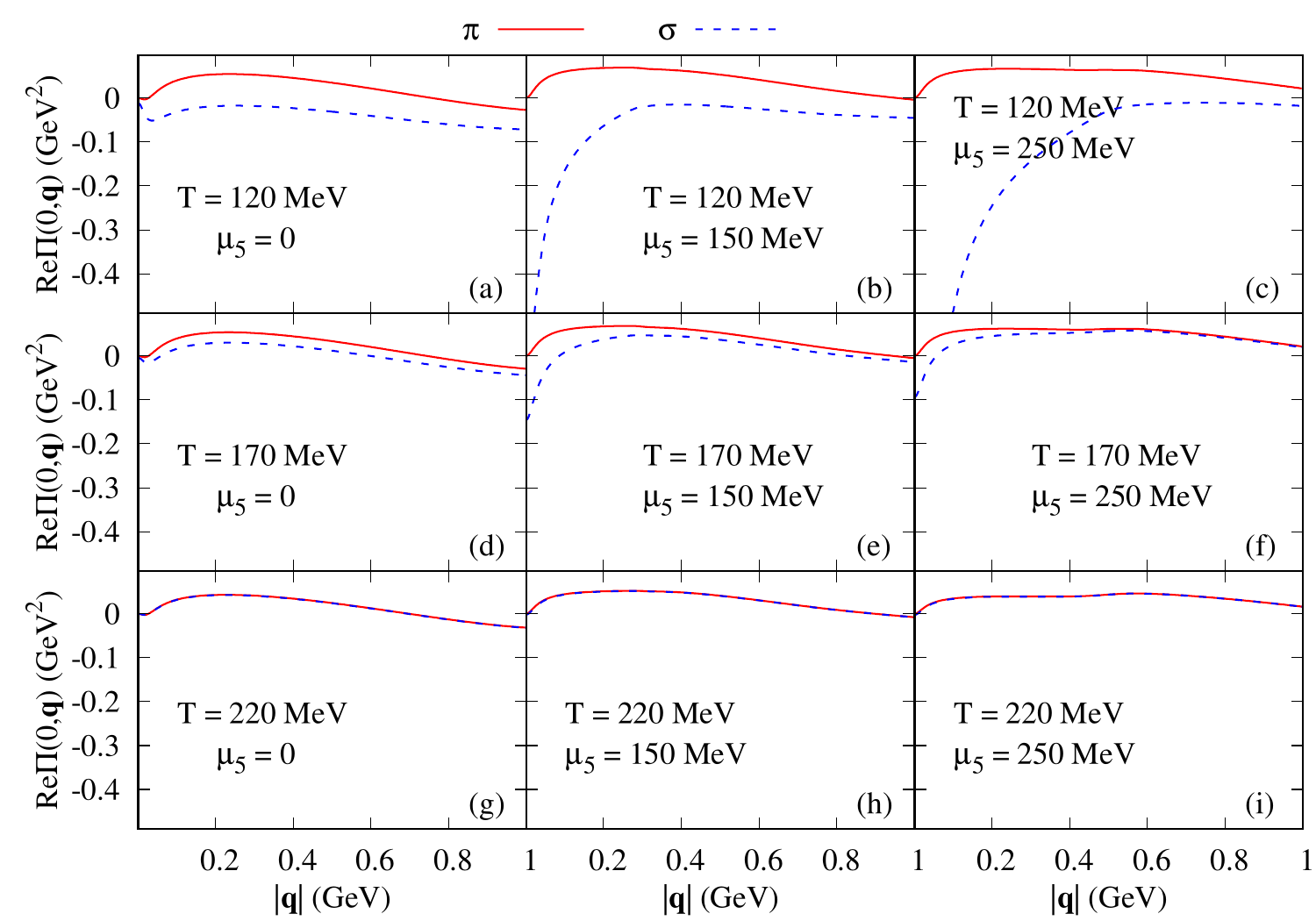} 
	\caption{(Color Online) The variation of polarization functions of the scalar ($\sigma$) and pseudo-scalar ($\pi$) mesons as a function of spatial momenta $|\bm{q}|$ at vanishing mesonic energy and at different values of temperature and CCP.}
	\label{fig.qv}
\end{figure}

We now turn our attention to the polarization functions of the mesonic excitation in scalar ($\sigma$) and pseudo-scalar ($\pi$) channels. As already mentioned in Secs.~\ref{sec.width} and \ref{sec.pola}, in the calculation of scattering cross section (followed by thermal width and electrical conductivity), we require the quantities $\RE\Pi_h(q^0,|\bm{q}|=0)$, $\IM\Pi_h(q^0,|\bm{q}|=0)$ and $\RE\Pi_h(q^0=0,|\bm{q}|)$ since $\IM\Pi_h(q^0=0,|\bm{q}|)=0$. In Figs.~\ref{fig.q0}(a)-(i) we have depicted the variations of $\RE\Pi_h(q^0,|\bm{q}|=0)$ and $\IM\Pi_h(q^0,|\bm{q}|=0)$ as a function of $q^0$ for nine different representative combinations of temperature (with $T=120$, 170, 220 MeV) and CCP (with $\mu_5=0$, 150, 250 MeV).  This figure comprises a grid of nine panels arranged in a $3\times3$ format where $\mu_5$ increases horizontally from left to right and $T$ increases vertically from top to bottom. These combinations of $T$ and $\mu_5$ are chosen in such a way to capture different stages of chiral phase transition. As can be understood from Fig.~\ref{fig.M}, the upper panel of Fig~\ref{fig.q0} with $T=120$ MeV corresponds to the chiral symmetry broken phase, the middle panel with $T=170$ MeV shows the behaviour in the vicinity of chiral phase transition and the lower panel with $T=220$ MeV corresponds to the phase with partially restored chiral symmetry. In Fig.~\ref{fig.q0}, the Unitary cut threshold (at $q^0=2M$) and the Landau cut threshold (at $q^0=2\mu_5$) are also shown by grey and orange colored vertical lines respectively. Let us first describe the variation of $\IM\Pi_h(q^0,|\bm{q}|=0)$ as a function of $q^0$ in the region $q^0>0$. At $\mu_5=0$ the $\IM\Pi_h(q^0,|\bm{q}|=0)$'s do not have contributions from the Landau cut in the kinematic region $q^0>0$, so these curves start from the Unitary cut threshold $q^0=2M$. This behavior is evident in the red and green curves presented in the leftmost vertical panels of Figs~\ref{fig.q0}. Moreover, as we increase the temperature, the unitary cut thresholds move towards lower value of $q^0$ owing to the decrease in $M$ (indicative of partial restoration of chiral symmetry) as can be understood from Fig.~\ref{fig.M}. At non-zero CCP the $\IM\Pi_h(q^0,|\bm{q}|=0)$'s acquire contributions from the Landau cuts as well. However the magnitude of the Landau cut contributions are sub-leading as compared to the same from the Unitary cuts and so they are not visible for the ordinate ranges considered in the graphs. For non-zero values of CCP the $\IM\Pi_\pi(q^0,|\bm{q}|=0)$'s corresponding to the pseudoscalar channel exhibit a non-monotonic behavior near the threshold of the Unitary cut. Specifically, as $q^0$ increases $\IM\Pi_\pi(q^0,|\bm{q}|=0)$ initially begins to rise at the Unitary cut threshold, then it decreases to reach a local minimum and subsequently starts increasing monotonically again at higher values of $q^0$. This behavior is noticeable in the middle and rightmost panels of both the first and second row in Fig.~\ref{fig.q0}. In contrast, the $\IM\Pi_\sigma(q^0,|\bm{q}|=0)$'s corresponding to the scalar channel, consistently exhibits a monotonic increase after crossing the Unitary cut threshold. However, at $T=220$ MeV this non-monotonicity in the $q^0$ dependence of $\IM\Pi_\pi(q^0,|\bm{q}|=0)$ is considerably reduced as is evident from the red curves in Figs.\ref{fig.q0}(h) and (i). With the increase in temperature as well as CCP the overall magnitude of $\IM\Pi_h(q^0,|\bm{q}|=0)$'s increases by small amounts. This trend is noticeable by comparing the corresponding curves while moving vertically downward and horizontally rightward through the panels of Fig.~\ref{fig.q0}. Furthermore, with an increase in $ T $ (moving from the upper panels to the lower panels with $T=220$ MeV in the lower panels), the difference between $\IM\Pi_\pi(q^0,|\bm{q}|=0)$ and $\IM\Pi_\sigma(q^0,|\bm{q}|=0)$ diminishes. At higher temperatures, particularly in the lower panels at $T=220$ MeV, the curves corresponding to the chiral partners $\pi$ and $\sigma$ coincide indicating the partial restoration of chiral symmetry.

Now, let us delve into the discussion regarding the variation of $\RE\Pi_h(q^0,|\bm{q}|=0)$ as a function of $q^0$ at different temperatures and CCPs. In all nine panels of Fig.~\ref{fig.q0}, the $\RE\Pi_h(q^0,|\bm{q}|=0)$'s exhibit a non-monotonic behavior with increasing $q^0$. Additionally, in the vicinity of each Landau and Unitary cut threshold, $\RE\Pi_h(q^0,|\bm{q}|=0)$ experiences a sudden change in its curvature due to the analytic properties of the polarization function. As $ T $ increases the overall magnitude of $\RE\Pi_h(q^0,|\bm{q}|=0)$ decreases which is evident when compared with the corresponding curves while moving vertically downward through the panels of Fig.~\ref{fig.q0}. Similar to the imaginary parts, $\RE\Pi_h(q^0,|\bm{q}|=0)$'s corresponding to the chiral partners $\pi$ and $\sigma$ also become degenerate at high temperatures when chiral symmetry is partially restored. For a more rigorous discussion regarding the variation of $\RE\Pi_h(q^0,|\bm{q}|=0)$ and $\IM\Pi_h(q^0,|\bm{q}|=0)$ as functions of $q^0$ can be found in~\cite{Ghosh:2023rft}.

In Figs.~\ref{fig.qv}(a)-(i), we have depicted the variations of $\RE\Pi_h(q^0=0,\bm{q})$ as a function of $|\bm{q}|$ for the same nine different representative combinations of temperature and CCP as chosen in Fig.~\ref{fig.q0} in order to capture different stages of chiral phase transition. With the increase in $|\bm{q}|$ we notice a non-monotonic behaviour of $\RE\Pi_h(q^0=0,\bm{q})$. In particular, as we increase $|\bm{q}|$, $\RE\Pi_h(q^0=0,\bm{q})$'s increase at low $|\bm{q}|$, attain a local maxima and then decrease slowly. Comparing the curves corresponding to $\pi$ and $\sigma$-modes, we notice that they are well separated in the chiral symmetry broken phase (i.e. at $T=120$) and the curves for $ \pi $ meson are always at higher magnitude than the corresponding  ones for $\sigma$ meson. With the increase in temperature the difference between the curves of $\pi$ and $\sigma$-modes decrease (as we move vertically downward through the panels). Finally, at $T=220$ MeV (the lower panels) the curves corresponding to scalar and pseudoscalar modes merge with each other due to the restoration of chiral symmetry. With the increase in temperature we notice that $\RE\Pi_\pi(q^0=0,\bm{q})$'s decrease by small amounts as can be seen by comparing the red curves while moving vertically-downward direction through the panels of Fig.~\ref{fig.qv}, whereas an opposite trend is observed for $\RE\Pi_\sigma(q^0=0,\bm{q})$'s which increase with the increase in $T$. The magnitudes of $\RE\Pi_\pi(q^0=0,\bm{q})$'s also increase with the increase in CCP for all temperatures. This trend is noticeable when comparing the corresponding curves while moving horizontally right across the panels. On the other hand, in the symmetry broken phase (i.e. at $T=120$ and 170 MeV), with the increase in $\mu_5$, we notice in the top and middle panels that the magnitude of $\RE\Pi_\sigma(q^0=0,\bm{q})$ decreases (increases in the negative direction). 
\begin{figure}[h]
	\includegraphics[angle=0,scale=0.95]{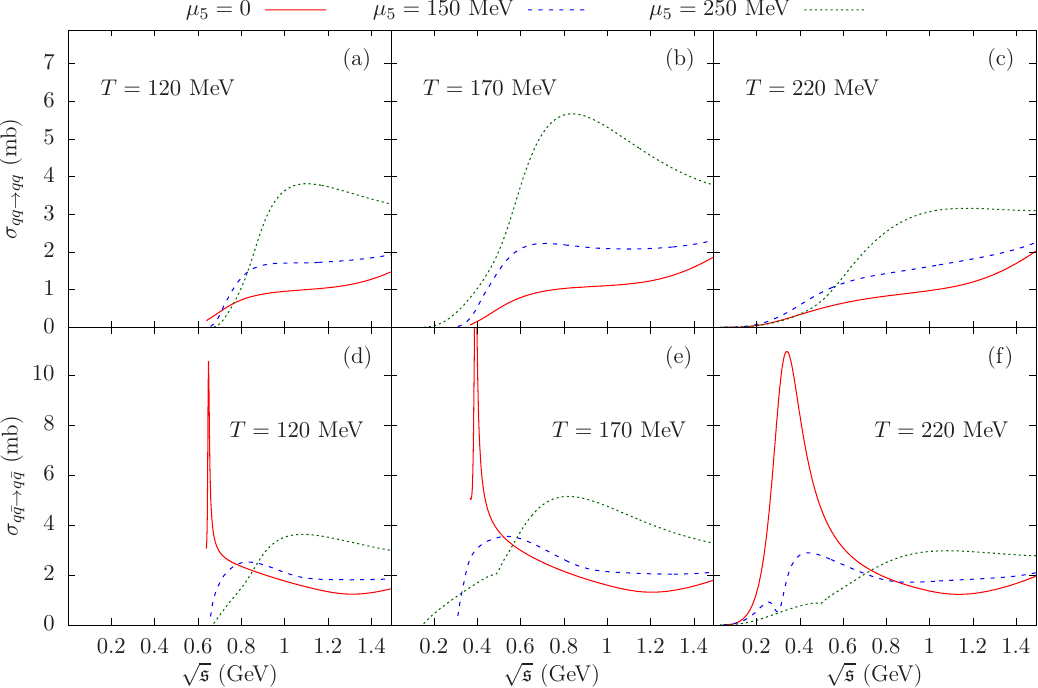} 
	\caption{(Color Online) The isospin-averaged total cross sections for the processes $qq\to qq$ and $q\qbar\to q\qbar$ as a function of centre of mass energy $\sqrt{\mans}$ for different values of temperature and CCP.}
	\label{fig.xsection}
\end{figure}

In Figs.\ref{fig.xsection}(a)-(f) we have plotted the isospin-averaged total elastic cross sections $\sigma_{qq\to qq}$ and $\sigma_{q\qbar\to q\qbar}$ as functions of the center of mass energy $(\sqrt{\mans})$. These plots correspond to the same nine representative combinations of temperature and CCP as chosen in Fig.~\ref{fig.q0} encompassing different stages of the chiral phase transition. In all the graphs of the cross section, the threshold centre of mass energy is $\sqrt{\mans}_\text{min}=2M$ which depends on the value of temperature and CCP. With the increase in $T$ the threshold $\sqrt{\mans}_\text{min}$ moves towards lower value of $\sqrt{\mans}$ due to the decrease in $M$ as can be understood from Fig.~\ref{fig.M}. Moreover, we can notice an overall Breit-Wigner like structure in the $\sqrt{\mans}$ dependence of $\sigma_{q\qbar\to q\qbar}$ where no such structure is present in $\sigma_{qq\to qq}$. This is due to the contribution from exchange of mesonic resonance mode in the $\mans$-channel diagrams present in $\sigma_{q\qbar\to q\qbar}$ which is absent in $\sigma_{qq\to qq}$ (which contains the non-resonant $t$ and $\manu$-channels). The temperature and CCP dependencies of the cross sections arise from two primary sources: firstly from the constituent quark mass $M=M(T,\mu_5)$ appearing in Eq.~\eqref{xsection}-\eqref{M2qqb} and secondly from the polarization functions $\Pi_h(q;T,\mu_5)$ appearing in Eq.~\eqref{Dh.f}. Additionally, it should be noted that $\Pi_h(q;T,\mu_5,M)$ also implicitly contains temperature and $\mu_5$ dependencies through its dependence on $M=M(T,\mu_5)$. These complexities in the cross section calculations make it challenging to straightforwardly analyze the $ T $ and $\mu_5$ dependencies of the cross sections. Let us first try to discuss the temperature and CCP dependence of $\sigma_{qq\to qq}$, which has contribution from terms like $\frac{2G}{1-2G\RE\Pi_h(q)}$, that contains only the real parts of the polarization functions (note that $\IM\Pi_h(q^0,|\bm{q}|=0)$'s contribute mainly to the $\mans$-channels where the exchange momentum is time-like). At vanishing CCP, on comparing the red curves in the upper panels of Fig~\ref{fig.xsection}, we find that $\sigma_{qq\to qq}$ has insignificant temperature dependence. This is due to the fact that both the $\RE\Pi_h(q^0,|\bm{q}|=0)$ and $\RE\Pi_h(q^0=0,\bm{q})$ depend mildly on the temperature as can be observed in the left panels of Figs.~\ref{fig.q0} and \ref{fig.qv}. At non-zero CCP $\sigma_{qq\to qq}$ is maximum near the chiral phase transition region ($T\simeq170$ MeV). Away from the scattering threshold region (i.e. high $\sqrt{\mans}$) $\sigma_{qq\to qq}$'s have almost monotonic behavior with CCP, in particular $\sigma_{qq\to qq}$ increases as we increase the value of $\mu_5$ for all the temperatures.

We now come to the temperature and CCP dependence of $\sigma_{q\qbar\to q\qbar}$ which is mainly controlled by the contributions from $\IM\Pi_h(q^0,|\bm{q}|=0)$'s  to the $\mans$-channel diagrams. With the increase in temperature we find broadening of the Breit-Wigner structure of $\sigma_{q\qbar\to q\qbar}$ due to the increase in $\IM\Pi_h(q^0,|\bm{q}|=0)$ with temperature as can be noticed in Fig.~\ref{fig.q0}. The CCP dependence of $\sigma_{q\qbar\to q\qbar}$ is found to be non-monotonic and the behaviour is different at different regions of center of mass energy. At low (high) $\sqrt{\mans}$ region, $\sigma_{q\qbar\to q\qbar}$ decreases (increases) with the increase in $\mu_5$. 
\begin{figure}[h]
	\includegraphics[angle=-90,scale=0.35]{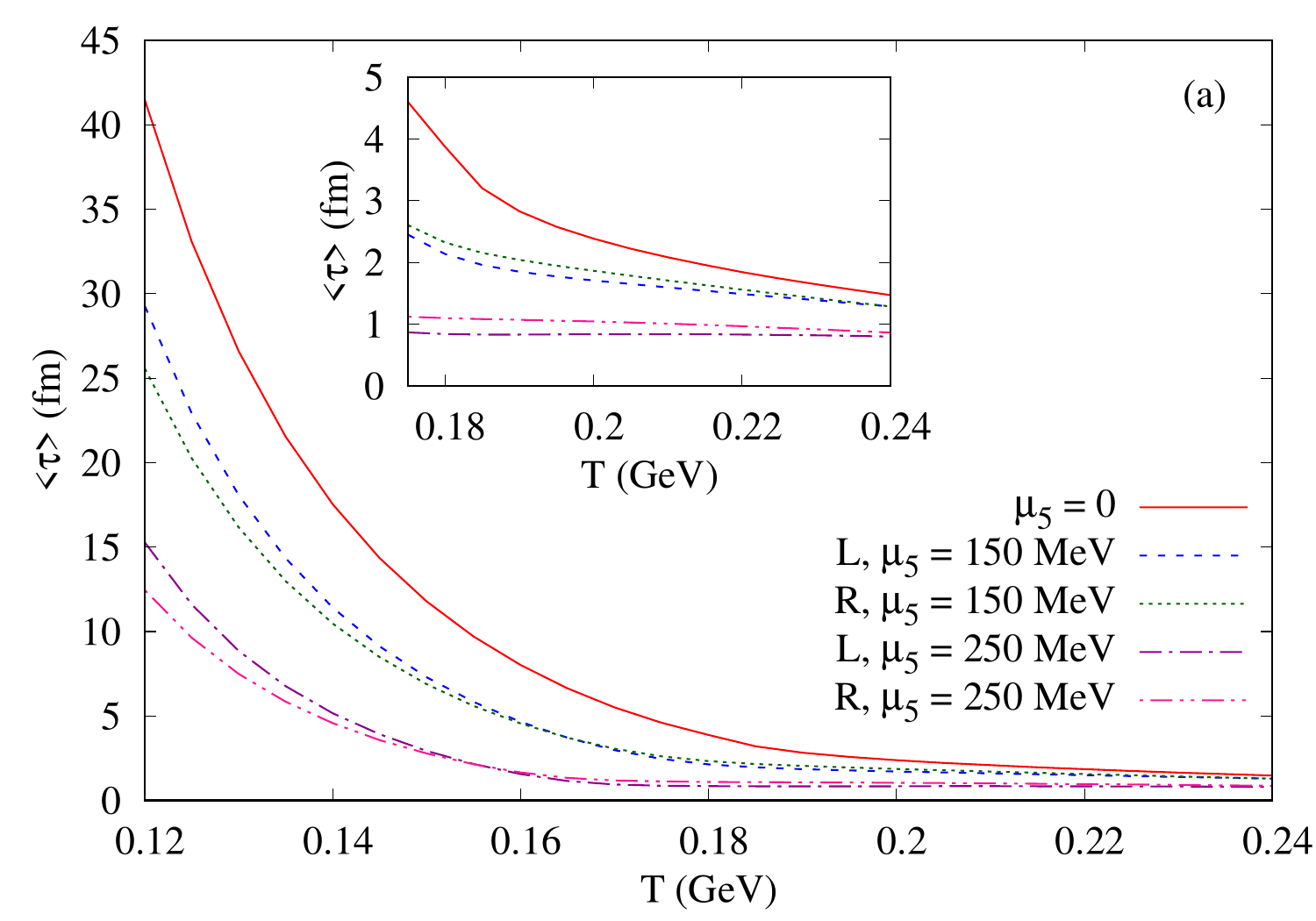}
	\includegraphics[angle=-90,scale=0.35]{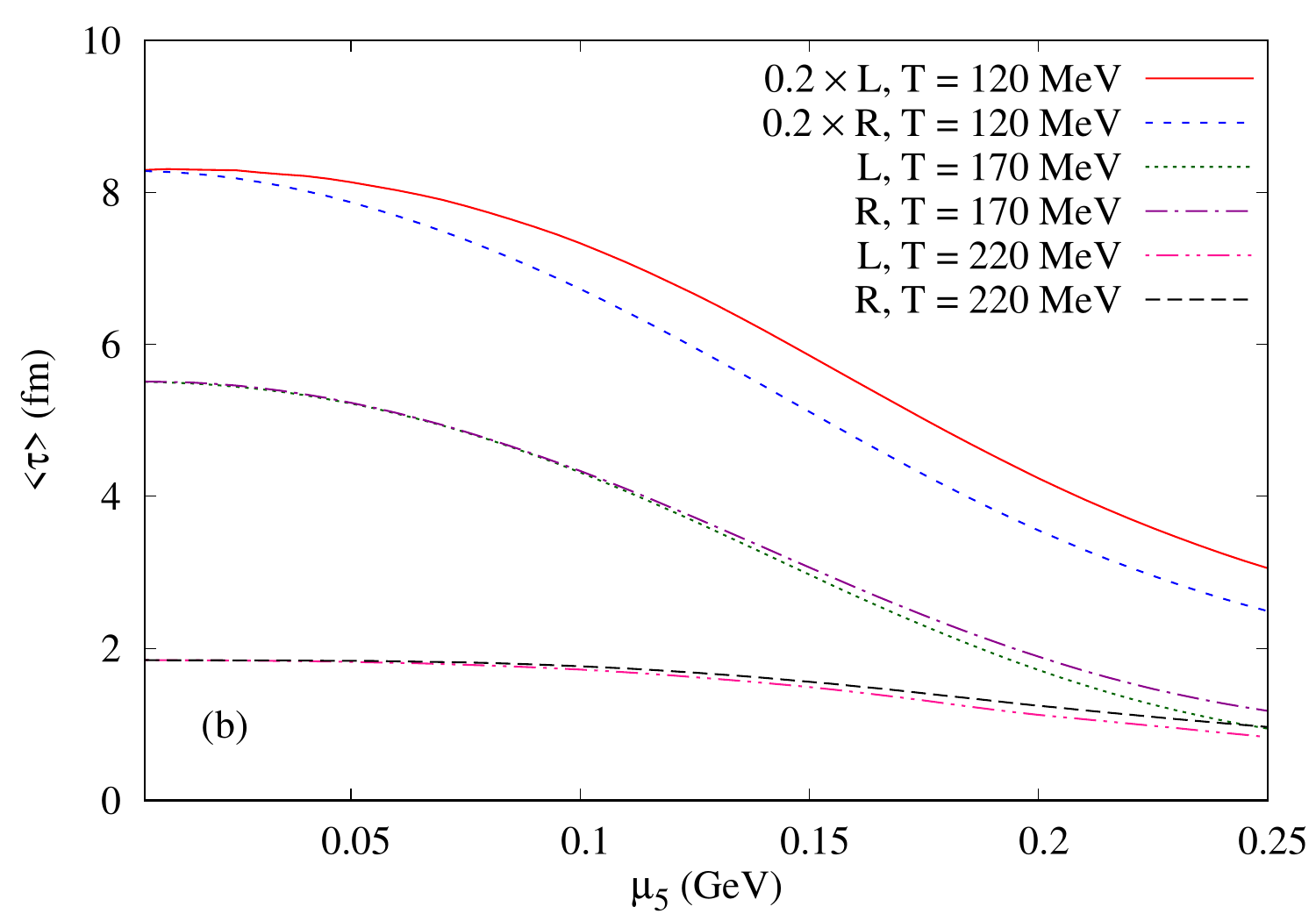}
	\caption{(Color Online) The variation of the average relaxation time $\ensembleaverage{\tau}$ of the quarks as a function of (a) temperature and (b) CCP. The inset plot in (a) zooms in the high temperature region.}
	\label{fig.tauavg}
\end{figure}
\begin{figure}[h]
	\includegraphics[angle=-90,scale=0.35]{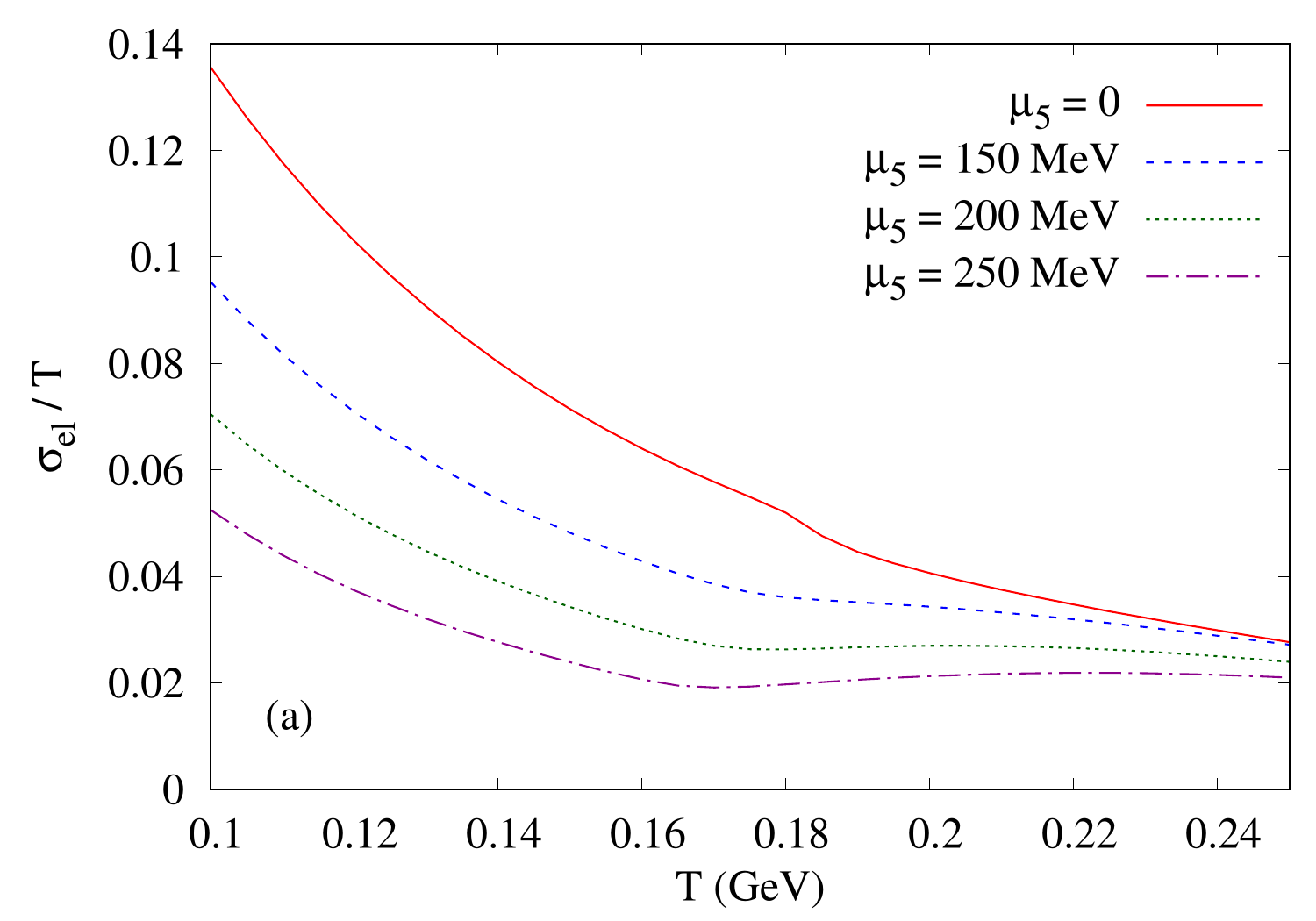}
	\includegraphics[angle=-90,scale=0.35]{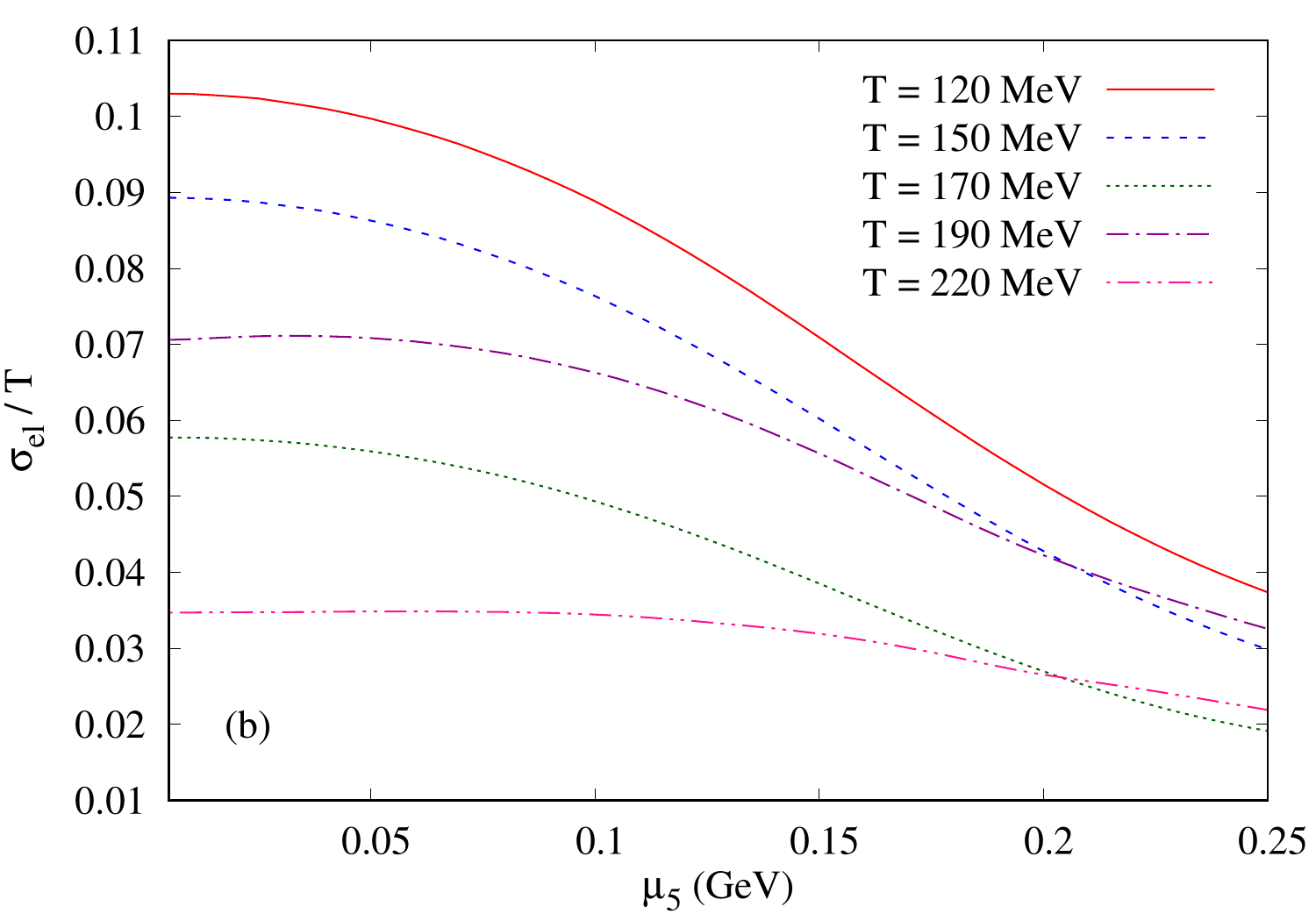}
	\caption{(Color Online) The variation of $\sigma_\text{el}/T$ as a function of (a) temperature, (b) CCP}
	\label{fig.cond}
\end{figure}

Having obtained the cross section, we will now show the results of temperature and CCP dependence of the momentum averaged relaxation time $\ensembleaverage{\tau}_r$ for the quarks/antiquarks with helicity `$r$' defined as
\begin{eqnarray}\label{Eq.tau.avg}
	\ensembleaverage{\tau}_r = \ensembleaverage{\frac{1}{\Gamma}}_r = \int d^3k f^{\bm{k}r} \frac{1}{\Gamma^r(\bm{k})} \Big/ \int d^3k f^{\bm{k}r} ~,
\end{eqnarray}
where $f^{\bm{k}r} = [e^{\wkr/T}+1]^{-1}$ is the Fermi-Dirac distribution function of the quarks/antiquarks at $\mu=0$. Note that in Eq.~\eqref{Eq.tau.avg} we have used the fact that as we are working with vanishing quark chemical potential i.e. $\mu=0$, Eqs.~\eqref{gam.p} and \eqref{gam.m} become identical so that $\Gamma_q^r(\bm{k})=\Gamma^r_{\bar q}(\bm{k})\equiv\Gamma^r(\bm{k})$. In Figs.~\ref{fig.tauavg}(a) and (b), we have depicted $\ensembleaverage{\tau}_\pm$ as a function of temperature (at different CCPs) and CCP (at different temperatures) respectively where `L(R)' corresponds to left (right) helical quarks/antiquarks with $r=-1$ ($r=1$).

As can be noticed in the figure, $\ensembleaverage{\tau}_\pm$ decreases monotonically as we increase temperature which can be understood by approximating $\ensembleaverage{\tau}_\pm \sim \dfrac{1}{n\sigma}$ from Eqs.~\eqref{gam.p} and \eqref{gam.m} where $n$ is the quark/antiquark density (they are same at $\mu=0$) at finite CCP and $\sigma$ is the value of typical $2\to2$ cross section in the medium. Thus, the leading temperature dependence of $\ensembleaverage{\tau}$ comes from the quark/antiquark density $n$ since the cross section $\sigma$ has weak temperature dependence as already noticed in Fig.~\ref{fig.xsection}. The quark/antiquark densities $n$ at finite CCP can analytically be calculated for massless quark/antiquark case as follows:
\begin{eqnarray}
	n(T,\mu_5,\mu=0) &=& 2N_c N_f\sum_{r\in \{\pm\} }^{} \int \!\! \frac{d^3k}{(2\pi)^3} \frac{1}{e^{||\bm{k}|+r\mu_5|/T}+1} = \frac{N_c N_f}{\pi^2} T^3 \TB{ 3\zeta(3) +\ln(4)\FB{\frac{\mu_5}{T}}^2 }  \label{density}
\end{eqnarray}
where $\zeta(x)$ is the Riemann zeta function. Therefore, at leading order the temperature dependence of the density is $n\sim T^3$ which in turn makes the average relaxation time to follow a trend of $\ensembleaverage{\tau}_\pm \sim \frac{1}{n\sigma} \sim \frac{1}{T^3}$ as evident from Fig.~\ref{fig.tauavg}(a). With the increase in CCP, $\ensembleaverage{\tau}_\pm$ also decreases monotonically for all temperature with can be attributed to mainly two factors: firstly the dominant CCP dependence of the density $n \sim \mu_5^2$ from Eq.~\eqref{density} which makes $\ensembleaverage{\tau}_\pm \sim \frac{1}{n\sigma} \sim \frac{1}{\mu_5^2}$ and secondly due to the mild increase in cross section with the increase in $\mu_5$ as observed in Fig.~\ref{fig.xsection}. We also notice in Fig.~\ref{fig.tauavg} that, the difference between the right and left helical relaxation times ($\ensembleaverage{\tau}_+ \sim \ensembleaverage{\tau}_-$) increases with the increase in CCP but decreases with the increase in temperature. This is obvious since the CCP is responsible for the imbalance of right and left helical quarks/antiquarks whereas the thermal fluctuations tries to diminish such an imbalance.

In Figs.~\ref{fig.cond}(a) and (b) we have presented the variations of the dimensionless quantity $\sigma_\text{el}/T$. Fig.~\ref{fig.cond}(a) shows the variation as a function of temperature for different values of CCP, while Fig.~\ref{fig.cond}(b) illustrates the variation with CCP for different temperatures. The temperature and CCP dependencies of $\sigma_\text{el}/T$ exhibit similarities with those of $\langle\tau\rangle$ in Fig.~\ref{fig.tauavg}. This correspondence can be understood from Eq.\eqref{cond.3}, where $\sigma_\text{el}$ is approximately proportional to the relaxation time $\tau=\Gamma^{-1}$. With the increase in temperature as well as CCP $\sigma_\text{el}/T$ shows a decreasing trend in the low temperature region, whereas in the high temperature region, the CCP dependence of $\sigma_\text{el}/T$ becomes slightly non-monotonic. We also note that, as the temperature increases, the CCP dependence of the $\sigma_{\rm el}$ tends to wash out. The physical reason could be the fact that at high temperature, the thermal fluctuations tries to diminish the chiral imbalance so that the CCP-effects becomes less prominent at high temperature.

\section{Summary \& Conclusion} \label{sec.summary}
In summary, we have evaluated the electrical conductivity of hot and dense strongly interacting matter in the presence of a chiral imbalance employing two-flavour gauged Nambu-Jona--Lasinio (NJL) model. The analysis of the temperature dependence of the constituent quark mass for various values of CCP reveals an interesting interplay. Specifically, CCP enhances the chiral condensate at low temperatures, indicative of a chiral catalysis effect. In contrast, at higher temperatures $\mu_5$ weakens the chiral condensate leading to the restoration of chiral symmetry at a relatively lower temperature compared to the scenario when $\mu_5=0$. This behavior can be characterized as an example of inverse chiral catalysis.

The real and imaginary parts of polarization functions for both scalar and pseudoscalar mesons are evaluated using the real time formulation of finite temperature field theory. Detailed analysis of the analytic structure reveals the existence of non-trivial Landau cut contribution in the kinematic region along with the usual Unitary cut contributions which is a purely finite CCP effect. The imaginary part of the pion polarization function is found to exhibit non-monotonic behavior specifically near the Unitary cut threshold. At higher values of temperature, independent of the values of CCP, both real and imaginary parts of polarization functions for $ \sigma $  and $ \pi $ coincide with each other indicating partial restoration of chiral symmetry.

The isospin-averaged total elastic cross sections $\sigma_{qq\to qq}$ and $\sigma_{q\qbar\to q\qbar}$ have been studied in detail as functions of the center of mass energy $(\sqrt{\mans})$. In the case of $\sigma_{q\qbar\to q\qbar}$ the $\sqrt{\mans}$ dependence exhibits a Breit-Wigner-like structure due to the contribution coming from the exchange of mesonic resonance modes in the $\mans$-channel diagrams. However, $\sigma_{qq\to qq}$ lacks such a structure since it contains non-resonant $t$ and $\manu$-channel diagrams. As the temperature increases the Breit-Wigner structure of $\sigma_{q\qbar\to q\qbar}$ broadens due to the increased imaginary part of the polarization functions which is a result of the enhancement of the available thermal phase space. The dependence of $\sigma_{q\qbar\to q\qbar}$ on CCP is non-monotonic and its behavior varies across different regions of the center of mass energy.

The momentum-averaged relaxation time $\AB{\tau}_\pm$ exhibits a monotonic decrease with increasing temperature and CCP. This behavior is primarily due to the leading temperature and CCP dependence of $\AB{\tau}_\pm$ which is dominated by the quark number density $n(T,\mu_5)$ which varies approximately as $n \sim T^3\TB{a+b \FB{\dfrac{\mu_5}{T}}^2 }$ (where $a$ and $b$ are dimensionless constants), while the cross-section has a weaker temperature and CCP dependence.

The ratio of electrical conductivity to temperature also shows a monotonic decrease as the temperature and CCP increase. The chiral imbalance has a pronounced effect on the electrical conductivity in the chiral symmetry broken phase as  $\sigma_{\rm el}/T$ decreases significantly as CCP increases.  In the high-temperature regime, the $\mu_5$ dependence of $\sigma_{\rm el}/T$ is relatively small and exhibits a weak non-monotonic behavior. This observation implies that the influence of chiral imbalance on the electrical conductivity becomes less significant as chiral symmetry is partially restored.

\section*{Acknowledgments}
S.G. is funded by the Department of Higher Education, Government of West Bengal, India. N.C., S.S. and P.R. are funded by the Department of Atomic Energy (DAE), Government of India. 
\section*{Data Availability Statement}
This manuscript has no associated data. [Authors' comment:  This is a theoretical study and thus the manuscript has no associated data. All the analytical expressions required to generate the plots are given in the manuscript.]

\section*{Code Availability Statement}
This manuscript has no associated code/software. [Authors comment: The numerical integrations are performed using the standard GNU Scientific Library (GSL).]
 

\appendix
\section{Invariant Amplitudes for $2\to2$ Scatterings in the NJL Model} \label{App.Amplitudes}
\begin{figure}[h]
	\includegraphics[angle=0,scale=0.5]{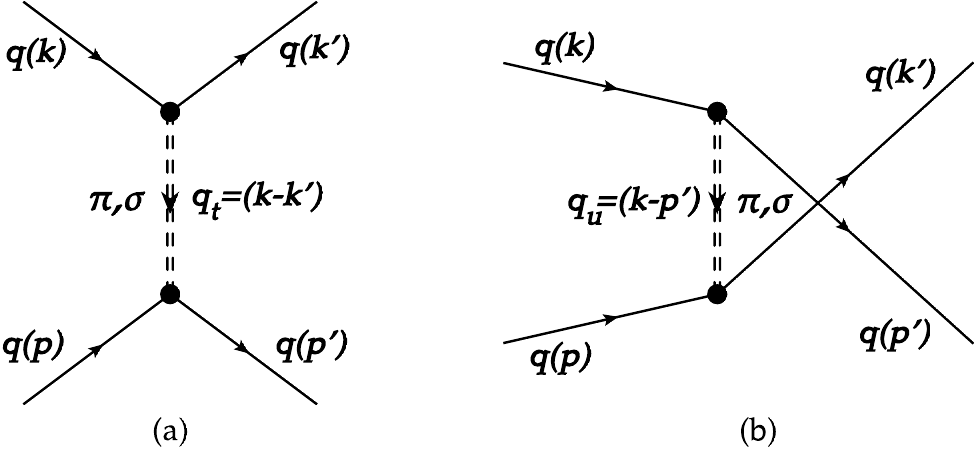} 
	\caption{(Color Online) The Feynman diagrams for $q(k)q(p)\to q(k')q(p')$ scattering via (a) $\mant$-channel and (b) $\manu$-channel in RPA. The arrows represent the direction of momentum.}
	\label{fig.qq2qq}
\end{figure}

In this appendix we will provide the calculations of invariant amplitudes for $2\to2$ scatterings in the 2-flavour NJL Model. Let us first consider the process $q(k)q(p)\to q(k')q(p')$ in which $q\in\{u,d\}$. Denoting the isospin states $\ket{I,I_z}$ of the up and down quarks by $\ket{u}=\ket{\frac{1}{2},\frac{1}{2}}$ and $\ket{d}=\ket{\frac{1}{2},-\frac{1}{2}}$ respectively, we can write the composite isospin states with total isospin 1 and 0 (using the table of Clebsch-Gordon coefficients) as
\begin{eqnarray}
	\ket{1,1} &=& \ket{u}\otimes\ket{u}, \label{composite.1}\\
	\ket{1,0} &=& \frac{1}{\sqrt{2}}\ket{u}\otimes\ket{d} + \frac{1}{\sqrt{2}}\ket{d}\otimes\ket{u}, \\
	\ket{1,-1} &=& \ket{d}\otimes\ket{d}, \\
	\ket{0,0} &=& \frac{1}{\sqrt{2}}\ket{u}\otimes\ket{d} - \frac{1}{\sqrt{2}}\ket{d}\otimes\ket{u}.
	\label{composite.2}
\end{eqnarray}
Making use of Eqs.~\eqref{composite.1}-\eqref{composite.2} we write the invariant amplitude $\scrM_{\ket{I,I_z}}$ in a particular isospin channel $\ket{I,I_z}$ in terms of amplitude of six different individual scattering processes namely, (i) $uu\to uu$, (ii) $ud\to ud$, (iii) $ud\to du$, (iv) $du\to ud$, (v) $du\to du$ and (vi) $dd\to dd$. Denoting the invariant amplitude of $i^\text{th}$ process by $\scrM_{(k)}$, we have from Eqs.~\eqref{composite.1}-\eqref{composite.2}
\begin{eqnarray}
	\scrM_{\ket{1,1}} &=& \scrM_{(1)}, \label{M.11}\\
	\scrM_{\ket{1,0}} &=& \frac{1}{2}\TB{ \scrM_{(2)}+\scrM_{(3)}+\scrM_{(4)}+\scrM_{(5)} }, \\
	\scrM_{\ket{1,-1}} &=& \scrM_{(6)}, \\
	\scrM_{\ket{0,0}} &=& \frac{1}{2}\TB{ \scrM_{(2)}-\scrM_{(3)}-\scrM_{(4)}+\scrM_{(5)} }. \label{M.00}
\end{eqnarray}
We now write the invariant amplitudes for all the six processes of $q(k)q(p)\to q(k')q(p')$ scattering using the (bosonized) NJL Lagrangian originating from Eq.~\eqref{lagrangian}. Each of the scattering process comes from two types of Feynman diagrams (namely the $\mant$-channel and $\manu$-channel) as shown in Fig.~\ref{fig.qq2qq}.
Making use of Eq.~\eqref{lagrangian} the invariant amplitudes for the six processes in the Random Phase Approximation (RPA) come out to be 
\begin{eqnarray}
	\scrM_{(1)} &=& \scrM_{(6)} = \FB{-\scrM_\mant^\pi - \scrM_\manu^\pi - \scrM_\mant^\sigma - \scrM_\manu^\sigma}, \label{M1}\\
	\scrM_{(2)} &=& \scrM_{(5)} = \FB{ \scrM_\mant^\pi - 2\scrM_\manu^\pi - \scrM_\mant^\sigma}, \\
	\scrM_{(3)} &=& \scrM_{(4)} = \FB{ -2\scrM_\mant^\pi + \scrM_\manu^\pi - \scrM_\manu^\sigma } \label{M3}~,
\end{eqnarray}
where
\begin{eqnarray}
	\scrM_\mant^h = \overline{U}(k')\Gamma_h U(k) D_\mant^h \overline{U}(p')\Gamma_h U(p), \\
	\scrM_\manu^h = \overline{U}(p')\Gamma_h U(k) D_\manu^h \overline{U}(k')\Gamma_h U(p) \label{z.1}
\end{eqnarray}
in which $h\in \{\pi,\sigma\}$, $\Gamma_\pi = i\gamma^5$, $\Gamma_\sigma = \identity$, $U(k)$ is the quark spinor (in which the color indices have been suppressed for brevity), $D_\mant^h = D^h (q_\mant=k-k')$, $D_\manu^h = D^h (q_\manu=k-p')$ and
\begin{eqnarray}
	D^h(q;T,\mu,\mu_5) = \frac{2G}{1-2G\Pi_h(q;T,\mu,\mu_5)}. \label{Dh}
\end{eqnarray}
In Eq.~\eqref{Dh}, $\Pi_h(q;T,\mu,\mu_5)$ is the polarization function for the meson $h$ which will be specified later. Making use of Eqs.~\eqref{M1}-\eqref{M3} in Eqs.~\eqref{M.11}-\eqref{M.00} one finds that the invariant amplitude $\scrM_{\ket{I,I_z}}$ in the particular isospin channel $\ket{I,I_z}$ does not depend on $I_z$ implying
\begin{eqnarray}
	\scrM_{\ket{1,1}} &=& \scrM_{\ket{1,0}} = \scrM_{\ket{1,-1}} = \FB{-\scrM_\mant^\pi - \scrM_\manu^\pi - \scrM_\mant^\sigma - \scrM_\manu^\sigma}, \label{M.11.f}\\
	\scrM_{\ket{0,0}} &=& \FB{3\scrM_\mant^\pi - 3\scrM_u^\pi - \scrM_\mant^\sigma + \scrM_u^\sigma}. \label{M.00.f}
\end{eqnarray}
Let us now calculate the absolute square of the invariant amplitudes of Eqs.~\eqref{M.11.f} and \eqref{M.00.f} by performing an average over initial and sum over final spin and color. The squared invariant amplitude in the isospin channel $\ket{I,I_z}$ is given by
\begin{eqnarray}
	\overline{|\scrM_{\ket{I,I_z}}|^2} = \frac{1}{4N_c^2} \sum_{\text{spin}}^{} ~ \sum_{\text{color}}^{} |\scrM_{\ket{I,I_z}}|^2.
\end{eqnarray}
The calculation of $\overline{|\scrM_{\ket{I,I_z}}|^2}$ is tedious but straightforward and the final result comes out to be
\begin{eqnarray}
		\overline{|\scrM_{\ket{1,I_z}}|^2} &=& \mant^2|D_\mant^\pi|^2 + \manu^2|D_\manu^\pi|^2 + (4M^2-\mant)^2|D_\mant^\sigma|^2 + (4M^2-\manu)^2|D_\manu^\sigma|^2 + \frac{1}{N_C} \RE \big[- \mant\manu D_\mant^\pi D_\manu^{\pi*} 
	- \mant(4M^2-\manu) D_\mant^\pi D_\manu^{\sigma*} \nn \\
	&&~~~ - \manu(4M^2-\mant) D_\mant^\sigma D_\manu^{\pi*} + \{(4M^2-\mant)(4M^2-\manu)+2\mant\manu\} D_\mant^\sigma D_\manu^{\sigma*}  \big], \label{M.1}\\
		\overline{|\scrM_{\ket{0,0}}|^2} &=& 9\mant^2|D_\mant^\pi|^2 + 9\manu^2|D_\manu^\pi|^2 + (4M^2-\mant)^2|D_\mant^\sigma|^2 + (4M^2-\manu)^2|D_\manu^\sigma|^2 + \frac{1}{N_C} \RE \big[9 \mant\manu D_\mant^\pi D_\manu^{\pi*} 
	- 3\mant(4M^2-\manu) D_\mant^\pi D_\manu^{\sigma*} \nn \\
	&&~~~ - 3\manu(4M^2-\mant) D_\mant^\sigma D_\manu^{\pi*} - \{(4M^2-\mant)(4M^2-\manu)+2\mant\manu\} D_\mant^\sigma D_\manu^{\sigma*}  \big], \label{M.0}
\end{eqnarray}
where, $\mant=(k-k')^2$ and $\manu=(k-p')^2$ are the Mandlestam variables. Making use of Eqs.~\eqref{M.1} and \eqref{M.0}, we obtain the isospin averaged $qq\to qq$ scattering amplitude (and also the $\qbar\qbar\to \qbar\qbar$ scattering amplitude using the crossing symmetry) as
\begin{eqnarray}
	\overline{\MB{\scrM_{qq}}^2}  = \overline{\MB{\scrM_{\qbar\qbar}}^2} &=& \frac{3}{4} \overline{|\scrM_{\ket{1,I_z}}|^2} + \frac{1}{4} \overline{|\scrM_{\ket{0,0}}|^2} \nn \\ &=& 
		  3\mant^2|D_\mant^\pi|^2 + 3\manu^2|D_\manu^\pi|^2 + (4M^2-\mant)^2|D_\mant^\sigma|^2 + (4M^2-\manu)^2|D_\manu^\sigma|^2 + \frac{1}{2N_C} \RE \big[ 3\mant\manu D_\mant^\pi D_\manu^{\pi*} \nn \\
		  && - 3\mant(4M^2-\manu) D_\mant^\pi D_\manu^{\sigma*}  - 3\manu(4M^2-\mant) D_\mant^\sigma D_\manu^{\pi*} + \{(4M^2-\mant)(4M^2-\manu)+2\mant\manu\} D_\mant^\sigma D_\manu^{\sigma*}  \big]. \label{M.qq}
\end{eqnarray}

\begin{figure}[h]
	\includegraphics[angle=0,scale=0.5]{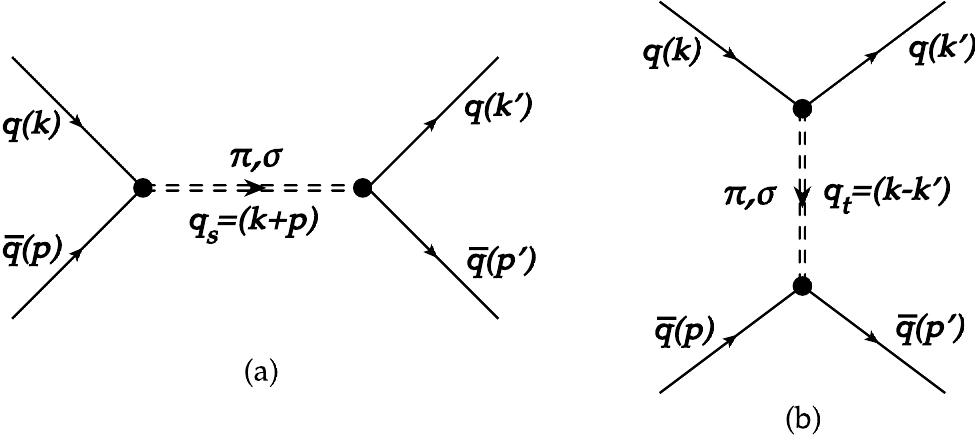} 
	\caption{(Color Online) The Feynman diagrams for $q(k)\qbar(p)\to q(k')\qbar(p')$ scattering via (a) $\mans$-channel and (b) $\mant$-channel in RPA. The arrows represent the direction of momentum.}
	\label{fig.qqb2qqb}
\end{figure}

The calculation of invariant amplitudes for the process $q(k)\qbar(p)\to q(k')\qbar(p')$ where $\qbar\in\{\ubar,\dbar\}$, is done in an analogous manner. In this case the isospin states $\ket{I,I_z}$ of the anti-up and anti-down quarks are given by $\ket{\ubar}=\ket{\frac{1}{2},-\frac{1}{2}}$ and $\ket{\dbar}=-\ket{\frac{1}{2},\frac{1}{2}}$ respectively so that the composite quark-antiquark isospin states with total isospin 1 and 0 are 
\begin{eqnarray}
	\widetilde{\ket{1,1}} &=& -\ket{u}\otimes\ket{\dbar}, \label{composite.3}\\
	\widetilde{\ket{1,0}} &=& \frac{1}{\sqrt{2}}\ket{u}\otimes\ket{\ubar} - \frac{1}{\sqrt{2}}\ket{d}\otimes\ket{
	\dbar}, \\
	\widetilde{\ket{1,-1}} &=& \ket{d}\otimes\ket{\ubar}, \\
	\widetilde{\ket{0,0}} &=& \frac{1}{\sqrt{2}}\ket{u}\otimes\ket{\ubar} + \frac{1}{\sqrt{2}}\ket{d}\otimes\ket{\dbar}.
	\label{composite.4}
\end{eqnarray}
As before, using Eqs.~\eqref{composite.3}-\eqref{composite.4}, we write the invariant amplitude $\scrM_{\widetilde{\ket{I,I_z}}}$ in a particular isospin channel $\widetilde{\ket{I,I_z}}$ in terms of amplitude of six different individual scattering processes namely, (i) $u\dbar\to u\dbar$, (ii) $u\ubar\to u\ubar$, (iii) $u\ubar\to d\dbar$, (iv) $d\dbar\to u\ubar$, (v) $d\dbar\to d\dbar$ and (vi) $d\ubar\to d\ubar$. Denoting the invariant amplitude of $i^\text{th}$ process by $\scrMtil_{(k)}$, we have from Eqs.~\eqref{composite.3}-\eqref{composite.4}
\begin{eqnarray}
	\scrM_{\widetilde{\ket{1,1}}} &=& \scrMtil_{(1)}, \label{Mtil.11}\\
	\scrM_{\widetilde{\ket{1,0}}} &=& \frac{1}{2}\TB{ \scrMtil_{(2)}-\scrMtil_{(3)}-\scrMtil_{(4)}+\scrMtil_{(5)} }, \\
	\scrM_{\widetilde{\ket{1,-1}}} &=& \scrMtil_{(6)}, \\
	\scrM_{\widetilde{\ket{0,0}}} &=& \frac{1}{2}\TB{ \scrMtil_{(2)}+\scrMtil_{(3)}+\scrMtil_{(4)}+\scrMtil_{(5)} }. \label{Mtil.00}
\end{eqnarray}
 Each of the six scattering processes of $q(k)\qbar(p)\to q(k')\qbar(p')$ comes from two types of Feynman diagrams (namely the $s$-channel and $t$-channel) as shown in Fig.~\ref{fig.qqb2qqb}.
Using Eq.~\eqref{lagrangian} the invariant amplitudes for the six processes in the RPA come out to be 
\begin{eqnarray}
	\scrMtil_{(1)} &=& \scrMtil_{(6)} = \FB{-2\scrMtil_\mans^\pi + \scrMtil_\mant^\pi - \scrMtil_\mant^\sigma}, \label{Mt1}\\
	\scrMtil_{(2)} &=& \scrMtil_{(5)} = \FB{ -\scrMtil_\mans^\pi - \scrMtil_\mant^\pi - \scrMtil_\mans^\sigma - \scrMtil_\mant^\sigma}, \\
	\scrMtil_{(3)} &=& \scrMtil_{(4)} = \FB{ \scrMtil_\mans^\pi - 2\scrMtil_\mant^\pi - \scrMtil_\mans^\sigma } \label{Mt3}~,
\end{eqnarray}
where
\begin{eqnarray}
	\scrMtil_\mans^h = \overline{U}(k')\Gamma_h V(p') D_\mans^h \overline{V}(p)\Gamma_h U(k), \\
	\scrMtil_\mant^h = \overline{U}(k')\Gamma_h U(k) D_\mant^h \overline{V}(p')\Gamma_h V(p)
\end{eqnarray}
$V(p)$ is the antiquark spinor (in which the color index has been suppressed for brevity), $D_\mans^h = D^h (q_\mans=k+p)$ and other terms are defined below Eq.~\eqref{z.1}. 

Making use of Eqs.~\eqref{Mt1}-\eqref{Mt3} in Eqs.~\eqref{Mtil.11}-\eqref{Mtil.00}, we again find that, the invariant amplitude $\scrM_{\widetilde{\ket{I,I_z}}}$ in the particular isospin channel $\widetilde{\ket{I,I_z}}$ does not depend on $I_z$ implying
\begin{eqnarray}
	\scrM_{\widetilde{\ket{1,1}}} &=& \scrM_{\widetilde{\ket{1,0}}} = \scrM_{\widetilde{\ket{1,-1}}} = \FB{-2\scrMtil_\mans^\pi + \scrMtil_\mant^\pi - \scrMtil_\mant^\sigma }, \label{Mt.11.f}\\
	\scrM_{\widetilde{\ket{0,0}}} &=& \FB{-3\scrMtil_\mant^\pi - 2\scrMtil_\mans^\sigma - \scrMtil_\mant^\sigma}. \label{Mt.00.f}
\end{eqnarray}
The squared invariant amplitude from Eqs.~\eqref{Mt.11.f} and \eqref{Mt.00.f} is given by the following expressions
\begin{eqnarray}
	\overline{|\scrM_{\widetilde{\ket{I,I_z}}}|^2} = \frac{1}{4N_c^2} \sum_{\text{spin}}^{} ~ \sum_{\text{color}}^{} |\scrM_{\widetilde{\ket{I,I_z}}}|^2.
\end{eqnarray}
Again a lengthy but straightforward calculation of $\overline{|\scrM_{\widetilde{\ket{I,I_z}}}|^2}$ gives the following  final expressions
\begin{eqnarray}
	\overline{|\scrM_{\widetilde{\ket{1,I_z}}}|^2} &=& 4\mans^2|D_\mans^\pi|^2 + \mant^2|D_\mant^\pi|^2 + (4M^2-\mant)^2|D_\mant^\sigma|^2 + \frac{2}{N_C} \RE \big[ \mans\mant D_\mans^\pi D_\mant^{\pi*} 
	- \mans(4M^2-\mant) D_\mans^\pi D_\mant^{\sigma*}  \big], \label{Mt.1}\\
	\overline{|\scrM_{\widetilde{\ket{0,0}}}|^2} &=& 9\mant^2|D_\mant^\pi|^2 + 4(4M^2-\mans)^2|D_\mans^\sigma|^2 + (4M^2-\mant)^2|D_\mant^\sigma|^2 \nn \\
	&&~~~ + \frac{2}{N_C} \RE \big[ - 3\mant(4M^2-\mans) D_\mans^\sigma D_\mant^{\pi*} + \{(4M^2-\mans)(4M^2-\mant)-2\mans\mant\} D_\mans^\sigma D_\mant^{\sigma*}  \big], \label{Mt.0}~.
\end{eqnarray}
Making use of Eqs.~\eqref{Mt.1} and \eqref{Mt.0} we finally obtain the isospin averaged $q\qbar\to q\qbar$ scattering amplitude (and also the $\qbar q\to \qbar q$ scattering amplitude using the crossing symmetry) as
\begin{eqnarray}
	\overline{\MB{\scrM_{q\qbar}}^2}  = \overline{\MB{\scrM_{\qbar q}}^2}  &=& \frac{3}{4} \overline{|\scrM_{\widetilde{\ket{1,I_z}}}|^2} + \frac{1}{4} \overline{|\scrM_{\widetilde{\ket{0,0}}}|^2} \nn \\ &=& 
	3\mans^2|D_\mans^\pi|^2 + 3\mant^2|D_\mant^\pi|^2 + (4M^2-\mant)^2|D_\mant^\sigma|^2 + (4M^2-\mans)^2|D_\mans^\sigma|^2 + \frac{1}{2N_C} \RE \big[ 3\mans\mant D_\mans^\pi D_\mant^{\pi*} \nn \\
	&&	- 3\mans(4M^2-\mant) D_\mans^\pi D_\mant^{\sigma*}  - 3\mant(4M^2-\mans) D_\mans^\sigma D_\mant^{\pi*} + \{(4M^2-\mans)(4M^2-\mant)-2\mans\mant\} D_\mans^\sigma D_\mant^{\sigma*}  \big]. \label{M.qqb}
\end{eqnarray}

\bibliography{z-Ref}

\begin{thebibliography}{113}%
\makeatletter
\providecommand \@ifxundefined [1]{%
 \@ifx{#1\undefined}
}%
\providecommand \@ifnum [1]{%
 \ifnum #1\expandafter \@firstoftwo
 \else \expandafter \@secondoftwo
 \fi
}%
\providecommand \@ifx [1]{%
 \ifx #1\expandafter \@firstoftwo
 \else \expandafter \@secondoftwo
 \fi
}%
\providecommand \natexlab [1]{#1}%
\providecommand \enquote  [1]{``#1''}%
\providecommand \bibnamefont  [1]{#1}%
\providecommand \bibfnamefont [1]{#1}%
\providecommand \citenamefont [1]{#1}%
\providecommand \href@noop [0]{\@secondoftwo}%
\providecommand \href [0]{\begingroup \@sanitize@url \@href}%
\providecommand \@href[1]{\@@startlink{#1}\@@href}%
\providecommand \@@href[1]{\endgroup#1\@@endlink}%
\providecommand \@sanitize@url [0]{\catcode `\\12\catcode `\$12\catcode
  `\&12\catcode `\#12\catcode `\^12\catcode `\_12\catcode `\%12\relax}%
\providecommand \@@startlink[1]{}%
\providecommand \@@endlink[0]{}%
\providecommand \url  [0]{\begingroup\@sanitize@url \@url }%
\providecommand \@url [1]{\endgroup\@href {#1}{\urlprefix }}%
\providecommand \urlprefix  [0]{URL }%
\providecommand \Eprint [0]{\href }%
\providecommand \doibase [0]{http://dx.doi.org/}%
\providecommand \selectlanguage [0]{\@gobble}%
\providecommand \bibinfo  [0]{\@secondoftwo}%
\providecommand \bibfield  [0]{\@secondoftwo}%
\providecommand \translation [1]{[#1]}%
\providecommand \BibitemOpen [0]{}%
\providecommand \bibitemStop [0]{}%
\providecommand \bibitemNoStop [0]{.\EOS\space}%
\providecommand \EOS [0]{\spacefactor3000\relax}%
\providecommand \BibitemShut  [1]{\csname bibitem#1\endcsname}%
\let\auto@bib@innerbib\@empty
\bibitem [{\citenamefont {Adams}\ \emph {et~al.}(2005)\citenamefont {Adams}
  \emph {et~al.}}]{STAR:2005gfr}%
  \BibitemOpen
  \bibfield  {author} {\bibinfo {author} {\bibfnamefont {J.}~\bibnamefont
  {Adams}} \emph {et~al.} (\bibinfo {collaboration} {STAR}),\ }\href {\doibase
  10.1016/j.nuclphysa.2005.03.085} {\bibfield  {journal} {\bibinfo  {journal}
  {Nucl. Phys. A}\ }\textbf {\bibinfo {volume} {757}},\ \bibinfo {pages} {102}
  (\bibinfo {year} {2005})},\ \Eprint {http://arxiv.org/abs/nucl-ex/0501009}
  {arXiv:nucl-ex/0501009} \BibitemShut {NoStop}%
\bibitem [{\citenamefont {Csernai}\ \emph {et~al.}(2006)\citenamefont
  {Csernai}, \citenamefont {Kapusta},\ and\ \citenamefont
  {McLerran}}]{Csernai:2006zz}%
  \BibitemOpen
  \bibfield  {author} {\bibinfo {author} {\bibfnamefont {L.~P.}\ \bibnamefont
  {Csernai}}, \bibinfo {author} {\bibfnamefont {J.~I.}\ \bibnamefont
  {Kapusta}}, \ and\ \bibinfo {author} {\bibfnamefont {L.~D.}\ \bibnamefont
  {McLerran}},\ }\href {\doibase 10.1103/PhysRevLett.97.152303} {\bibfield
  {journal} {\bibinfo  {journal} {Phys. Rev. Lett.}\ }\textbf {\bibinfo
  {volume} {97}},\ \bibinfo {pages} {152303} (\bibinfo {year} {2006})},\
  \Eprint {http://arxiv.org/abs/nucl-th/0604032} {arXiv:nucl-th/0604032}
  \BibitemShut {NoStop}%
\bibitem [{\citenamefont {Romatschke}\ and\ \citenamefont
  {Romatschke}(2007)}]{Romatschke:2007mq}%
  \BibitemOpen
  \bibfield  {author} {\bibinfo {author} {\bibfnamefont {P.}~\bibnamefont
  {Romatschke}}\ and\ \bibinfo {author} {\bibfnamefont {U.}~\bibnamefont
  {Romatschke}},\ }\href {\doibase 10.1103/PhysRevLett.99.172301} {\bibfield
  {journal} {\bibinfo  {journal} {Phys. Rev. Lett.}\ }\textbf {\bibinfo
  {volume} {99}},\ \bibinfo {pages} {172301} (\bibinfo {year} {2007})},\
  \Eprint {http://arxiv.org/abs/0706.1522} {arXiv:0706.1522 [nucl-th]}
  \BibitemShut {NoStop}%
\bibitem [{\citenamefont {Luzum}\ and\ \citenamefont
  {Romatschke}(2008)}]{Luzum:2008cw}%
  \BibitemOpen
  \bibfield  {author} {\bibinfo {author} {\bibfnamefont {M.}~\bibnamefont
  {Luzum}}\ and\ \bibinfo {author} {\bibfnamefont {P.}~\bibnamefont
  {Romatschke}},\ }\href {\doibase 10.1103/PhysRevC.78.034915} {\bibfield
  {journal} {\bibinfo  {journal} {Phys. Rev. C}\ }\textbf {\bibinfo {volume}
  {78}},\ \bibinfo {pages} {034915} (\bibinfo {year} {2008})},\ \bibinfo {note}
  {[Erratum: Phys.Rev.C 79, 039903 (2009)]},\ \Eprint
  {http://arxiv.org/abs/0804.4015} {arXiv:0804.4015 [nucl-th]} \BibitemShut
  {NoStop}%
\bibitem [{\citenamefont {Kovtun}\ \emph {et~al.}(2005)\citenamefont {Kovtun},
  \citenamefont {Son},\ and\ \citenamefont {Starinets}}]{Kovtun:2004de}%
  \BibitemOpen
  \bibfield  {author} {\bibinfo {author} {\bibfnamefont {P.}~\bibnamefont
  {Kovtun}}, \bibinfo {author} {\bibfnamefont {D.~T.}\ \bibnamefont {Son}}, \
  and\ \bibinfo {author} {\bibfnamefont {A.~O.}\ \bibnamefont {Starinets}},\
  }\href {\doibase 10.1103/PhysRevLett.94.111601} {\bibfield  {journal}
  {\bibinfo  {journal} {Phys. Rev. Lett.}\ }\textbf {\bibinfo {volume} {94}},\
  \bibinfo {pages} {111601} (\bibinfo {year} {2005})},\ \Eprint
  {http://arxiv.org/abs/hep-th/0405231} {arXiv:hep-th/0405231} \BibitemShut
  {NoStop}%
\bibitem [{\citenamefont {Shifman}(1989)}]{Shifman:1988zk}%
  \BibitemOpen
  \bibfield  {author} {\bibinfo {author} {\bibfnamefont {M.~A.}\ \bibnamefont
  {Shifman}},\ }\href {\doibase 10.1016/0370-1573(91)90020-M} {\bibfield
  {journal} {\bibinfo  {journal} {Sov. Phys. Usp.}\ }\textbf {\bibinfo {volume}
  {32}},\ \bibinfo {pages} {289} (\bibinfo {year} {1989})}\BibitemShut
  {NoStop}%
\bibitem [{\citenamefont {Belavin}\ \emph {et~al.}(1975)\citenamefont
  {Belavin}, \citenamefont {Polyakov}, \citenamefont {Schwartz},\ and\
  \citenamefont {Tyupkin}}]{Belavin:1975fg}%
  \BibitemOpen
  \bibfield  {author} {\bibinfo {author} {\bibfnamefont {A.~A.}\ \bibnamefont
  {Belavin}}, \bibinfo {author} {\bibfnamefont {A.~M.}\ \bibnamefont
  {Polyakov}}, \bibinfo {author} {\bibfnamefont {A.~S.}\ \bibnamefont
  {Schwartz}}, \ and\ \bibinfo {author} {\bibfnamefont {Y.~S.}\ \bibnamefont
  {Tyupkin}},\ }\href {\doibase 10.1016/0370-2693(75)90163-X} {\bibfield
  {journal} {\bibinfo  {journal} {Phys. Lett. B}\ }\textbf {\bibinfo {volume}
  {59}},\ \bibinfo {pages} {85} (\bibinfo {year} {1975})}\BibitemShut {NoStop}%
\bibitem [{\citenamefont {'t~Hooft}(1976{\natexlab{a}})}]{tHooft:1976rip}%
  \BibitemOpen
  \bibfield  {author} {\bibinfo {author} {\bibfnamefont {G.}~\bibnamefont
  {'t~Hooft}},\ }\href {\doibase 10.1103/PhysRevLett.37.8} {\bibfield
  {journal} {\bibinfo  {journal} {Phys. Rev. Lett.}\ }\textbf {\bibinfo
  {volume} {37}},\ \bibinfo {pages} {8} (\bibinfo {year}
  {1976}{\natexlab{a}})}\BibitemShut {NoStop}%
\bibitem [{\citenamefont {'t~Hooft}(1976{\natexlab{b}})}]{tHooft:1976snw}%
  \BibitemOpen
  \bibfield  {author} {\bibinfo {author} {\bibfnamefont {G.}~\bibnamefont
  {'t~Hooft}},\ }\href {\doibase 10.1103/PhysRevD.14.3432} {\bibfield
  {journal} {\bibinfo  {journal} {Phys. Rev. D}\ }\textbf {\bibinfo {volume}
  {14}},\ \bibinfo {pages} {3432} (\bibinfo {year} {1976}{\natexlab{b}})},\
  \bibinfo {note} {[Erratum: Phys.Rev.D 18, 2199 (1978)]}\BibitemShut {NoStop}%
\bibitem [{\citenamefont {Manton}(1983)}]{Manton:1983nd}%
  \BibitemOpen
  \bibfield  {author} {\bibinfo {author} {\bibfnamefont {N.~S.}\ \bibnamefont
  {Manton}},\ }\href {\doibase 10.1103/PhysRevD.28.2019} {\bibfield  {journal}
  {\bibinfo  {journal} {Phys. Rev. D}\ }\textbf {\bibinfo {volume} {28}},\
  \bibinfo {pages} {2019} (\bibinfo {year} {1983})}\BibitemShut {NoStop}%
\bibitem [{\citenamefont {Klinkhamer}\ and\ \citenamefont
  {Manton}(1984)}]{Klinkhamer:1984di}%
  \BibitemOpen
  \bibfield  {author} {\bibinfo {author} {\bibfnamefont {F.~R.}\ \bibnamefont
  {Klinkhamer}}\ and\ \bibinfo {author} {\bibfnamefont {N.~S.}\ \bibnamefont
  {Manton}},\ }\href {\doibase 10.1103/PhysRevD.30.2212} {\bibfield  {journal}
  {\bibinfo  {journal} {Phys. Rev. D}\ }\textbf {\bibinfo {volume} {30}},\
  \bibinfo {pages} {2212} (\bibinfo {year} {1984})}\BibitemShut {NoStop}%
\bibitem [{\citenamefont {Adler}(1969)}]{Adler:1969gk}%
  \BibitemOpen
  \bibfield  {author} {\bibinfo {author} {\bibfnamefont {S.~L.}\ \bibnamefont
  {Adler}},\ }\href {\doibase 10.1103/PhysRev.177.2426} {\bibfield  {journal}
  {\bibinfo  {journal} {Phys. Rev.}\ }\textbf {\bibinfo {volume} {177}},\
  \bibinfo {pages} {2426} (\bibinfo {year} {1969})}\BibitemShut {NoStop}%
\bibitem [{\citenamefont {Bell}\ and\ \citenamefont
  {Jackiw}(1969)}]{Bell:1969ts}%
  \BibitemOpen
  \bibfield  {author} {\bibinfo {author} {\bibfnamefont {J.~S.}\ \bibnamefont
  {Bell}}\ and\ \bibinfo {author} {\bibfnamefont {R.}~\bibnamefont {Jackiw}},\
  }\href {\doibase 10.1007/BF02823296} {\bibfield  {journal} {\bibinfo
  {journal} {Nuovo Cim. A}\ }\textbf {\bibinfo {volume} {60}},\ \bibinfo
  {pages} {47} (\bibinfo {year} {1969})}\BibitemShut {NoStop}%
\bibitem [{\citenamefont {Tuchin}(2013)}]{Tuchin:2013ie}%
  \BibitemOpen
  \bibfield  {author} {\bibinfo {author} {\bibfnamefont {K.}~\bibnamefont
  {Tuchin}},\ }\href {\doibase 10.1155/2013/490495} {\bibfield  {journal}
  {\bibinfo  {journal} {Adv. High Energy Phys.}\ }\textbf {\bibinfo {volume}
  {2013}},\ \bibinfo {pages} {490495} (\bibinfo {year} {2013})},\ \Eprint
  {http://arxiv.org/abs/1301.0099} {arXiv:1301.0099 [hep-ph]} \BibitemShut
  {NoStop}%
\bibitem [{\citenamefont {Gursoy}\ \emph {et~al.}(2014)\citenamefont {Gursoy},
  \citenamefont {Kharzeev},\ and\ \citenamefont {Rajagopal}}]{Gursoy:2014aka}%
  \BibitemOpen
  \bibfield  {author} {\bibinfo {author} {\bibfnamefont {U.}~\bibnamefont
  {Gursoy}}, \bibinfo {author} {\bibfnamefont {D.}~\bibnamefont {Kharzeev}}, \
  and\ \bibinfo {author} {\bibfnamefont {K.}~\bibnamefont {Rajagopal}},\ }\href
  {\doibase 10.1103/PhysRevC.89.054905} {\bibfield  {journal} {\bibinfo
  {journal} {Phys. Rev.}\ }\textbf {\bibinfo {volume} {C89}},\ \bibinfo {pages}
  {054905} (\bibinfo {year} {2014})},\ \Eprint {http://arxiv.org/abs/1401.3805}
  {arXiv:1401.3805 [hep-ph]} \BibitemShut {NoStop}%
\bibitem [{\citenamefont {McLerran}\ and\ \citenamefont
  {Skokov}(2014)}]{McLerran:2013hla}%
  \BibitemOpen
  \bibfield  {author} {\bibinfo {author} {\bibfnamefont {L.}~\bibnamefont
  {McLerran}}\ and\ \bibinfo {author} {\bibfnamefont {V.}~\bibnamefont
  {Skokov}},\ }\href {\doibase 10.1016/j.nuclphysa.2014.05.008} {\bibfield
  {journal} {\bibinfo  {journal} {Nucl. Phys. A}\ }\textbf {\bibinfo {volume}
  {929}},\ \bibinfo {pages} {184} (\bibinfo {year} {2014})},\ \Eprint
  {http://arxiv.org/abs/1305.0774} {arXiv:1305.0774 [hep-ph]} \BibitemShut
  {NoStop}%
\bibitem [{\citenamefont {Satow}(2014)}]{Satow:2014lia}%
  \BibitemOpen
  \bibfield  {author} {\bibinfo {author} {\bibfnamefont {D.}~\bibnamefont
  {Satow}},\ }\href {\doibase 10.1103/PhysRevD.90.034018} {\bibfield  {journal}
  {\bibinfo  {journal} {Phys. Rev. D}\ }\textbf {\bibinfo {volume} {90}},\
  \bibinfo {pages} {034018} (\bibinfo {year} {2014})},\ \Eprint
  {http://arxiv.org/abs/1406.7032} {arXiv:1406.7032 [hep-ph]} \BibitemShut
  {NoStop}%
\bibitem [{\citenamefont {Ling}\ \emph {et~al.}(2014)\citenamefont {Ling},
  \citenamefont {Springer},\ and\ \citenamefont {Stephanov}}]{Ling:2013ksb}%
  \BibitemOpen
  \bibfield  {author} {\bibinfo {author} {\bibfnamefont {B.}~\bibnamefont
  {Ling}}, \bibinfo {author} {\bibfnamefont {T.}~\bibnamefont {Springer}}, \
  and\ \bibinfo {author} {\bibfnamefont {M.}~\bibnamefont {Stephanov}},\ }\href
  {\doibase 10.1103/PhysRevC.89.064901} {\bibfield  {journal} {\bibinfo
  {journal} {Phys. Rev. C}\ }\textbf {\bibinfo {volume} {89}},\ \bibinfo
  {pages} {064901} (\bibinfo {year} {2014})},\ \Eprint
  {http://arxiv.org/abs/1310.6036} {arXiv:1310.6036 [nucl-th]} \BibitemShut
  {NoStop}%
\bibitem [{\citenamefont {Hirono}\ \emph {et~al.}(2014)\citenamefont {Hirono},
  \citenamefont {Hongo},\ and\ \citenamefont {Hirano}}]{Hirono:2012rt}%
  \BibitemOpen
  \bibfield  {author} {\bibinfo {author} {\bibfnamefont {Y.}~\bibnamefont
  {Hirono}}, \bibinfo {author} {\bibfnamefont {M.}~\bibnamefont {Hongo}}, \
  and\ \bibinfo {author} {\bibfnamefont {T.}~\bibnamefont {Hirano}},\ }\href
  {\doibase 10.1103/PhysRevC.90.021903} {\bibfield  {journal} {\bibinfo
  {journal} {Phys. Rev. C}\ }\textbf {\bibinfo {volume} {90}},\ \bibinfo
  {pages} {021903} (\bibinfo {year} {2014})},\ \Eprint
  {http://arxiv.org/abs/1211.1114} {arXiv:1211.1114 [nucl-th]} \BibitemShut
  {NoStop}%
\bibitem [{\citenamefont {Deng}\ and\ \citenamefont
  {Huang}(2012)}]{Deng:2012pc}%
  \BibitemOpen
  \bibfield  {author} {\bibinfo {author} {\bibfnamefont {W.-T.}\ \bibnamefont
  {Deng}}\ and\ \bibinfo {author} {\bibfnamefont {X.-G.}\ \bibnamefont
  {Huang}},\ }\href {\doibase 10.1103/PhysRevC.85.044907} {\bibfield  {journal}
  {\bibinfo  {journal} {Phys. Rev. C}\ }\textbf {\bibinfo {volume} {85}},\
  \bibinfo {pages} {044907} (\bibinfo {year} {2012})},\ \Eprint
  {http://arxiv.org/abs/1201.5108} {arXiv:1201.5108 [nucl-th]} \BibitemShut
  {NoStop}%
\bibitem [{\citenamefont {Yin}(2014)}]{Yin:2013kya}%
  \BibitemOpen
  \bibfield  {author} {\bibinfo {author} {\bibfnamefont {Y.}~\bibnamefont
  {Yin}},\ }\href {\doibase 10.1103/PhysRevC.90.044903} {\bibfield  {journal}
  {\bibinfo  {journal} {Phys. Rev. C}\ }\textbf {\bibinfo {volume} {90}},\
  \bibinfo {pages} {044903} (\bibinfo {year} {2014})},\ \Eprint
  {http://arxiv.org/abs/1312.4434} {arXiv:1312.4434 [nucl-th]} \BibitemShut
  {NoStop}%
\bibitem [{\citenamefont {Kharzeev}\ \emph {et~al.}(2016)\citenamefont
  {Kharzeev}, \citenamefont {Liao}, \citenamefont {Voloshin},\ and\
  \citenamefont {Wang}}]{Kharzeev:2015znc}%
  \BibitemOpen
  \bibfield  {author} {\bibinfo {author} {\bibfnamefont {D.~E.}\ \bibnamefont
  {Kharzeev}}, \bibinfo {author} {\bibfnamefont {J.}~\bibnamefont {Liao}},
  \bibinfo {author} {\bibfnamefont {S.~A.}\ \bibnamefont {Voloshin}}, \ and\
  \bibinfo {author} {\bibfnamefont {G.}~\bibnamefont {Wang}},\ }\href {\doibase
  10.1016/j.ppnp.2016.01.001} {\bibfield  {journal} {\bibinfo  {journal} {Prog.
  Part. Nucl. Phys.}\ }\textbf {\bibinfo {volume} {88}},\ \bibinfo {pages} {1}
  (\bibinfo {year} {2016})},\ \Eprint {http://arxiv.org/abs/1511.04050}
  {arXiv:1511.04050 [hep-ph]} \BibitemShut {NoStop}%
\bibitem [{\citenamefont {Pratt}\ and\ \citenamefont
  {Plumberg}(2020)}]{Pratt:2019pnd}%
  \BibitemOpen
  \bibfield  {author} {\bibinfo {author} {\bibfnamefont {S.}~\bibnamefont
  {Pratt}}\ and\ \bibinfo {author} {\bibfnamefont {C.}~\bibnamefont
  {Plumberg}},\ }\href {\doibase 10.1103/PhysRevC.102.044909} {\bibfield
  {journal} {\bibinfo  {journal} {Phys. Rev. C}\ }\textbf {\bibinfo {volume}
  {102}},\ \bibinfo {pages} {044909} (\bibinfo {year} {2020})},\ \Eprint
  {http://arxiv.org/abs/1904.11459} {arXiv:1904.11459 [nucl-th]} \BibitemShut
  {NoStop}%
\bibitem [{\citenamefont {Aarts}\ and\ \citenamefont
  {Martinez~Resco}(2002)}]{Aarts:2002tn}%
  \BibitemOpen
  \bibfield  {author} {\bibinfo {author} {\bibfnamefont {G.}~\bibnamefont
  {Aarts}}\ and\ \bibinfo {author} {\bibfnamefont {J.~M.}\ \bibnamefont
  {Martinez~Resco}},\ }\href {\doibase 10.1088/1126-6708/2002/11/022}
  {\bibfield  {journal} {\bibinfo  {journal} {JHEP}\ }\textbf {\bibinfo
  {volume} {11}},\ \bibinfo {pages} {022} (\bibinfo {year} {2002})},\ \Eprint
  {http://arxiv.org/abs/hep-ph/0209048} {arXiv:hep-ph/0209048} \BibitemShut
  {NoStop}%
\bibitem [{\citenamefont {Gagnon}\ and\ \citenamefont
  {Jeon}(2007)}]{Gagnon:2006hi}%
  \BibitemOpen
  \bibfield  {author} {\bibinfo {author} {\bibfnamefont {J.-S.}\ \bibnamefont
  {Gagnon}}\ and\ \bibinfo {author} {\bibfnamefont {S.}~\bibnamefont {Jeon}},\
  }\href {\doibase 10.1103/PhysRevD.75.025014} {\bibfield  {journal} {\bibinfo
  {journal} {Phys. Rev. D}\ }\textbf {\bibinfo {volume} {75}},\ \bibinfo
  {pages} {025014} (\bibinfo {year} {2007})},\ \bibinfo {note} {[Erratum:
  Phys.Rev.D 76, 089902 (2007)]},\ \Eprint
  {http://arxiv.org/abs/hep-ph/0610235} {arXiv:hep-ph/0610235} \BibitemShut
  {NoStop}%
\bibitem [{\citenamefont {Arnold}\ \emph {et~al.}(2000)\citenamefont {Arnold},
  \citenamefont {Moore},\ and\ \citenamefont {Yaffe}}]{Arnold:2000dr}%
  \BibitemOpen
  \bibfield  {author} {\bibinfo {author} {\bibfnamefont {P.~B.}\ \bibnamefont
  {Arnold}}, \bibinfo {author} {\bibfnamefont {G.~D.}\ \bibnamefont {Moore}}, \
  and\ \bibinfo {author} {\bibfnamefont {L.~G.}\ \bibnamefont {Yaffe}},\ }\href
  {\doibase 10.1088/1126-6708/2000/11/001} {\bibfield  {journal} {\bibinfo
  {journal} {JHEP}\ }\textbf {\bibinfo {volume} {11}},\ \bibinfo {pages} {001}
  (\bibinfo {year} {2000})},\ \Eprint {http://arxiv.org/abs/hep-ph/0010177}
  {arXiv:hep-ph/0010177} \BibitemShut {NoStop}%
\bibitem [{\citenamefont {Arnold}\ \emph {et~al.}(2003)\citenamefont {Arnold},
  \citenamefont {Moore},\ and\ \citenamefont {Yaffe}}]{Arnold:2003zc}%
  \BibitemOpen
  \bibfield  {author} {\bibinfo {author} {\bibfnamefont {P.~B.}\ \bibnamefont
  {Arnold}}, \bibinfo {author} {\bibfnamefont {G.~D.}\ \bibnamefont {Moore}}, \
  and\ \bibinfo {author} {\bibfnamefont {L.~G.}\ \bibnamefont {Yaffe}},\ }\href
  {\doibase 10.1088/1126-6708/2003/05/051} {\bibfield  {journal} {\bibinfo
  {journal} {JHEP}\ }\textbf {\bibinfo {volume} {05}},\ \bibinfo {pages} {051}
  (\bibinfo {year} {2003})},\ \Eprint {http://arxiv.org/abs/hep-ph/0302165}
  {arXiv:hep-ph/0302165} \BibitemShut {NoStop}%
\bibitem [{\citenamefont {Cassing}\ \emph {et~al.}(2013)\citenamefont
  {Cassing}, \citenamefont {Linnyk}, \citenamefont {Steinert},\ and\
  \citenamefont {Ozvenchuk}}]{Cassing:2013iz}%
  \BibitemOpen
  \bibfield  {author} {\bibinfo {author} {\bibfnamefont {W.}~\bibnamefont
  {Cassing}}, \bibinfo {author} {\bibfnamefont {O.}~\bibnamefont {Linnyk}},
  \bibinfo {author} {\bibfnamefont {T.}~\bibnamefont {Steinert}}, \ and\
  \bibinfo {author} {\bibfnamefont {V.}~\bibnamefont {Ozvenchuk}},\ }\href
  {\doibase 10.1103/PhysRevLett.110.182301} {\bibfield  {journal} {\bibinfo
  {journal} {Phys. Rev. Lett.}\ }\textbf {\bibinfo {volume} {110}},\ \bibinfo
  {pages} {182301} (\bibinfo {year} {2013})},\ \Eprint
  {http://arxiv.org/abs/1302.0906} {arXiv:1302.0906 [hep-ph]} \BibitemShut
  {NoStop}%
\bibitem [{\citenamefont {Marty}\ \emph {et~al.}(2013)\citenamefont {Marty},
  \citenamefont {Bratkovskaya}, \citenamefont {Cassing}, \citenamefont
  {Aichelin},\ and\ \citenamefont {Berrehrah}}]{Marty:2013ita}%
  \BibitemOpen
  \bibfield  {author} {\bibinfo {author} {\bibfnamefont {R.}~\bibnamefont
  {Marty}}, \bibinfo {author} {\bibfnamefont {E.}~\bibnamefont {Bratkovskaya}},
  \bibinfo {author} {\bibfnamefont {W.}~\bibnamefont {Cassing}}, \bibinfo
  {author} {\bibfnamefont {J.}~\bibnamefont {Aichelin}}, \ and\ \bibinfo
  {author} {\bibfnamefont {H.}~\bibnamefont {Berrehrah}},\ }\href {\doibase
  10.1103/PhysRevC.88.045204} {\bibfield  {journal} {\bibinfo  {journal} {Phys.
  Rev. C}\ }\textbf {\bibinfo {volume} {88}},\ \bibinfo {pages} {045204}
  (\bibinfo {year} {2013})},\ \Eprint {http://arxiv.org/abs/1305.7180}
  {arXiv:1305.7180 [hep-ph]} \BibitemShut {NoStop}%
\bibitem [{\citenamefont {Mitra}\ and\ \citenamefont
  {Chandra}(2016)}]{Mitra:2016zdw}%
  \BibitemOpen
  \bibfield  {author} {\bibinfo {author} {\bibfnamefont {S.}~\bibnamefont
  {Mitra}}\ and\ \bibinfo {author} {\bibfnamefont {V.}~\bibnamefont
  {Chandra}},\ }\href {\doibase 10.1103/PhysRevD.94.034025} {\bibfield
  {journal} {\bibinfo  {journal} {Phys. Rev. D}\ }\textbf {\bibinfo {volume}
  {94}},\ \bibinfo {pages} {034025} (\bibinfo {year} {2016})},\ \Eprint
  {http://arxiv.org/abs/1606.08556} {arXiv:1606.08556 [nucl-th]} \BibitemShut
  {NoStop}%
\bibitem [{\citenamefont {Policastro}\ \emph {et~al.}(2002)\citenamefont
  {Policastro}, \citenamefont {Son},\ and\ \citenamefont
  {Starinets}}]{Policastro:2002se}%
  \BibitemOpen
  \bibfield  {author} {\bibinfo {author} {\bibfnamefont {G.}~\bibnamefont
  {Policastro}}, \bibinfo {author} {\bibfnamefont {D.~T.}\ \bibnamefont {Son}},
  \ and\ \bibinfo {author} {\bibfnamefont {A.~O.}\ \bibnamefont {Starinets}},\
  }\href {\doibase 10.1088/1126-6708/2002/09/043} {\bibfield  {journal}
  {\bibinfo  {journal} {JHEP}\ }\textbf {\bibinfo {volume} {09}},\ \bibinfo
  {pages} {043} (\bibinfo {year} {2002})},\ \Eprint
  {http://arxiv.org/abs/hep-th/0205052} {arXiv:hep-th/0205052} \BibitemShut
  {NoStop}%
\bibitem [{\citenamefont {Teaney}(2006)}]{Teaney:2006nc}%
  \BibitemOpen
  \bibfield  {author} {\bibinfo {author} {\bibfnamefont {D.}~\bibnamefont
  {Teaney}},\ }\href {\doibase 10.1103/PhysRevD.74.045025} {\bibfield
  {journal} {\bibinfo  {journal} {Phys. Rev. D}\ }\textbf {\bibinfo {volume}
  {74}},\ \bibinfo {pages} {045025} (\bibinfo {year} {2006})},\ \Eprint
  {http://arxiv.org/abs/hep-ph/0602044} {arXiv:hep-ph/0602044} \BibitemShut
  {NoStop}%
\bibitem [{\citenamefont {Gupta}(2004)}]{Gupta:2003zh}%
  \BibitemOpen
  \bibfield  {author} {\bibinfo {author} {\bibfnamefont {S.}~\bibnamefont
  {Gupta}},\ }\href {\doibase 10.1016/j.physletb.2004.05.079} {\bibfield
  {journal} {\bibinfo  {journal} {Phys. Lett. B}\ }\textbf {\bibinfo {volume}
  {597}},\ \bibinfo {pages} {57} (\bibinfo {year} {2004})},\ \Eprint
  {http://arxiv.org/abs/hep-lat/0301006} {arXiv:hep-lat/0301006} \BibitemShut
  {NoStop}%
\bibitem [{\citenamefont {Aarts}\ \emph {et~al.}(2007)\citenamefont {Aarts},
  \citenamefont {Allton}, \citenamefont {Foley}, \citenamefont {Hands},\ and\
  \citenamefont {Kim}}]{Aarts:2007wj}%
  \BibitemOpen
  \bibfield  {author} {\bibinfo {author} {\bibfnamefont {G.}~\bibnamefont
  {Aarts}}, \bibinfo {author} {\bibfnamefont {C.}~\bibnamefont {Allton}},
  \bibinfo {author} {\bibfnamefont {J.}~\bibnamefont {Foley}}, \bibinfo
  {author} {\bibfnamefont {S.}~\bibnamefont {Hands}}, \ and\ \bibinfo {author}
  {\bibfnamefont {S.}~\bibnamefont {Kim}},\ }\href {\doibase
  10.1103/PhysRevLett.99.022002} {\bibfield  {journal} {\bibinfo  {journal}
  {Phys. Rev. Lett.}\ }\textbf {\bibinfo {volume} {99}},\ \bibinfo {pages}
  {022002} (\bibinfo {year} {2007})},\ \Eprint
  {http://arxiv.org/abs/hep-lat/0703008} {arXiv:hep-lat/0703008} \BibitemShut
  {NoStop}%
\bibitem [{\citenamefont {Ding}\ \emph {et~al.}(2011)\citenamefont {Ding},
  \citenamefont {Francis}, \citenamefont {Kaczmarek}, \citenamefont {Karsch},
  \citenamefont {Laermann},\ and\ \citenamefont {Soeldner}}]{Ding:2010ga}%
  \BibitemOpen
  \bibfield  {author} {\bibinfo {author} {\bibfnamefont {H.~T.}\ \bibnamefont
  {Ding}}, \bibinfo {author} {\bibfnamefont {A.}~\bibnamefont {Francis}},
  \bibinfo {author} {\bibfnamefont {O.}~\bibnamefont {Kaczmarek}}, \bibinfo
  {author} {\bibfnamefont {F.}~\bibnamefont {Karsch}}, \bibinfo {author}
  {\bibfnamefont {E.}~\bibnamefont {Laermann}}, \ and\ \bibinfo {author}
  {\bibfnamefont {W.}~\bibnamefont {Soeldner}},\ }\href {\doibase
  10.1103/PhysRevD.83.034504} {\bibfield  {journal} {\bibinfo  {journal} {Phys.
  Rev. D}\ }\textbf {\bibinfo {volume} {83}},\ \bibinfo {pages} {034504}
  (\bibinfo {year} {2011})},\ \Eprint {http://arxiv.org/abs/1012.4963}
  {arXiv:1012.4963 [hep-lat]} \BibitemShut {NoStop}%
\bibitem [{\citenamefont {Francis}\ and\ \citenamefont
  {Kaczmarek}(2012)}]{Francis:2011bt}%
  \BibitemOpen
  \bibfield  {author} {\bibinfo {author} {\bibfnamefont {A.}~\bibnamefont
  {Francis}}\ and\ \bibinfo {author} {\bibfnamefont {O.}~\bibnamefont
  {Kaczmarek}},\ }\href {\doibase 10.1016/j.ppnp.2011.12.020} {\bibfield
  {journal} {\bibinfo  {journal} {Prog. Part. Nucl. Phys.}\ }\textbf {\bibinfo
  {volume} {67}},\ \bibinfo {pages} {212} (\bibinfo {year} {2012})},\ \Eprint
  {http://arxiv.org/abs/1112.4802} {arXiv:1112.4802 [hep-lat]} \BibitemShut
  {NoStop}%
\bibitem [{\citenamefont {Amato}\ \emph {et~al.}(2013)\citenamefont {Amato},
  \citenamefont {Aarts}, \citenamefont {Allton}, \citenamefont {Giudice},
  \citenamefont {Hands},\ and\ \citenamefont {Skullerud}}]{Amato:2013naa}%
  \BibitemOpen
  \bibfield  {author} {\bibinfo {author} {\bibfnamefont {A.}~\bibnamefont
  {Amato}}, \bibinfo {author} {\bibfnamefont {G.}~\bibnamefont {Aarts}},
  \bibinfo {author} {\bibfnamefont {C.}~\bibnamefont {Allton}}, \bibinfo
  {author} {\bibfnamefont {P.}~\bibnamefont {Giudice}}, \bibinfo {author}
  {\bibfnamefont {S.}~\bibnamefont {Hands}}, \ and\ \bibinfo {author}
  {\bibfnamefont {J.-I.}\ \bibnamefont {Skullerud}},\ }\href {\doibase
  10.1103/PhysRevLett.111.172001} {\bibfield  {journal} {\bibinfo  {journal}
  {Phys. Rev. Lett.}\ }\textbf {\bibinfo {volume} {111}},\ \bibinfo {pages}
  {172001} (\bibinfo {year} {2013})},\ \Eprint {http://arxiv.org/abs/1307.6763}
  {arXiv:1307.6763 [hep-lat]} \BibitemShut {NoStop}%
\bibitem [{\citenamefont {Aarts}\ \emph {et~al.}(2015)\citenamefont {Aarts},
  \citenamefont {Allton}, \citenamefont {Amato}, \citenamefont {Giudice},
  \citenamefont {Hands},\ and\ \citenamefont {Skullerud}}]{Aarts:2014nba}%
  \BibitemOpen
  \bibfield  {author} {\bibinfo {author} {\bibfnamefont {G.}~\bibnamefont
  {Aarts}}, \bibinfo {author} {\bibfnamefont {C.}~\bibnamefont {Allton}},
  \bibinfo {author} {\bibfnamefont {A.}~\bibnamefont {Amato}}, \bibinfo
  {author} {\bibfnamefont {P.}~\bibnamefont {Giudice}}, \bibinfo {author}
  {\bibfnamefont {S.}~\bibnamefont {Hands}}, \ and\ \bibinfo {author}
  {\bibfnamefont {J.-I.}\ \bibnamefont {Skullerud}},\ }\href {\doibase
  10.1007/JHEP02(2015)186} {\bibfield  {journal} {\bibinfo  {journal} {JHEP}\
  }\textbf {\bibinfo {volume} {02}},\ \bibinfo {pages} {186} (\bibinfo {year}
  {2015})},\ \Eprint {http://arxiv.org/abs/1412.6411} {arXiv:1412.6411
  [hep-lat]} \BibitemShut {NoStop}%
\bibitem [{\citenamefont {Floerchinger}\ \emph {et~al.}(2023)\citenamefont
  {Floerchinger}, \citenamefont {Gebhardt},\ and\ \citenamefont
  {Reygers}}]{Floerchinger:2021xhb}%
  \BibitemOpen
  \bibfield  {author} {\bibinfo {author} {\bibfnamefont {S.}~\bibnamefont
  {Floerchinger}}, \bibinfo {author} {\bibfnamefont {C.}~\bibnamefont
  {Gebhardt}}, \ and\ \bibinfo {author} {\bibfnamefont {K.}~\bibnamefont
  {Reygers}},\ }\href {\doibase 10.1016/j.physletb.2022.137647} {\bibfield
  {journal} {\bibinfo  {journal} {Phys. Lett. B}\ }\textbf {\bibinfo {volume}
  {837}},\ \bibinfo {pages} {137647} (\bibinfo {year} {2023})},\ \Eprint
  {http://arxiv.org/abs/2112.12497} {arXiv:2112.12497 [nucl-th]} \BibitemShut
  {NoStop}%
\bibitem [{\citenamefont {Atchison}\ \emph {et~al.}(2024)\citenamefont
  {Atchison}, \citenamefont {Han},\ and\ \citenamefont
  {Geurts}}]{Atchison:2024lmf}%
  \BibitemOpen
  \bibfield  {author} {\bibinfo {author} {\bibfnamefont {J.}~\bibnamefont
  {Atchison}}, \bibinfo {author} {\bibfnamefont {Y.}~\bibnamefont {Han}}, \
  and\ \bibinfo {author} {\bibfnamefont {F.}~\bibnamefont {Geurts}},\
  }\href@noop {} {\  (\bibinfo {year} {2024})},\ \Eprint
  {http://arxiv.org/abs/2408.10176} {arXiv:2408.10176 [nucl-th]} \BibitemShut
  {NoStop}%
\bibitem [{\citenamefont {Kharzeev}\ \emph {et~al.}(2008)\citenamefont
  {Kharzeev}, \citenamefont {McLerran},\ and\ \citenamefont
  {Warringa}}]{Kharzeev:2007jp}%
  \BibitemOpen
  \bibfield  {author} {\bibinfo {author} {\bibfnamefont {D.~E.}\ \bibnamefont
  {Kharzeev}}, \bibinfo {author} {\bibfnamefont {L.~D.}\ \bibnamefont
  {McLerran}}, \ and\ \bibinfo {author} {\bibfnamefont {H.~J.}\ \bibnamefont
  {Warringa}},\ }\href {\doibase 10.1016/j.nuclphysa.2008.02.298} {\bibfield
  {journal} {\bibinfo  {journal} {Nucl. Phys.}\ }\textbf {\bibinfo {volume}
  {A803}},\ \bibinfo {pages} {227} (\bibinfo {year} {2008})},\ \Eprint
  {http://arxiv.org/abs/0711.0950} {arXiv:0711.0950 [hep-ph]} \BibitemShut
  {NoStop}%
\bibitem [{\citenamefont {Skokov}\ \emph {et~al.}(2009)\citenamefont {Skokov},
  \citenamefont {Illarionov},\ and\ \citenamefont {Toneev}}]{Skokov:2009qp}%
  \BibitemOpen
  \bibfield  {author} {\bibinfo {author} {\bibfnamefont {V.}~\bibnamefont
  {Skokov}}, \bibinfo {author} {\bibfnamefont {A.~{\relax Yu}.}\ \bibnamefont
  {Illarionov}}, \ and\ \bibinfo {author} {\bibfnamefont {V.}~\bibnamefont
  {Toneev}},\ }\href {\doibase 10.1142/S0217751X09047570} {\bibfield  {journal}
  {\bibinfo  {journal} {Int. J. Mod. Phys.}\ }\textbf {\bibinfo {volume}
  {A24}},\ \bibinfo {pages} {5925} (\bibinfo {year} {2009})},\ \Eprint
  {http://arxiv.org/abs/0907.1396} {arXiv:0907.1396 [nucl-th]} \BibitemShut
  {NoStop}%
\bibitem [{\citenamefont {Abdallah}\ \emph {et~al.}(2021)\citenamefont
  {Abdallah} \emph {et~al.}}]{STAR:2021mii}%
  \BibitemOpen
  \bibfield  {author} {\bibinfo {author} {\bibfnamefont {M.}~\bibnamefont
  {Abdallah}} \emph {et~al.} (\bibinfo {collaboration} {STAR}),\ }\href@noop {}
  {\  (\bibinfo {year} {2021})},\ \Eprint {http://arxiv.org/abs/2109.00131}
  {arXiv:2109.00131 [nucl-ex]} \BibitemShut {NoStop}%
\bibitem [{\citenamefont {An}\ \emph {et~al.}(2022)\citenamefont {An} \emph
  {et~al.}}]{An:2021wof}%
  \BibitemOpen
  \bibfield  {author} {\bibinfo {author} {\bibfnamefont {X.}~\bibnamefont {An}}
  \emph {et~al.},\ }\href {\doibase 10.1016/j.nuclphysa.2021.122343} {\bibfield
   {journal} {\bibinfo  {journal} {Nucl. Phys. A}\ }\textbf {\bibinfo {volume}
  {1017}},\ \bibinfo {pages} {122343} (\bibinfo {year} {2022})},\ \Eprint
  {http://arxiv.org/abs/2108.13867} {arXiv:2108.13867 [nucl-th]} \BibitemShut
  {NoStop}%
\bibitem [{\citenamefont {Abdulhamid}\ \emph {et~al.}(2023)\citenamefont
  {Abdulhamid} \emph {et~al.}}]{STAR:2022zpv}%
  \BibitemOpen
  \bibfield  {author} {\bibinfo {author} {\bibfnamefont {M.~I.}\ \bibnamefont
  {Abdulhamid}} \emph {et~al.} (\bibinfo {collaboration} {STAR}),\ }\href
  {\doibase 10.1103/PhysRevC.108.014908} {\bibfield  {journal} {\bibinfo
  {journal} {Phys. Rev. C}\ }\textbf {\bibinfo {volume} {108}},\ \bibinfo
  {pages} {014908} (\bibinfo {year} {2023})},\ \Eprint
  {http://arxiv.org/abs/2210.14027} {arXiv:2210.14027 [nucl-ex]} \BibitemShut
  {NoStop}%
\bibitem [{\citenamefont {Vilenkin}(1979)}]{Vilenkin:1979ui}%
  \BibitemOpen
  \bibfield  {author} {\bibinfo {author} {\bibfnamefont {A.}~\bibnamefont
  {Vilenkin}},\ }\href {\doibase 10.1103/PhysRevD.20.1807} {\bibfield
  {journal} {\bibinfo  {journal} {Phys. Rev. D}\ }\textbf {\bibinfo {volume}
  {20}},\ \bibinfo {pages} {1807} (\bibinfo {year} {1979})}\BibitemShut
  {NoStop}%
\bibitem [{\citenamefont {Vilenkin}(1980)}]{Vilenkin:1980fu}%
  \BibitemOpen
  \bibfield  {author} {\bibinfo {author} {\bibfnamefont {A.}~\bibnamefont
  {Vilenkin}},\ }\href {\doibase 10.1103/PhysRevD.22.3080} {\bibfield
  {journal} {\bibinfo  {journal} {Phys. Rev. D}\ }\textbf {\bibinfo {volume}
  {22}},\ \bibinfo {pages} {3080} (\bibinfo {year} {1980})}\BibitemShut
  {NoStop}%
\bibitem [{\citenamefont {Fukushima}\ \emph {et~al.}(2008)\citenamefont
  {Fukushima}, \citenamefont {Kharzeev},\ and\ \citenamefont
  {Warringa}}]{Fukushima:2008xe}%
  \BibitemOpen
  \bibfield  {author} {\bibinfo {author} {\bibfnamefont {K.}~\bibnamefont
  {Fukushima}}, \bibinfo {author} {\bibfnamefont {D.~E.}\ \bibnamefont
  {Kharzeev}}, \ and\ \bibinfo {author} {\bibfnamefont {H.~J.}\ \bibnamefont
  {Warringa}},\ }\href {\doibase 10.1103/PhysRevD.78.074033} {\bibfield
  {journal} {\bibinfo  {journal} {Phys. Rev.}\ }\textbf {\bibinfo {volume}
  {D78}},\ \bibinfo {pages} {074033} (\bibinfo {year} {2008})},\ \Eprint
  {http://arxiv.org/abs/0808.3382} {arXiv:0808.3382 [hep-ph]} \BibitemShut
  {NoStop}%
\bibitem [{\citenamefont {Son}\ and\ \citenamefont
  {Surowka}(2009)}]{Son:2009tf}%
  \BibitemOpen
  \bibfield  {author} {\bibinfo {author} {\bibfnamefont {D.~T.}\ \bibnamefont
  {Son}}\ and\ \bibinfo {author} {\bibfnamefont {P.}~\bibnamefont {Surowka}},\
  }\href {\doibase 10.1103/PhysRevLett.103.191601} {\bibfield  {journal}
  {\bibinfo  {journal} {Phys. Rev. Lett.}\ }\textbf {\bibinfo {volume} {103}},\
  \bibinfo {pages} {191601} (\bibinfo {year} {2009})},\ \Eprint
  {http://arxiv.org/abs/0906.5044} {arXiv:0906.5044 [hep-th]} \BibitemShut
  {NoStop}%
\bibitem [{\citenamefont {Akamatsu}\ and\ \citenamefont
  {Yamamoto}(2013)}]{Akamatsu:2013pjd}%
  \BibitemOpen
  \bibfield  {author} {\bibinfo {author} {\bibfnamefont {Y.}~\bibnamefont
  {Akamatsu}}\ and\ \bibinfo {author} {\bibfnamefont {N.}~\bibnamefont
  {Yamamoto}},\ }\href {\doibase 10.1103/PhysRevLett.111.052002} {\bibfield
  {journal} {\bibinfo  {journal} {Phys. Rev. Lett.}\ }\textbf {\bibinfo
  {volume} {111}},\ \bibinfo {pages} {052002} (\bibinfo {year} {2013})},\
  \Eprint {http://arxiv.org/abs/1302.2125} {arXiv:1302.2125 [nucl-th]}
  \BibitemShut {NoStop}%
\bibitem [{\citenamefont {Carignano}\ and\ \citenamefont
  {Manuel}(2019)}]{Carignano:2018thu}%
  \BibitemOpen
  \bibfield  {author} {\bibinfo {author} {\bibfnamefont {S.}~\bibnamefont
  {Carignano}}\ and\ \bibinfo {author} {\bibfnamefont {C.}~\bibnamefont
  {Manuel}},\ }\href {\doibase 10.1103/PhysRevD.99.096022} {\bibfield
  {journal} {\bibinfo  {journal} {Phys. Rev. D}\ }\textbf {\bibinfo {volume}
  {99}},\ \bibinfo {pages} {096022} (\bibinfo {year} {2019})},\ \Eprint
  {http://arxiv.org/abs/1811.06394} {arXiv:1811.06394 [hep-ph]} \BibitemShut
  {NoStop}%
\bibitem [{\citenamefont {Carignano}\ and\ \citenamefont
  {Manuel}(2021)}]{Carignano:2021mrn}%
  \BibitemOpen
  \bibfield  {author} {\bibinfo {author} {\bibfnamefont {S.}~\bibnamefont
  {Carignano}}\ and\ \bibinfo {author} {\bibfnamefont {C.}~\bibnamefont
  {Manuel}},\ }\href {\doibase 10.1103/PhysRevD.103.116002} {\bibfield
  {journal} {\bibinfo  {journal} {Phys. Rev. D}\ }\textbf {\bibinfo {volume}
  {103}},\ \bibinfo {pages} {116002} (\bibinfo {year} {2021})},\ \Eprint
  {http://arxiv.org/abs/2103.02491} {arXiv:2103.02491 [hep-ph]} \BibitemShut
  {NoStop}%
\bibitem [{\citenamefont {Carignano}\ and\ \citenamefont
  {Buballa}(2020)}]{Carignano:2019ivp}%
  \BibitemOpen
  \bibfield  {author} {\bibinfo {author} {\bibfnamefont {S.}~\bibnamefont
  {Carignano}}\ and\ \bibinfo {author} {\bibfnamefont {M.}~\bibnamefont
  {Buballa}},\ }\href {\doibase 10.1103/PhysRevD.101.014026} {\bibfield
  {journal} {\bibinfo  {journal} {Phys. Rev. D}\ }\textbf {\bibinfo {volume}
  {101}},\ \bibinfo {pages} {014026} (\bibinfo {year} {2020})},\ \Eprint
  {http://arxiv.org/abs/1910.03604} {arXiv:1910.03604 [hep-ph]} \BibitemShut
  {NoStop}%
\bibitem [{\citenamefont {Ruggieri}\ and\ \citenamefont
  {Peng}(2016{\natexlab{a}})}]{Ruggieri:2016lrn}%
  \BibitemOpen
  \bibfield  {author} {\bibinfo {author} {\bibfnamefont {M.}~\bibnamefont
  {Ruggieri}}\ and\ \bibinfo {author} {\bibfnamefont {G.~X.}\ \bibnamefont
  {Peng}},\ }\href {\doibase 10.1103/PhysRevD.93.094021} {\bibfield  {journal}
  {\bibinfo  {journal} {Phys. Rev. D}\ }\textbf {\bibinfo {volume} {93}},\
  \bibinfo {pages} {094021} (\bibinfo {year} {2016}{\natexlab{a}})},\ \Eprint
  {http://arxiv.org/abs/1602.08994} {arXiv:1602.08994 [hep-ph]} \BibitemShut
  {NoStop}%
\bibitem [{\citenamefont {Ruggieri}\ \emph {et~al.}(2016)\citenamefont
  {Ruggieri}, \citenamefont {Peng},\ and\ \citenamefont
  {Chernodub}}]{Ruggieri:2016asg}%
  \BibitemOpen
  \bibfield  {author} {\bibinfo {author} {\bibfnamefont {M.}~\bibnamefont
  {Ruggieri}}, \bibinfo {author} {\bibfnamefont {G.~X.}\ \bibnamefont {Peng}},
  \ and\ \bibinfo {author} {\bibfnamefont {M.}~\bibnamefont {Chernodub}},\
  }\href {\doibase 10.1103/PhysRevD.94.054011} {\bibfield  {journal} {\bibinfo
  {journal} {Phys. Rev. D}\ }\textbf {\bibinfo {volume} {94}},\ \bibinfo
  {pages} {054011} (\bibinfo {year} {2016})},\ \Eprint
  {http://arxiv.org/abs/1606.03287} {arXiv:1606.03287 [hep-ph]} \BibitemShut
  {NoStop}%
\bibitem [{\citenamefont {Ruggieri}\ \emph {et~al.}(2020)\citenamefont
  {Ruggieri}, \citenamefont {Chernodub},\ and\ \citenamefont
  {Lu}}]{Ruggieri:2020qtq}%
  \BibitemOpen
  \bibfield  {author} {\bibinfo {author} {\bibfnamefont {M.}~\bibnamefont
  {Ruggieri}}, \bibinfo {author} {\bibfnamefont {M.~N.}\ \bibnamefont
  {Chernodub}}, \ and\ \bibinfo {author} {\bibfnamefont {Z.-Y.}\ \bibnamefont
  {Lu}},\ }\href {\doibase 10.1103/PhysRevD.102.014031} {\bibfield  {journal}
  {\bibinfo  {journal} {Phys. Rev. D}\ }\textbf {\bibinfo {volume} {102}},\
  \bibinfo {pages} {014031} (\bibinfo {year} {2020})},\ \Eprint
  {http://arxiv.org/abs/2004.09393} {arXiv:2004.09393 [hep-ph]} \BibitemShut
  {NoStop}%
\bibitem [{\citenamefont {Chaudhuri}\ \emph {et~al.}(2021)\citenamefont
  {Chaudhuri}, \citenamefont {Mukherjee}, \citenamefont {Ghosh}, \citenamefont
  {Sarkar},\ and\ \citenamefont {Roy}}]{Chaudhuri:2021lui}%
  \BibitemOpen
  \bibfield  {author} {\bibinfo {author} {\bibfnamefont {N.}~\bibnamefont
  {Chaudhuri}}, \bibinfo {author} {\bibfnamefont {A.}~\bibnamefont
  {Mukherjee}}, \bibinfo {author} {\bibfnamefont {S.}~\bibnamefont {Ghosh}},
  \bibinfo {author} {\bibfnamefont {S.}~\bibnamefont {Sarkar}}, \ and\ \bibinfo
  {author} {\bibfnamefont {P.}~\bibnamefont {Roy}},\ }\href@noop {} {\
  (\bibinfo {year} {2021})},\ \Eprint {http://arxiv.org/abs/2111.12058}
  {arXiv:2111.12058 [hep-ph]} \BibitemShut {NoStop}%
\bibitem [{\citenamefont {Fukushima}\ \emph {et~al.}(2010)\citenamefont
  {Fukushima}, \citenamefont {Ruggieri},\ and\ \citenamefont
  {Gatto}}]{Fukushima:2010fe}%
  \BibitemOpen
  \bibfield  {author} {\bibinfo {author} {\bibfnamefont {K.}~\bibnamefont
  {Fukushima}}, \bibinfo {author} {\bibfnamefont {M.}~\bibnamefont {Ruggieri}},
  \ and\ \bibinfo {author} {\bibfnamefont {R.}~\bibnamefont {Gatto}},\ }\href
  {\doibase 10.1103/PhysRevD.81.114031} {\bibfield  {journal} {\bibinfo
  {journal} {Phys. Rev. D}\ }\textbf {\bibinfo {volume} {81}},\ \bibinfo
  {pages} {114031} (\bibinfo {year} {2010})},\ \Eprint
  {http://arxiv.org/abs/1003.0047} {arXiv:1003.0047 [hep-ph]} \BibitemShut
  {NoStop}%
\bibitem [{\citenamefont {Andrianov}\ \emph {et~al.}(2014)\citenamefont
  {Andrianov}, \citenamefont {Espriu},\ and\ \citenamefont
  {Planells}}]{Andrianov:2013dta}%
  \BibitemOpen
  \bibfield  {author} {\bibinfo {author} {\bibfnamefont {A.~A.}\ \bibnamefont
  {Andrianov}}, \bibinfo {author} {\bibfnamefont {D.}~\bibnamefont {Espriu}}, \
  and\ \bibinfo {author} {\bibfnamefont {X.}~\bibnamefont {Planells}},\ }\href
  {\doibase 10.1140/epjc/s10052-014-2776-8} {\bibfield  {journal} {\bibinfo
  {journal} {Eur. Phys. J. C}\ }\textbf {\bibinfo {volume} {74}},\ \bibinfo
  {pages} {2776} (\bibinfo {year} {2014})},\ \Eprint
  {http://arxiv.org/abs/1310.4416} {arXiv:1310.4416 [hep-ph]} \BibitemShut
  {NoStop}%
\bibitem [{\citenamefont {Braguta}\ and\ \citenamefont
  {Kotov}(2016)}]{Braguta:2016aov}%
  \BibitemOpen
  \bibfield  {author} {\bibinfo {author} {\bibfnamefont {V.~V.}\ \bibnamefont
  {Braguta}}\ and\ \bibinfo {author} {\bibfnamefont {A.~Y.}\ \bibnamefont
  {Kotov}},\ }\href {\doibase 10.1103/PhysRevD.93.105025} {\bibfield  {journal}
  {\bibinfo  {journal} {Phys. Rev. D}\ }\textbf {\bibinfo {volume} {93}},\
  \bibinfo {pages} {105025} (\bibinfo {year} {2016})},\ \Eprint
  {http://arxiv.org/abs/1601.04957} {arXiv:1601.04957 [hep-th]} \BibitemShut
  {NoStop}%
\bibitem [{\citenamefont {Azeredo}\ \emph {et~al.}(2024)\citenamefont
  {Azeredo}, \citenamefont {Duarte}, \citenamefont {Farias}, \citenamefont
  {Krein},\ and\ \citenamefont {O.~Ramos}}]{Azeredo:2024sqc}%
  \BibitemOpen
  \bibfield  {author} {\bibinfo {author} {\bibfnamefont {F.~X.}\ \bibnamefont
  {Azeredo}}, \bibinfo {author} {\bibfnamefont {D.~C.}\ \bibnamefont {Duarte}},
  \bibinfo {author} {\bibfnamefont {R.~L.~S.}\ \bibnamefont {Farias}}, \bibinfo
  {author} {\bibfnamefont {G.~a.}\ \bibnamefont {Krein}}, \ and\ \bibinfo
  {author} {\bibfnamefont {R.}~\bibnamefont {O.~Ramos}},\ }\href@noop {} {\
  (\bibinfo {year} {2024})},\ \Eprint {http://arxiv.org/abs/2406.04900}
  {arXiv:2406.04900 [hep-ph]} \BibitemShut {NoStop}%
\bibitem [{\citenamefont {Espriu}\ \emph {et~al.}(2020)\citenamefont {Espriu},
  \citenamefont {G\'omez~Nicola},\ and\ \citenamefont
  {Vioque-Rodr\'\i{}guez}}]{Espriu:2020dge}%
  \BibitemOpen
  \bibfield  {author} {\bibinfo {author} {\bibfnamefont {D.}~\bibnamefont
  {Espriu}}, \bibinfo {author} {\bibfnamefont {A.}~\bibnamefont
  {G\'omez~Nicola}}, \ and\ \bibinfo {author} {\bibfnamefont {A.}~\bibnamefont
  {Vioque-Rodr\'\i{}guez}},\ }\href {\doibase 10.1007/JHEP06(2020)062}
  {\bibfield  {journal} {\bibinfo  {journal} {JHEP}\ }\textbf {\bibinfo
  {volume} {06}},\ \bibinfo {pages} {062} (\bibinfo {year} {2020})},\ \Eprint
  {http://arxiv.org/abs/2002.11696} {arXiv:2002.11696 [hep-ph]} \BibitemShut
  {NoStop}%
\bibitem [{\citenamefont {Braguta}\ \emph {et~al.}(2015)\citenamefont
  {Braguta}, \citenamefont {Goy}, \citenamefont {Ilgenfritz}, \citenamefont
  {Kotov}, \citenamefont {Molochkov}, \citenamefont {Muller-Preussker},\ and\
  \citenamefont {Petersson}}]{Braguta:2015zta}%
  \BibitemOpen
  \bibfield  {author} {\bibinfo {author} {\bibfnamefont {V.~V.}\ \bibnamefont
  {Braguta}}, \bibinfo {author} {\bibfnamefont {V.~A.}\ \bibnamefont {Goy}},
  \bibinfo {author} {\bibfnamefont {E.~M.}\ \bibnamefont {Ilgenfritz}},
  \bibinfo {author} {\bibfnamefont {A.~Y.}\ \bibnamefont {Kotov}}, \bibinfo
  {author} {\bibfnamefont {A.~V.}\ \bibnamefont {Molochkov}}, \bibinfo {author}
  {\bibfnamefont {M.}~\bibnamefont {Muller-Preussker}}, \ and\ \bibinfo
  {author} {\bibfnamefont {B.}~\bibnamefont {Petersson}},\ }\href {\doibase
  10.1007/JHEP06(2015)094} {\bibfield  {journal} {\bibinfo  {journal} {JHEP}\
  }\textbf {\bibinfo {volume} {06}},\ \bibinfo {pages} {094} (\bibinfo {year}
  {2015})},\ \Eprint {http://arxiv.org/abs/1503.06670} {arXiv:1503.06670
  [hep-lat]} \BibitemShut {NoStop}%
\bibitem [{\citenamefont {Braguta}\ \emph {et~al.}(2016)\citenamefont
  {Braguta}, \citenamefont {Ilgenfritz}, \citenamefont {Kotov}, \citenamefont
  {Petersson},\ and\ \citenamefont {Skinderev}}]{Braguta:2015owi}%
  \BibitemOpen
  \bibfield  {author} {\bibinfo {author} {\bibfnamefont {V.~V.}\ \bibnamefont
  {Braguta}}, \bibinfo {author} {\bibfnamefont {E.~M.}\ \bibnamefont
  {Ilgenfritz}}, \bibinfo {author} {\bibfnamefont {A.~Y.}\ \bibnamefont
  {Kotov}}, \bibinfo {author} {\bibfnamefont {B.}~\bibnamefont {Petersson}}, \
  and\ \bibinfo {author} {\bibfnamefont {S.~A.}\ \bibnamefont {Skinderev}},\
  }\href {\doibase 10.1103/PhysRevD.93.034509} {\bibfield  {journal} {\bibinfo
  {journal} {Phys. Rev. D}\ }\textbf {\bibinfo {volume} {93}},\ \bibinfo
  {pages} {034509} (\bibinfo {year} {2016})},\ \Eprint
  {http://arxiv.org/abs/1512.05873} {arXiv:1512.05873 [hep-lat]} \BibitemShut
  {NoStop}%
\bibitem [{\citenamefont {Ghosh}\ \emph {et~al.}(2022)\citenamefont {Ghosh},
  \citenamefont {Chaudhuri}, \citenamefont {Sarkar},\ and\ \citenamefont
  {Roy}}]{Ghosh:2022xbf}%
  \BibitemOpen
  \bibfield  {author} {\bibinfo {author} {\bibfnamefont {S.}~\bibnamefont
  {Ghosh}}, \bibinfo {author} {\bibfnamefont {N.}~\bibnamefont {Chaudhuri}},
  \bibinfo {author} {\bibfnamefont {S.}~\bibnamefont {Sarkar}}, \ and\ \bibinfo
  {author} {\bibfnamefont {P.}~\bibnamefont {Roy}},\ }\href {\doibase
  10.1103/PhysRevD.105.096005} {\bibfield  {journal} {\bibinfo  {journal}
  {Phys. Rev. D}\ }\textbf {\bibinfo {volume} {105}},\ \bibinfo {pages}
  {096005} (\bibinfo {year} {2022})},\ \Eprint
  {http://arxiv.org/abs/2201.06473} {arXiv:2201.06473 [hep-ph]} \BibitemShut
  {NoStop}%
\bibitem [{\citenamefont {Chaudhuri}\ \emph {et~al.}(2022)\citenamefont
  {Chaudhuri}, \citenamefont {Ghosh}, \citenamefont {Sarkar},\ and\
  \citenamefont {Roy}}]{Chaudhuri:2022rwo}%
  \BibitemOpen
  \bibfield  {author} {\bibinfo {author} {\bibfnamefont {N.}~\bibnamefont
  {Chaudhuri}}, \bibinfo {author} {\bibfnamefont {S.}~\bibnamefont {Ghosh}},
  \bibinfo {author} {\bibfnamefont {S.}~\bibnamefont {Sarkar}}, \ and\ \bibinfo
  {author} {\bibfnamefont {P.}~\bibnamefont {Roy}},\ }\href {\doibase
  10.1103/PhysRevD.105.096001} {\bibfield  {journal} {\bibinfo  {journal}
  {Phys. Rev. D}\ }\textbf {\bibinfo {volume} {105}},\ \bibinfo {pages}
  {096001} (\bibinfo {year} {2022})},\ \Eprint
  {http://arxiv.org/abs/2204.00892} {arXiv:2204.00892 [hep-ph]} \BibitemShut
  {NoStop}%
\bibitem [{\citenamefont {Andrianov}\ \emph {et~al.}(2013)\citenamefont
  {Andrianov}, \citenamefont {Espriu},\ and\ \citenamefont
  {Planells}}]{Andrianov:2012dj}%
  \BibitemOpen
  \bibfield  {author} {\bibinfo {author} {\bibfnamefont {A.~A.}\ \bibnamefont
  {Andrianov}}, \bibinfo {author} {\bibfnamefont {D.}~\bibnamefont {Espriu}}, \
  and\ \bibinfo {author} {\bibfnamefont {X.}~\bibnamefont {Planells}},\ }\href
  {\doibase 10.1140/epjc/s10052-013-2294-0} {\bibfield  {journal} {\bibinfo
  {journal} {Eur. Phys. J. C}\ }\textbf {\bibinfo {volume} {73}},\ \bibinfo
  {pages} {2294} (\bibinfo {year} {2013})},\ \Eprint
  {http://arxiv.org/abs/1210.7712} {arXiv:1210.7712 [hep-ph]} \BibitemShut
  {NoStop}%
\bibitem [{\citenamefont {G\'omez~Nicola}\ \emph {et~al.}(2024)\citenamefont
  {G\'omez~Nicola}, \citenamefont {Roa-Bravo},\ and\ \citenamefont
  {Vioque-Rodr\'\i{}guez}}]{GomezNicola:2023ghi}%
  \BibitemOpen
  \bibfield  {author} {\bibinfo {author} {\bibfnamefont {A.}~\bibnamefont
  {G\'omez~Nicola}}, \bibinfo {author} {\bibfnamefont {P.}~\bibnamefont
  {Roa-Bravo}}, \ and\ \bibinfo {author} {\bibfnamefont {A.}~\bibnamefont
  {Vioque-Rodr\'\i{}guez}},\ }\href {\doibase 10.1103/PhysRevD.109.034011}
  {\bibfield  {journal} {\bibinfo  {journal} {Phys. Rev. D}\ }\textbf {\bibinfo
  {volume} {109}},\ \bibinfo {pages} {034011} (\bibinfo {year} {2024})},\
  \Eprint {http://arxiv.org/abs/2311.01917} {arXiv:2311.01917 [hep-ph]}
  \BibitemShut {NoStop}%
\bibitem [{\citenamefont {Ghosh}\ \emph {et~al.}(2023)\citenamefont {Ghosh},
  \citenamefont {Chaudhuri}, \citenamefont {Sarkar},\ and\ \citenamefont
  {Roy}}]{Ghosh:2023rft}%
  \BibitemOpen
  \bibfield  {author} {\bibinfo {author} {\bibfnamefont {S.}~\bibnamefont
  {Ghosh}}, \bibinfo {author} {\bibfnamefont {N.}~\bibnamefont {Chaudhuri}},
  \bibinfo {author} {\bibfnamefont {S.}~\bibnamefont {Sarkar}}, \ and\ \bibinfo
  {author} {\bibfnamefont {P.}~\bibnamefont {Roy}},\ }\href@noop {} {\
  (\bibinfo {year} {2023})},\ \Eprint {http://arxiv.org/abs/2312.08652}
  {arXiv:2312.08652 [hep-ph]} \BibitemShut {NoStop}%
\bibitem [{\citenamefont {Gorbar}\ and\ \citenamefont
  {Shovkovy}(2021)}]{Gorbar:2021tnw}%
  \BibitemOpen
  \bibfield  {author} {\bibinfo {author} {\bibfnamefont {E.~V.}\ \bibnamefont
  {Gorbar}}\ and\ \bibinfo {author} {\bibfnamefont {I.~A.}\ \bibnamefont
  {Shovkovy}},\ }\href@noop {} {\  (\bibinfo {year} {2021})},\ \Eprint
  {http://arxiv.org/abs/2110.11380} {arXiv:2110.11380 [astro-ph.HE]}
  \BibitemShut {NoStop}%
\bibitem [{\citenamefont {Charbonneau}\ and\ \citenamefont
  {Zhitnitsky}(2010)}]{Charbonneau:2009ax}%
  \BibitemOpen
  \bibfield  {author} {\bibinfo {author} {\bibfnamefont {J.}~\bibnamefont
  {Charbonneau}}\ and\ \bibinfo {author} {\bibfnamefont {A.}~\bibnamefont
  {Zhitnitsky}},\ }\href {\doibase 10.1088/1475-7516/2010/08/010} {\bibfield
  {journal} {\bibinfo  {journal} {JCAP}\ }\textbf {\bibinfo {volume} {08}},\
  \bibinfo {pages} {010} (\bibinfo {year} {2010})},\ \Eprint
  {http://arxiv.org/abs/0903.4450} {arXiv:0903.4450 [astro-ph.HE]} \BibitemShut
  {NoStop}%
\bibitem [{\citenamefont {Yamamoto}(2016)}]{Yamamoto:2015gzz}%
  \BibitemOpen
  \bibfield  {author} {\bibinfo {author} {\bibfnamefont {N.}~\bibnamefont
  {Yamamoto}},\ }\href {\doibase 10.1103/PhysRevD.93.065017} {\bibfield
  {journal} {\bibinfo  {journal} {Phys. Rev. D}\ }\textbf {\bibinfo {volume}
  {93}},\ \bibinfo {pages} {065017} (\bibinfo {year} {2016})},\ \Eprint
  {http://arxiv.org/abs/1511.00933} {arXiv:1511.00933 [astro-ph.HE]}
  \BibitemShut {NoStop}%
\bibitem [{\citenamefont {Shovkovy}(2021)}]{Shovkovy:2021yyw}%
  \BibitemOpen
  \bibfield  {author} {\bibinfo {author} {\bibfnamefont {I.~A.}\ \bibnamefont
  {Shovkovy}},\ }\href@noop {} {\  (\bibinfo {year} {2021})},\ \Eprint
  {http://arxiv.org/abs/2111.11416} {arXiv:2111.11416 [nucl-th]} \BibitemShut
  {NoStop}%
\bibitem [{\citenamefont {Li}\ \emph {et~al.}(2016)\citenamefont {Li},
  \citenamefont {Kharzeev}, \citenamefont {Zhang}, \citenamefont {Huang},
  \citenamefont {Pletikosic}, \citenamefont {Fedorov}, \citenamefont {Zhong},
  \citenamefont {Schneeloch}, \citenamefont {Gu},\ and\ \citenamefont
  {Valla}}]{Li:2014bha}%
  \BibitemOpen
  \bibfield  {author} {\bibinfo {author} {\bibfnamefont {Q.}~\bibnamefont
  {Li}}, \bibinfo {author} {\bibfnamefont {D.~E.}\ \bibnamefont {Kharzeev}},
  \bibinfo {author} {\bibfnamefont {C.}~\bibnamefont {Zhang}}, \bibinfo
  {author} {\bibfnamefont {Y.}~\bibnamefont {Huang}}, \bibinfo {author}
  {\bibfnamefont {I.}~\bibnamefont {Pletikosic}}, \bibinfo {author}
  {\bibfnamefont {A.~V.}\ \bibnamefont {Fedorov}}, \bibinfo {author}
  {\bibfnamefont {R.~D.}\ \bibnamefont {Zhong}}, \bibinfo {author}
  {\bibfnamefont {J.~A.}\ \bibnamefont {Schneeloch}}, \bibinfo {author}
  {\bibfnamefont {G.~D.}\ \bibnamefont {Gu}}, \ and\ \bibinfo {author}
  {\bibfnamefont {T.}~\bibnamefont {Valla}},\ }\href {\doibase
  10.1038/nphys3648} {\bibfield  {journal} {\bibinfo  {journal} {Nature Phys.}\
  }\textbf {\bibinfo {volume} {12}},\ \bibinfo {pages} {550} (\bibinfo {year}
  {2016})},\ \Eprint {http://arxiv.org/abs/1412.6543} {arXiv:1412.6543
  [cond-mat.str-el]} \BibitemShut {NoStop}%
\bibitem [{\citenamefont {de~Forcrand}\ and\ \citenamefont
  {Philipsen}(2007)}]{deForcrand:2006pv}%
  \BibitemOpen
  \bibfield  {author} {\bibinfo {author} {\bibfnamefont {P.}~\bibnamefont
  {de~Forcrand}}\ and\ \bibinfo {author} {\bibfnamefont {O.}~\bibnamefont
  {Philipsen}},\ }\href {\doibase 10.1088/1126-6708/2007/01/077} {\bibfield
  {journal} {\bibinfo  {journal} {JHEP}\ }\textbf {\bibinfo {volume} {01}},\
  \bibinfo {pages} {077} (\bibinfo {year} {2007})},\ \Eprint
  {http://arxiv.org/abs/hep-lat/0607017} {arXiv:hep-lat/0607017} \BibitemShut
  {NoStop}%
\bibitem [{\citenamefont {Aoki}\ \emph {et~al.}(2006)\citenamefont {Aoki},
  \citenamefont {Fodor}, \citenamefont {Katz},\ and\ \citenamefont
  {Szabo}}]{Aoki:2006br}%
  \BibitemOpen
  \bibfield  {author} {\bibinfo {author} {\bibfnamefont {Y.}~\bibnamefont
  {Aoki}}, \bibinfo {author} {\bibfnamefont {Z.}~\bibnamefont {Fodor}},
  \bibinfo {author} {\bibfnamefont {S.~D.}\ \bibnamefont {Katz}}, \ and\
  \bibinfo {author} {\bibfnamefont {K.~K.}\ \bibnamefont {Szabo}},\ }\href
  {\doibase 10.1016/j.physletb.2006.10.021} {\bibfield  {journal} {\bibinfo
  {journal} {Phys. Lett. B}\ }\textbf {\bibinfo {volume} {643}},\ \bibinfo
  {pages} {46} (\bibinfo {year} {2006})},\ \Eprint
  {http://arxiv.org/abs/hep-lat/0609068} {arXiv:hep-lat/0609068} \BibitemShut
  {NoStop}%
\bibitem [{\citenamefont {Aoki}\ \emph {et~al.}(2009)\citenamefont {Aoki},
  \citenamefont {Borsanyi}, \citenamefont {Durr}, \citenamefont {Fodor},
  \citenamefont {Katz}, \citenamefont {Krieg},\ and\ \citenamefont
  {Szabo}}]{Aoki:2009sc}%
  \BibitemOpen
  \bibfield  {author} {\bibinfo {author} {\bibfnamefont {Y.}~\bibnamefont
  {Aoki}}, \bibinfo {author} {\bibfnamefont {S.}~\bibnamefont {Borsanyi}},
  \bibinfo {author} {\bibfnamefont {S.}~\bibnamefont {Durr}}, \bibinfo {author}
  {\bibfnamefont {Z.}~\bibnamefont {Fodor}}, \bibinfo {author} {\bibfnamefont
  {S.~D.}\ \bibnamefont {Katz}}, \bibinfo {author} {\bibfnamefont
  {S.}~\bibnamefont {Krieg}}, \ and\ \bibinfo {author} {\bibfnamefont {K.~K.}\
  \bibnamefont {Szabo}},\ }\href {\doibase 10.1088/1126-6708/2009/06/088}
  {\bibfield  {journal} {\bibinfo  {journal} {JHEP}\ }\textbf {\bibinfo
  {volume} {06}},\ \bibinfo {pages} {088} (\bibinfo {year} {2009})},\ \Eprint
  {http://arxiv.org/abs/0903.4155} {arXiv:0903.4155 [hep-lat]} \BibitemShut
  {NoStop}%
\bibitem [{\citenamefont {Bazavov}\ \emph {et~al.}(2009)\citenamefont {Bazavov}
  \emph {et~al.}}]{Bazavov:2009zn}%
  \BibitemOpen
  \bibfield  {author} {\bibinfo {author} {\bibfnamefont {A.}~\bibnamefont
  {Bazavov}} \emph {et~al.},\ }\href {\doibase 10.1103/PhysRevD.80.014504}
  {\bibfield  {journal} {\bibinfo  {journal} {Phys. Rev. D}\ }\textbf {\bibinfo
  {volume} {80}},\ \bibinfo {pages} {014504} (\bibinfo {year} {2009})},\
  \Eprint {http://arxiv.org/abs/0903.4379} {arXiv:0903.4379 [hep-lat]}
  \BibitemShut {NoStop}%
\bibitem [{\citenamefont {Cheng}\ \emph {et~al.}(2008)\citenamefont {Cheng}
  \emph {et~al.}}]{Cheng:2007jq}%
  \BibitemOpen
  \bibfield  {author} {\bibinfo {author} {\bibfnamefont {M.}~\bibnamefont
  {Cheng}} \emph {et~al.},\ }\href {\doibase 10.1103/PhysRevD.77.014511}
  {\bibfield  {journal} {\bibinfo  {journal} {Phys. Rev. D}\ }\textbf {\bibinfo
  {volume} {77}},\ \bibinfo {pages} {014511} (\bibinfo {year} {2008})},\
  \Eprint {http://arxiv.org/abs/0710.0354} {arXiv:0710.0354 [hep-lat]}
  \BibitemShut {NoStop}%
\bibitem [{\citenamefont {Muroya}\ \emph {et~al.}(2003)\citenamefont {Muroya},
  \citenamefont {Nakamura}, \citenamefont {Nonaka},\ and\ \citenamefont
  {Takaishi}}]{Muroya:2003qs}%
  \BibitemOpen
  \bibfield  {author} {\bibinfo {author} {\bibfnamefont {S.}~\bibnamefont
  {Muroya}}, \bibinfo {author} {\bibfnamefont {A.}~\bibnamefont {Nakamura}},
  \bibinfo {author} {\bibfnamefont {C.}~\bibnamefont {Nonaka}}, \ and\ \bibinfo
  {author} {\bibfnamefont {T.}~\bibnamefont {Takaishi}},\ }\href {\doibase
  10.1143/PTP.110.615} {\bibfield  {journal} {\bibinfo  {journal} {Prog. Theor.
  Phys.}\ }\textbf {\bibinfo {volume} {110}},\ \bibinfo {pages} {615} (\bibinfo
  {year} {2003})},\ \Eprint {http://arxiv.org/abs/hep-lat/0306031}
  {arXiv:hep-lat/0306031} \BibitemShut {NoStop}%
\bibitem [{\citenamefont {Nambu}\ and\ \citenamefont
  {Jona-Lasinio}(1961{\natexlab{a}})}]{Nambu:1961fr}%
  \BibitemOpen
  \bibfield  {author} {\bibinfo {author} {\bibfnamefont {Y.}~\bibnamefont
  {Nambu}}\ and\ \bibinfo {author} {\bibfnamefont {G.}~\bibnamefont
  {Jona-Lasinio}},\ }\href {\doibase 10.1103/PhysRev.124.246} {\bibfield
  {journal} {\bibinfo  {journal} {Phys. Rev.}\ }\textbf {\bibinfo {volume}
  {124}},\ \bibinfo {pages} {246} (\bibinfo {year}
  {1961}{\natexlab{a}})}\BibitemShut {NoStop}%
\bibitem [{\citenamefont {Nambu}\ and\ \citenamefont
  {Jona-Lasinio}(1961{\natexlab{b}})}]{Nambu:1961tp}%
  \BibitemOpen
  \bibfield  {author} {\bibinfo {author} {\bibfnamefont {Y.}~\bibnamefont
  {Nambu}}\ and\ \bibinfo {author} {\bibfnamefont {G.}~\bibnamefont
  {Jona-Lasinio}},\ }\href {\doibase 10.1103/PhysRev.122.345} {\bibfield
  {journal} {\bibinfo  {journal} {Phys. Rev.}\ }\textbf {\bibinfo {volume}
  {122}},\ \bibinfo {pages} {345} (\bibinfo {year}
  {1961}{\natexlab{b}})}\BibitemShut {NoStop}%
\bibitem [{\citenamefont {Klevansky}(1992)}]{Klevansky:1992qe}%
  \BibitemOpen
  \bibfield  {author} {\bibinfo {author} {\bibfnamefont {S.~P.}\ \bibnamefont
  {Klevansky}},\ }\href {\doibase 10.1103/RevModPhys.64.649} {\bibfield
  {journal} {\bibinfo  {journal} {Rev. Mod. Phys.}\ }\textbf {\bibinfo {volume}
  {64}},\ \bibinfo {pages} {649} (\bibinfo {year} {1992})}\BibitemShut
  {NoStop}%
\bibitem [{\citenamefont {Vogl}\ and\ \citenamefont
  {Weise}(1991)}]{Vogl:1991qt}%
  \BibitemOpen
  \bibfield  {author} {\bibinfo {author} {\bibfnamefont {U.}~\bibnamefont
  {Vogl}}\ and\ \bibinfo {author} {\bibfnamefont {W.}~\bibnamefont {Weise}},\
  }\href {\doibase 10.1016/0146-6410(91)90005-9} {\bibfield  {journal}
  {\bibinfo  {journal} {Prog. Part. Nucl. Phys.}\ }\textbf {\bibinfo {volume}
  {27}},\ \bibinfo {pages} {195} (\bibinfo {year} {1991})}\BibitemShut
  {NoStop}%
\bibitem [{\citenamefont {Buballa}(2005)}]{Buballa:2003qv}%
  \BibitemOpen
  \bibfield  {author} {\bibinfo {author} {\bibfnamefont {M.}~\bibnamefont
  {Buballa}},\ }\href {\doibase 10.1016/j.physrep.2004.11.004} {\bibfield
  {journal} {\bibinfo  {journal} {Phys. Rept.}\ }\textbf {\bibinfo {volume}
  {407}},\ \bibinfo {pages} {205} (\bibinfo {year} {2005})},\ \Eprint
  {http://arxiv.org/abs/hep-ph/0402234} {arXiv:hep-ph/0402234} \BibitemShut
  {NoStop}%
\bibitem [{\citenamefont {Volkov}\ and\ \citenamefont
  {Radzhabov}(2006)}]{Volkov:2005kw}%
  \BibitemOpen
  \bibfield  {author} {\bibinfo {author} {\bibfnamefont {M.~K.}\ \bibnamefont
  {Volkov}}\ and\ \bibinfo {author} {\bibfnamefont {A.~E.}\ \bibnamefont
  {Radzhabov}},\ }\href {\doibase 10.1070/PU2006v049n06ABEH005905} {\bibfield
  {journal} {\bibinfo  {journal} {Phys. Usp.}\ }\textbf {\bibinfo {volume}
  {49}},\ \bibinfo {pages} {551} (\bibinfo {year} {2006})},\ \Eprint
  {http://arxiv.org/abs/hep-ph/0508263} {arXiv:hep-ph/0508263} \BibitemShut
  {NoStop}%
\bibitem [{\citenamefont {Lang}\ and\ \citenamefont
  {Weise}(2014)}]{Lang:2013lla}%
  \BibitemOpen
  \bibfield  {author} {\bibinfo {author} {\bibfnamefont {R.}~\bibnamefont
  {Lang}}\ and\ \bibinfo {author} {\bibfnamefont {W.}~\bibnamefont {Weise}},\
  }\href {\doibase 10.1140/epja/i2014-14063-4} {\bibfield  {journal} {\bibinfo
  {journal} {Eur. Phys. J. A}\ }\textbf {\bibinfo {volume} {50}},\ \bibinfo
  {pages} {63} (\bibinfo {year} {2014})},\ \Eprint
  {http://arxiv.org/abs/1311.4628} {arXiv:1311.4628 [hep-ph]} \BibitemShut
  {NoStop}%
\bibitem [{\citenamefont {Lang}\ \emph {et~al.}(2015)\citenamefont {Lang},
  \citenamefont {Kaiser},\ and\ \citenamefont {Weise}}]{Lang:2015nca}%
  \BibitemOpen
  \bibfield  {author} {\bibinfo {author} {\bibfnamefont {R.}~\bibnamefont
  {Lang}}, \bibinfo {author} {\bibfnamefont {N.}~\bibnamefont {Kaiser}}, \ and\
  \bibinfo {author} {\bibfnamefont {W.}~\bibnamefont {Weise}},\ }\href
  {\doibase 10.1140/epja/i2015-15127-7} {\bibfield  {journal} {\bibinfo
  {journal} {Eur. Phys. J. A}\ }\textbf {\bibinfo {volume} {51}},\ \bibinfo
  {pages} {127} (\bibinfo {year} {2015})},\ \Eprint
  {http://arxiv.org/abs/1506.02459} {arXiv:1506.02459 [hep-ph]} \BibitemShut
  {NoStop}%
\bibitem [{\citenamefont {Iwasaki}\ \emph {et~al.}(2008)\citenamefont
  {Iwasaki}, \citenamefont {Ohnishi},\ and\ \citenamefont
  {Fukutome}}]{Iwasaki:2007iv}%
  \BibitemOpen
  \bibfield  {author} {\bibinfo {author} {\bibfnamefont {M.}~\bibnamefont
  {Iwasaki}}, \bibinfo {author} {\bibfnamefont {H.}~\bibnamefont {Ohnishi}}, \
  and\ \bibinfo {author} {\bibfnamefont {T.}~\bibnamefont {Fukutome}},\ }\href
  {\doibase 10.1088/0954-3899/35/3/035003} {\bibfield  {journal} {\bibinfo
  {journal} {J. Phys. G}\ }\textbf {\bibinfo {volume} {35}},\ \bibinfo {pages}
  {035003} (\bibinfo {year} {2008})},\ \Eprint
  {http://arxiv.org/abs/hep-ph/0703271} {arXiv:hep-ph/0703271} \BibitemShut
  {NoStop}%
\bibitem [{\citenamefont {Alberico}\ \emph {et~al.}(2008)\citenamefont
  {Alberico}, \citenamefont {Chiacchiera}, \citenamefont {Hansen},
  \citenamefont {Molinari},\ and\ \citenamefont {Nardi}}]{Alberico:2007fu}%
  \BibitemOpen
  \bibfield  {author} {\bibinfo {author} {\bibfnamefont {W.~M.}\ \bibnamefont
  {Alberico}}, \bibinfo {author} {\bibfnamefont {S.}~\bibnamefont
  {Chiacchiera}}, \bibinfo {author} {\bibfnamefont {H.}~\bibnamefont {Hansen}},
  \bibinfo {author} {\bibfnamefont {A.}~\bibnamefont {Molinari}}, \ and\
  \bibinfo {author} {\bibfnamefont {M.}~\bibnamefont {Nardi}},\ }\href
  {\doibase 10.1140/epja/i2008-10648-8} {\bibfield  {journal} {\bibinfo
  {journal} {Eur. Phys. J. A}\ }\textbf {\bibinfo {volume} {38}},\ \bibinfo
  {pages} {97} (\bibinfo {year} {2008})},\ \Eprint
  {http://arxiv.org/abs/0707.4442} {arXiv:0707.4442 [hep-ph]} \BibitemShut
  {NoStop}%
\bibitem [{\citenamefont {Ghosh}\ \emph {et~al.}(2016)\citenamefont {Ghosh},
  \citenamefont {Peixoto}, \citenamefont {Roy}, \citenamefont {Serna},\ and\
  \citenamefont {Krein}}]{Ghosh:2015mda}%
  \BibitemOpen
  \bibfield  {author} {\bibinfo {author} {\bibfnamefont {S.}~\bibnamefont
  {Ghosh}}, \bibinfo {author} {\bibfnamefont {T.~C.}\ \bibnamefont {Peixoto}},
  \bibinfo {author} {\bibfnamefont {V.}~\bibnamefont {Roy}}, \bibinfo {author}
  {\bibfnamefont {F.~E.}\ \bibnamefont {Serna}}, \ and\ \bibinfo {author}
  {\bibfnamefont {G.~a.}\ \bibnamefont {Krein}},\ }\href {\doibase
  10.1103/PhysRevC.93.045205} {\bibfield  {journal} {\bibinfo  {journal} {Phys.
  Rev. C}\ }\textbf {\bibinfo {volume} {93}},\ \bibinfo {pages} {045205}
  (\bibinfo {year} {2016})},\ \Eprint {http://arxiv.org/abs/1507.08798}
  {arXiv:1507.08798 [nucl-th]} \BibitemShut {NoStop}%
\bibitem [{\citenamefont {Mallik}\ and\ \citenamefont
  {Sarkar}(2016)}]{Mallik:2016anp}%
  \BibitemOpen
  \bibfield  {author} {\bibinfo {author} {\bibfnamefont {S.}~\bibnamefont
  {Mallik}}\ and\ \bibinfo {author} {\bibfnamefont {S.}~\bibnamefont
  {Sarkar}},\ }\href {\doibase 10.1017/9781316535585} {\emph {\bibinfo {title}
  {{Hadrons at Finite Temperature}}}}\ (\bibinfo  {publisher} {Cambridge
  University Press},\ \bibinfo {address} {Cambridge},\ \bibinfo {year}
  {2016})\BibitemShut {NoStop}%
\bibitem [{\citenamefont {Quack}\ and\ \citenamefont
  {Klevansky}(1994)}]{Quack:1993ie}%
  \BibitemOpen
  \bibfield  {author} {\bibinfo {author} {\bibfnamefont {E.}~\bibnamefont
  {Quack}}\ and\ \bibinfo {author} {\bibfnamefont {S.~P.}\ \bibnamefont
  {Klevansky}},\ }\href {\doibase 10.1103/PhysRevC.49.3283} {\bibfield
  {journal} {\bibinfo  {journal} {Phys. Rev. C}\ }\textbf {\bibinfo {volume}
  {49}},\ \bibinfo {pages} {3283} (\bibinfo {year} {1994})}\BibitemShut
  {NoStop}%
\bibitem [{\citenamefont {Harutyunyan}\ \emph {et~al.}(2017)\citenamefont
  {Harutyunyan}, \citenamefont {Rischke},\ and\ \citenamefont
  {Sedrakian}}]{Harutyunyan:2017ttz}%
  \BibitemOpen
  \bibfield  {author} {\bibinfo {author} {\bibfnamefont {A.}~\bibnamefont
  {Harutyunyan}}, \bibinfo {author} {\bibfnamefont {D.~H.}\ \bibnamefont
  {Rischke}}, \ and\ \bibinfo {author} {\bibfnamefont {A.}~\bibnamefont
  {Sedrakian}},\ }\href {\doibase 10.1103/PhysRevD.95.114021} {\bibfield
  {journal} {\bibinfo  {journal} {Phys. Rev. D}\ }\textbf {\bibinfo {volume}
  {95}},\ \bibinfo {pages} {114021} (\bibinfo {year} {2017})},\ \Eprint
  {http://arxiv.org/abs/1702.04291} {arXiv:1702.04291 [nucl-th]} \BibitemShut
  {NoStop}%
\bibitem [{\citenamefont {Hosoya}\ \emph {et~al.}(1984)\citenamefont {Hosoya},
  \citenamefont {Sakagami},\ and\ \citenamefont {Takao}}]{Hosoya:1983id}%
  \BibitemOpen
  \bibfield  {author} {\bibinfo {author} {\bibfnamefont {A.}~\bibnamefont
  {Hosoya}}, \bibinfo {author} {\bibfnamefont {M.-a.}\ \bibnamefont
  {Sakagami}}, \ and\ \bibinfo {author} {\bibfnamefont {M.}~\bibnamefont
  {Takao}},\ }\href {\doibase 10.1016/0003-4916(84)90144-1} {\bibfield
  {journal} {\bibinfo  {journal} {Annals Phys.}\ }\textbf {\bibinfo {volume}
  {154}},\ \bibinfo {pages} {229} (\bibinfo {year} {1984})}\BibitemShut
  {NoStop}%
\bibitem [{\citenamefont {Harutyunyan}\ and\ \citenamefont
  {Sedrakian}(2016)}]{Harutyunyan:2016rxm}%
  \BibitemOpen
  \bibfield  {author} {\bibinfo {author} {\bibfnamefont {A.}~\bibnamefont
  {Harutyunyan}}\ and\ \bibinfo {author} {\bibfnamefont {A.}~\bibnamefont
  {Sedrakian}},\ }\href {\doibase 10.1103/PhysRevC.94.025805} {\bibfield
  {journal} {\bibinfo  {journal} {Phys. Rev. C}\ }\textbf {\bibinfo {volume}
  {94}},\ \bibinfo {pages} {025805} (\bibinfo {year} {2016})},\ \Eprint
  {http://arxiv.org/abs/1605.07612} {arXiv:1605.07612 [astro-ph.HE]}
  \BibitemShut {NoStop}%
\bibitem [{\citenamefont {Fernandez-Fraile}\ and\ \citenamefont
  {Gomez~Nicola}(2006)}]{Fernandez-Fraile:2005bew}%
  \BibitemOpen
  \bibfield  {author} {\bibinfo {author} {\bibfnamefont {D.}~\bibnamefont
  {Fernandez-Fraile}}\ and\ \bibinfo {author} {\bibfnamefont {A.}~\bibnamefont
  {Gomez~Nicola}},\ }\href {\doibase 10.1103/PhysRevD.73.045025} {\bibfield
  {journal} {\bibinfo  {journal} {Phys. Rev. D}\ }\textbf {\bibinfo {volume}
  {73}},\ \bibinfo {pages} {045025} (\bibinfo {year} {2006})},\ \Eprint
  {http://arxiv.org/abs/hep-ph/0512283} {arXiv:hep-ph/0512283} \BibitemShut
  {NoStop}%
\bibitem [{\citenamefont {Jeon}(1995)}]{Jeon:1994if}%
  \BibitemOpen
  \bibfield  {author} {\bibinfo {author} {\bibfnamefont {S.}~\bibnamefont
  {Jeon}},\ }\href {\doibase 10.1103/PhysRevD.52.3591} {\bibfield  {journal}
  {\bibinfo  {journal} {Phys. Rev. D}\ }\textbf {\bibinfo {volume} {52}},\
  \bibinfo {pages} {3591} (\bibinfo {year} {1995})},\ \Eprint
  {http://arxiv.org/abs/hep-ph/9409250} {arXiv:hep-ph/9409250} \BibitemShut
  {NoStop}%
\bibitem [{\citenamefont {Jeon}\ and\ \citenamefont
  {Yaffe}(1996)}]{Jeon:1995zm}%
  \BibitemOpen
  \bibfield  {author} {\bibinfo {author} {\bibfnamefont {S.}~\bibnamefont
  {Jeon}}\ and\ \bibinfo {author} {\bibfnamefont {L.~G.}\ \bibnamefont
  {Yaffe}},\ }\href {\doibase 10.1103/PhysRevD.53.5799} {\bibfield  {journal}
  {\bibinfo  {journal} {Phys. Rev. D}\ }\textbf {\bibinfo {volume} {53}},\
  \bibinfo {pages} {5799} (\bibinfo {year} {1996})},\ \Eprint
  {http://arxiv.org/abs/hep-ph/9512263} {arXiv:hep-ph/9512263} \BibitemShut
  {NoStop}%
\bibitem [{\citenamefont {Valle~Basagoiti}(2002)}]{ValleBasagoiti:2002ir}%
  \BibitemOpen
  \bibfield  {author} {\bibinfo {author} {\bibfnamefont {M.~A.}\ \bibnamefont
  {Valle~Basagoiti}},\ }\href {\doibase 10.1103/PhysRevD.66.045005} {\bibfield
  {journal} {\bibinfo  {journal} {Phys. Rev. D}\ }\textbf {\bibinfo {volume}
  {66}},\ \bibinfo {pages} {045005} (\bibinfo {year} {2002})},\ \Eprint
  {http://arxiv.org/abs/hep-ph/0204334} {arXiv:hep-ph/0204334} \BibitemShut
  {NoStop}%
\bibitem [{\citenamefont {Carrington}\ \emph {et~al.}(2001)\citenamefont
  {Carrington}, \citenamefont {Hou},\ and\ \citenamefont
  {Kobes}}]{Carrington:2001ms}%
  \BibitemOpen
  \bibfield  {author} {\bibinfo {author} {\bibfnamefont {M.~E.}\ \bibnamefont
  {Carrington}}, \bibinfo {author} {\bibfnamefont {D.-f.}\ \bibnamefont {Hou}},
  \ and\ \bibinfo {author} {\bibfnamefont {R.}~\bibnamefont {Kobes}},\ }\href
  {\doibase 10.1103/PhysRevD.64.025001} {\bibfield  {journal} {\bibinfo
  {journal} {Phys. Rev. D}\ }\textbf {\bibinfo {volume} {64}},\ \bibinfo
  {pages} {025001} (\bibinfo {year} {2001})},\ \Eprint
  {http://arxiv.org/abs/hep-ph/0102256} {arXiv:hep-ph/0102256} \BibitemShut
  {NoStop}%
\bibitem [{\citenamefont {Chakraborty}\ and\ \citenamefont
  {Kapusta}(2011)}]{Chakraborty:2010fr}%
  \BibitemOpen
  \bibfield  {author} {\bibinfo {author} {\bibfnamefont {P.}~\bibnamefont
  {Chakraborty}}\ and\ \bibinfo {author} {\bibfnamefont {J.~I.}\ \bibnamefont
  {Kapusta}},\ }\href {\doibase 10.1103/PhysRevC.83.014906} {\bibfield
  {journal} {\bibinfo  {journal} {Phys. Rev. C}\ }\textbf {\bibinfo {volume}
  {83}},\ \bibinfo {pages} {014906} (\bibinfo {year} {2011})},\ \Eprint
  {http://arxiv.org/abs/1006.0257} {arXiv:1006.0257 [nucl-th]} \BibitemShut
  {NoStop}%
\bibitem [{\citenamefont {Zhuang}\ \emph {et~al.}(1995)\citenamefont {Zhuang},
  \citenamefont {Hufner}, \citenamefont {Klevansky},\ and\ \citenamefont
  {Neise}}]{Zhuang:1995uf}%
  \BibitemOpen
  \bibfield  {author} {\bibinfo {author} {\bibfnamefont {P.}~\bibnamefont
  {Zhuang}}, \bibinfo {author} {\bibfnamefont {J.}~\bibnamefont {Hufner}},
  \bibinfo {author} {\bibfnamefont {S.~P.}\ \bibnamefont {Klevansky}}, \ and\
  \bibinfo {author} {\bibfnamefont {L.}~\bibnamefont {Neise}},\ }\href
  {\doibase 10.1103/PhysRevD.51.3728} {\bibfield  {journal} {\bibinfo
  {journal} {Phys. Rev. D}\ }\textbf {\bibinfo {volume} {51}},\ \bibinfo
  {pages} {3728} (\bibinfo {year} {1995})}\BibitemShut {NoStop}%
\bibitem [{\citenamefont {Schulze}(1995)}]{Schulze:1995rb}%
  \BibitemOpen
  \bibfield  {author} {\bibinfo {author} {\bibfnamefont {H.~J.}\ \bibnamefont
  {Schulze}},\ }\href {\doibase 10.1088/0954-3899/21/2/006} {\bibfield
  {journal} {\bibinfo  {journal} {J. Phys. G}\ }\textbf {\bibinfo {volume}
  {21}},\ \bibinfo {pages} {185} (\bibinfo {year} {1995})}\BibitemShut
  {NoStop}%
\bibitem [{\citenamefont {Bernard}\ \emph {et~al.}(1991)\citenamefont
  {Bernard}, \citenamefont {Meissner}, \citenamefont {Blin},\ and\
  \citenamefont {Hiller}}]{Bernard:1990ye}%
  \BibitemOpen
  \bibfield  {author} {\bibinfo {author} {\bibfnamefont {V.}~\bibnamefont
  {Bernard}}, \bibinfo {author} {\bibfnamefont {U.~G.}\ \bibnamefont
  {Meissner}}, \bibinfo {author} {\bibfnamefont {A.}~\bibnamefont {Blin}}, \
  and\ \bibinfo {author} {\bibfnamefont {B.}~\bibnamefont {Hiller}},\ }\href
  {\doibase 10.1016/0370-2693(91)91749-L} {\bibfield  {journal} {\bibinfo
  {journal} {Phys. Lett. B}\ }\textbf {\bibinfo {volume} {253}},\ \bibinfo
  {pages} {443} (\bibinfo {year} {1991})}\BibitemShut {NoStop}%
\bibitem [{\citenamefont {Kohyama}\ \emph {et~al.}(2015)\citenamefont
  {Kohyama}, \citenamefont {Kimura},\ and\ \citenamefont
  {Inagaki}}]{Kohyama:2015hix}%
  \BibitemOpen
  \bibfield  {author} {\bibinfo {author} {\bibfnamefont {H.}~\bibnamefont
  {Kohyama}}, \bibinfo {author} {\bibfnamefont {D.}~\bibnamefont {Kimura}}, \
  and\ \bibinfo {author} {\bibfnamefont {T.}~\bibnamefont {Inagaki}},\ }\href
  {\doibase 10.1016/j.nuclphysb.2015.05.015} {\bibfield  {journal} {\bibinfo
  {journal} {Nucl. Phys. B}\ }\textbf {\bibinfo {volume} {896}},\ \bibinfo
  {pages} {682} (\bibinfo {year} {2015})},\ \Eprint
  {http://arxiv.org/abs/1501.00449} {arXiv:1501.00449 [hep-ph]} \BibitemShut
  {NoStop}%
\bibitem [{\citenamefont {Gatto}\ and\ \citenamefont
  {Ruggieri}(2012)}]{Gatto:2011wc}%
  \BibitemOpen
  \bibfield  {author} {\bibinfo {author} {\bibfnamefont {R.}~\bibnamefont
  {Gatto}}\ and\ \bibinfo {author} {\bibfnamefont {M.}~\bibnamefont
  {Ruggieri}},\ }\href {\doibase 10.1103/PhysRevD.85.054013} {\bibfield
  {journal} {\bibinfo  {journal} {Phys. Rev. D}\ }\textbf {\bibinfo {volume}
  {85}},\ \bibinfo {pages} {054013} (\bibinfo {year} {2012})},\ \Eprint
  {http://arxiv.org/abs/1110.4904} {arXiv:1110.4904 [hep-ph]} \BibitemShut
  {NoStop}%
\bibitem [{\citenamefont {Chernodub}\ and\ \citenamefont
  {Nedelin}(2011)}]{Chernodub:2011fr}%
  \BibitemOpen
  \bibfield  {author} {\bibinfo {author} {\bibfnamefont {M.~N.}\ \bibnamefont
  {Chernodub}}\ and\ \bibinfo {author} {\bibfnamefont {A.~S.}\ \bibnamefont
  {Nedelin}},\ }\href {\doibase 10.1103/PhysRevD.83.105008} {\bibfield
  {journal} {\bibinfo  {journal} {Phys. Rev. D}\ }\textbf {\bibinfo {volume}
  {83}},\ \bibinfo {pages} {105008} (\bibinfo {year} {2011})},\ \Eprint
  {http://arxiv.org/abs/1102.0188} {arXiv:1102.0188 [hep-ph]} \BibitemShut
  {NoStop}%
\bibitem [{\citenamefont {Ruggieri}(2011)}]{Ruggieri:2011xc}%
  \BibitemOpen
  \bibfield  {author} {\bibinfo {author} {\bibfnamefont {M.}~\bibnamefont
  {Ruggieri}},\ }\href {\doibase 10.1103/PhysRevD.84.014011} {\bibfield
  {journal} {\bibinfo  {journal} {Phys. Rev. D}\ }\textbf {\bibinfo {volume}
  {84}},\ \bibinfo {pages} {014011} (\bibinfo {year} {2011})},\ \Eprint
  {http://arxiv.org/abs/1103.6186} {arXiv:1103.6186 [hep-ph]} \BibitemShut
  {NoStop}%
\bibitem [{\citenamefont {Yu}\ \emph {et~al.}(2016)\citenamefont {Yu},
  \citenamefont {Liu},\ and\ \citenamefont {Huang}}]{Yu:2015hym}%
  \BibitemOpen
  \bibfield  {author} {\bibinfo {author} {\bibfnamefont {L.}~\bibnamefont
  {Yu}}, \bibinfo {author} {\bibfnamefont {H.}~\bibnamefont {Liu}}, \ and\
  \bibinfo {author} {\bibfnamefont {M.}~\bibnamefont {Huang}},\ }\href
  {\doibase 10.1103/PhysRevD.94.014026} {\bibfield  {journal} {\bibinfo
  {journal} {Phys. Rev. D}\ }\textbf {\bibinfo {volume} {94}},\ \bibinfo
  {pages} {014026} (\bibinfo {year} {2016})},\ \Eprint
  {http://arxiv.org/abs/1511.03073} {arXiv:1511.03073 [hep-ph]} \BibitemShut
  {NoStop}%
\bibitem [{\citenamefont {Ruggieri}\ and\ \citenamefont
  {Peng}(2016{\natexlab{b}})}]{Ruggieri:2016ejz}%
  \BibitemOpen
  \bibfield  {author} {\bibinfo {author} {\bibfnamefont {M.}~\bibnamefont
  {Ruggieri}}\ and\ \bibinfo {author} {\bibfnamefont {G.~X.}\ \bibnamefont
  {Peng}},\ }\href {\doibase 10.1088/0954-3899/43/12/125101} {\bibfield
  {journal} {\bibinfo  {journal} {J. Phys. G}\ }\textbf {\bibinfo {volume}
  {43}},\ \bibinfo {pages} {125101} (\bibinfo {year} {2016}{\natexlab{b}})},\
  \Eprint {http://arxiv.org/abs/1602.05250} {arXiv:1602.05250 [hep-ph]}
  \BibitemShut {NoStop}%
\bibitem [{\citenamefont {Farias}\ \emph {et~al.}(2017)\citenamefont {Farias},
  \citenamefont {Timoteo}, \citenamefont {Avancini}, \citenamefont {Pinto},\
  and\ \citenamefont {Krein}}]{Farias:2016gmy}%
  \BibitemOpen
  \bibfield  {author} {\bibinfo {author} {\bibfnamefont {R.~L.~S.}\
  \bibnamefont {Farias}}, \bibinfo {author} {\bibfnamefont {V.~S.}\
  \bibnamefont {Timoteo}}, \bibinfo {author} {\bibfnamefont {S.~S.}\
  \bibnamefont {Avancini}}, \bibinfo {author} {\bibfnamefont {M.~B.}\
  \bibnamefont {Pinto}}, \ and\ \bibinfo {author} {\bibfnamefont
  {G.}~\bibnamefont {Krein}},\ }\href {\doibase 10.1140/epja/i2017-12320-8}
  {\bibfield  {journal} {\bibinfo  {journal} {Eur. Phys. J. A}\ }\textbf
  {\bibinfo {volume} {53}},\ \bibinfo {pages} {101} (\bibinfo {year} {2017})},\
  \Eprint {http://arxiv.org/abs/1603.03847} {arXiv:1603.03847 [hep-ph]}
  \BibitemShut {NoStop}%
\bibitem [{\citenamefont {Ferreira}\ \emph {et~al.}(2014)\citenamefont
  {Ferreira}, \citenamefont {Costa}, \citenamefont {Louren\c{c}o},
  \citenamefont {Frederico},\ and\ \citenamefont
  {Provid\^encia}}]{Ferreira:2014kpa}%
  \BibitemOpen
  \bibfield  {author} {\bibinfo {author} {\bibfnamefont {M.}~\bibnamefont
  {Ferreira}}, \bibinfo {author} {\bibfnamefont {P.}~\bibnamefont {Costa}},
  \bibinfo {author} {\bibfnamefont {O.}~\bibnamefont {Louren\c{c}o}}, \bibinfo
  {author} {\bibfnamefont {T.}~\bibnamefont {Frederico}}, \ and\ \bibinfo
  {author} {\bibfnamefont {C.}~\bibnamefont {Provid\^encia}},\ }\href {\doibase
  10.1103/PhysRevD.89.116011} {\bibfield  {journal} {\bibinfo  {journal} {Phys.
  Rev. D}\ }\textbf {\bibinfo {volume} {89}},\ \bibinfo {pages} {116011}
  (\bibinfo {year} {2014})},\ \Eprint {http://arxiv.org/abs/1404.5577}
  {arXiv:1404.5577 [hep-ph]} \BibitemShut {NoStop}%
\end{thebibliography}%

\end{document}